\documentclass[onecolumn]{aastex6}

\usepackage{commands}
\usepackage{amsmath}
\usepackage{amssymb}
\usepackage[retainorgcmds]{IEEEtrantools}
\usepackage{savesym}
\savesymbol{tablenum}
\usepackage{siunitx}
\restoresymbol{SIX}{tablenum}
\usepackage{siunitx}     
\begin{document}

\title{A reduced-order NLTE kinetic model for radiating plasmas of outer envelopes of stellar atmospheres}

\author{Alessandro Munaf\`{o}\altaffilmark{1}}
\affil{Aerospace Engineering Department, University of Illinois at Urbana-Champaign, \\
206A Talbot Lab., 104 S. Wright St. Urbana, IL 61801, USA}

\author{Nagi N. Mansour\altaffilmark{2}}
\affil{NASA Ames Research Center, Moffett Field, 94035 CA, USA}

\author{Marco Panesi\altaffilmark{3}}
\affil{Aerospace Engineering Department, University of Illinois at Urbana-Champaign, \\
306 Talbot Lab., 104 S. Wright St. Urbana, IL 61801, USA}

\altaffiltext{1}{Post-Doctoral Research Associate, Aerospace Engineering Department, University of Illinois at Urbana-Champaign, \\
206A Talbot Lab., 104 S. Wright St. Urbana, IL 61801, USA, munafo@illinois.edu}
\altaffiltext{2}{Branch Chief, NASA Ames Research Center, Moffett Field, 94035 CA, USA, nagi.n.mansour@nasa.gov}
\altaffiltext{3}{Assistant Professor, Aerospace Engineering Department, University of Illinois at Urbana-Champaign, \\
306 Talbot Lab., 104 S. Wright St. Urbana, IL 61801, USA, m.panesi@illinois.edu}

\begin{abstract}
The present work proposes a self-consistent reduced-order NLTE kinetic model for radiating plasmas such as are found in the outer layers of stellar atmospheres. Starting from the most up-to-date 
set of \emph{ab-initio} and experimental data, the highly complex collisional-radiative kinetic mechanism is simplified by lumping the bound energy states in groups. Different grouping strategies are investigated, such as uniform and Maxwell-Boltzmann.
The reduced set of governing equations for the material gas and the radiation field is obtained based on a moment method. Applications consider the steady flow across a shock wave in partially ionized hydrogen. The results show that adopting a Maxwell-Boltzmann grouping allows, on the one hand, for a substantial reduction of the number of unknowns and, on the other, to maintain accuracy for both gas and radiation quantities. It is observed that, when neglecting line radiation, the use of two groups already leads to a very accurate resolution of the photo-ionization precursor, internal relaxation and radiative cooling regions. The inclusion of line radiation requires adopting just one additional group to account for optically thin losses in the $\alpha$, $\beta$ and $\gamma$ lines of the Balmer and Paschen series. This trend has been observed for a wide range of shock wave velocities.                   
\end{abstract}
\section{Introduction}\label{sec:intro}
Stellar atmospheres are the outer gaseous layers of stars and are the locus of a broad domain of physical phenomena such as shock waves, winds, flares, coronal mass ejections and magnetic reconnection \citep{Priest_book,Mihalas_RadHyd_1984}. The accurate prediction of the behavior of stellar atmospheres is challenging for a variety of reasons. First, stellar atmospheres are dynamical objects whose state is continuously changing with time. This was recognized since the early days of observations from the asymmetric profiles of atomic lines in observed spectra, which revealed the existence of massive gas motions \citep{Priest_book}. Second, the low values of pressure and density are such that collisional rates among gas particles (\emph{e.g.} atoms, molecules, free-electrons) are not sufficient to ensure Local-Thermodynamic-Equilibrium (LTE) \citep{Mihalas_StAtm_1978}. A further, and major, complexity comes from energy transfer by radiation, which plays a dominant role. Radiative transfer introduces a global coupling between the stellar material at all points in the atmosphere.
 
Within the context of a fluid description, the modeling of stellar atmospheres should be attacked from the governing equations of radiation hydrodynamics. Under conditions where Non-LTE (NLTE) prevails and relativistic effects can be ignored, these equations comprise: the continuity equations for each gaseous species, the global momentum and energy equations, and the radiative transfer equation (\emph{i.e.} kinetic equation for photons) \citep{Oxenius_book,Mihalas_RadHyd_1984}. In the most accurate formulation, each bound state of atomic and molecular components is treated as a separate \emph{pseudo-species} to allow for departures from equilibrium (\emph{i.e.} Maxwell-Boltzmann distribution). This approach is sometimes referred to as State-to-State (StS) approach \citep{Capitelli_book,Kustova_book}. The radiation field is treated based on a line-by-line method with the purpose of capturing the finest details of radiative processes leading to line and continuous spectra. It is immediately clear that this accurate modeling strategy, despite its use, with substantial simplifications, in one-dimensional situations \citep{Kneer_1976,Klein_1976,Klein_1978,Carlsson_1992,Carlsson_1995,Carlsson_1997,Carlsson_2002}, becomes unfeasible when moving to multi-dimensional configurations and when accounting for all possible opacity sources, not to mention the coupling with electro-magnetic fields. 

The present work is the first step of a long-term effort aimed at developing reduced-order models for astrophysical plasmas, with the final goal of enabling accurate predictions in unsteady NLTE multi-dimensional simulations. In this paper a self-consistent and systematic method to reduce the complexity of the NLTE collisional-radiative kinetics is proposed. The formulation is based on the general Maximum Entropy principle framework recently developed by \cite{Yen_2015}, which has already been applied with success to the study of collisional excitation, dissociation and ionization in atomic and molecular gases \citep{Panesi_Lani_CR_2013,Munafo_PRE_2014,Yen_2015,Munafo_POF_2015}. The starting point is the set of StS governing equations for a two-temperature plasma, where the mass and energy source terms due to collisional and radiative processes are obtained self-consistently from the Boltzmann equation (\emph{i.e.} kinetic equation for gas particles) \citep{Oxenius_book}. Kinetic and thermodynamic data are taken from the most recent and accurate \emph{ab-initio} calculations and/or experiments. The complexity reduction of the NLTE kinetic mechanism is achieved by lumping the bound energy states in groups. Different grouping strategies are investigated, such as uniform and Maxwell-Boltzmann. The reduced set of governing equations are obtained by taking moments with respect to internal energy of the StS governing equations. The grouping is not applied, for the moment, to the radiative transfer problem which is still treated in a line-by-line fashion. 

Applications consider the steady flow across radiative shock waves in partially ionized hydrogen. Hydrogen was selected as it represents the most abundant element in stellar atmospheres and the inter-stellar medium \citep{Mihalas_StAtm_1978}. A standing shock was chosen as testcase since: i) its numerical calculation requires much less computational time compared to time-dependent simulations \citep{Carlsson_1992}, and ii) shock waves are common in astrophysical flows and play important roles, for instance, in pulsating stars, accretion disks and the chromosphere of the Sun \citep{Priest_book}. Hence, their accurate prediction is of great interest, especially for comparison with observations. At the same time, if a reduced-order model already performs poorly when applied to a standing shock, it is almost certain that the same situation (if not worse) will occur under time-dependent conditions.          

Radiative shock waves have been investigated for a long time (and still are) and were among the first fluid mechanical benchmarks to be studied based on the methods of NLTE radiation hydrodynamics \citep{Mihalas_RadHyd_1984}. Some examples are: the early works by \cite{Skalafuris_ApJ_1963}, \cite{Skalafuris_ApJ_1965,Skalafuris_ApJ_1968,Skalafuris_ApJ_1968_II}, \cite{Murty_JQSRT_1971}, Clarke and coworkers (\citeyear{Clarke_Ferrari_POF_1965},\citeyear{Clarke_POF_1971},\citeyear{Foley_POF_1973}), and the more recent papers by \cite{Fadeyev_AA_1998,Fadeyev_AA_2000,Fadeyev_AA_2001}, \cite{Fadeyev_AA_2002}, \cite{Panesi_JTHT_2009,Panesi_JTHT_2011}, \cite{Kapper_2011_1,Kapper_2011_2} and \cite{Dammando_2013}. Within the context of the inviscid flow approximation, the structure of a radiative shock wave in an atomic plasma can be, in general, subdivided in three distinct regions: i) a radiative precursor ahead of the gasdynamic \emph{jump}, ii) an internal relaxation region behind the shock dominated by collisional excitation and ionization, and iii) a radiative relaxation (or cooling) region where the temperature decreases due to optically thin radiation produced, for instance, by radiative recombination. The radiative precursor can be further divided in: i) far precursor and ii) near precursor \citep{Foley_POF_1973,Liberman_Book}. In the former non-equilibrium excitation may occur due to absorption of resonant radiation in atomic line wings. The  latter is characterized instead by photo-ionization from the ground-state (\emph{e.g.} Lyman continuum) by photons emitted in the hot gas layers of the radiative cooling region which propagate upstream. The extent (and also the existence) of the aforementioned zones strongly depends on the shock velocity, and the gas pressure and temperature in the free-stream \citep{Foley_POF_1973}.        

The present paper is structured as follows. Section \ref{sec:StS} describes the State-to-State model for partially ionized hydrogen. Section \ref{sec:Red} introdducess the reduced-order modeling technique with the aim of simplifying the complexity of the NLTE kinetic mechanism. The numerical method to solve the material gas and the radiation field governing equations is outlined in Sec. \ref{sec:Num}. Computational results are discussed and analyzed in Sec. \ref{sec:Res}. Conclusions are summarized in Sec. \ref{sec:Concl}. MKS units are used throughout the manuscript.   
\section{State-to-State physico-chemical modeling}\label{sec:StS}
This Section describes the State-to-State (StS) model for partially ionized hydrogen adopted in this work. Section \ref{sec:StSAss} introduces the notation used throughout the manuscript and the main simplifying assumptions. The computation of thermodynamic properties is discussed in Sec. \ref{sec:StSThermo}. Section \ref{sec:StSKin} describes the set of kinetic processes (collisional and radiative) and the related calculation of mass/energy production terms for the material gas and emission/absorption coefficients for the radiation field. The governing equations for the problem under investigation (\emph{i.e.} radiative shock-wave in a plane-parallel medium) are given in Sec. \ref{sec:StSEqs}.

Thermodynamic and kinetic data (\emph{e.g} energy levels, oscillator strengths, Einstein coefficients) have been taken from the NIST database for atomic lines and energy levels \citep{Kramida_2010,NIST_ASD}. The former provides data for \num{94} hydrogen bound energy states. The total number of atomic lines (with transition probabilities) is \num{374}, accounting for electric dipole, electric quadrupole, and magnetic dipole transitions \citep{Herzberg_book}.
\subsection{Notation and assumptions}\label{sec:StSAss}
The material gas under investigation is made of three chemical components: free-electrons $\text{e}^-$, hydrogen atoms $\text{H}$ and protons $\text{H}^+$. These components constitute the set $\mathcal{C} = \{\text{e}^-, \, \text{H}, \, \text{H}^+ \}$. The heavy-particle components (\emph{i.e.} hydrogen atoms and protons) are stored in the subset $\mathcal{C}_{\text{h}} = \{\text{H}, \, \text{H}^+ \}$. The hydrogen bound-states (sorted by increasing energy) are denoted by the global index $i$ and stored in set $\mathcal{I}$. Based on a StS approach, the bound-states $\text{H}(i)$ are treated as separate \emph{pseudo-species} \citep{Bates_1962I,Bates_1962II,Mihalas_RadHyd_1984,Capitelli_book}. The related statistical weights and formation energies are indicated, respectively, by the symbols $\gi{i}$ and $\Ei{i}$. The corresponding quantities for the protons are $g_+ = 1$ and $E_+ = \SI{13.59}{eV}$. The StS species are stored in the \emph{species} set $\mathcal{S} = \{\text{e}^-, \, \text{H}(i), \, \text{H}^+; \, i \in \mathcal{I}\}$. 

In order to make the problem tractable, simplifying assumptions are introduced. For the adopted pressure and temperature conditions, it is possible to use Boltzmann statistics. Each species is modeled as a thermally perfect gas by disregarding collective plasma effects such as pressure ionization and multiple charge-charge interactions \citep{Rogers_PhysRevA_1974,Mihalas_StAtm_1978,Rogers_ApJ1996}. Further, the velocity distribution function of each species is taken to be Maxwellian at its own temperature. For heavy-particles, this temperature is taken to be a common heavy-particle temperature $\Ti{h}$. This assumption is justified in view of i) the efficient energy transfer in collisions between particles with similar masses and ii) the large cross-section for resonant charge transfer in hydrogen-proton collisions (\emph{i.e.} $\text{H} + \text{H}^+ = \text{H}^+ + \text{H}$) \citep{Mihalas_RadHyd_1984}. On the other hand, collisions between heavy-particles and free-electrons are energetically inefficient, due to the large mass disparity. This motivates the introduction of a separate free-electron temperature $\Te$. The adopted StS model is therefore a two-temperature plasma model where thermal non-equilibrium effects between heavy-particles and free-electrons are taken into account through the macroscopic \emph{parameters} $\Ti{h}$ and $\Ti{e}$, respectively \citep{App_Bray_1964,Murty_JQSRT_1971}. In this work, it is further assumed that the plasma is locally neutral. Electromagnetic fields and transport phenomena are also neglected.          
\subsection{Thermodynamics}\label{sec:StSThermo}
The gas pressure is obtained based on Dalton's law, $p = \pe + \ph$, where the free-electron and heavy-particle partial pressures are, respectively, $\pe = \ni{\text{e}} \kb \Ti{e}$ and $\ph = \ni{\text{h}} \kb \Ti{h}$, with the symbol $\kb$ denoting Boltzmann's constant. Quantities $\ni{\text{e}}$ and $\ni{\text{h}}$ stand, respectively, for the free-electron and heavy-particle number densities. The latter is obtained by summing the contributions of the hydrogen bound-states and protons, $\ni{\text{h}} = \sum_{i \in \mathcal{I}} \ni{i} + \np$, where $\np = \ni{\text{e}}$ in view of the charge neutrality assumption introduced in Sec. \ref{sec:StSAss}. 

The heavy-particle and free-electron thermal energy densities are:
\be \label{eq:rhoe_he}
\rho e_{\text{h}} = \fr{3}{2} \ph + \sum_{i \in \mathcal{I}} n_i E_i + \np \Ep, \quad \rho e_{\text{e}} = \fr{3}{2} \pe.
\ee
The gas total energy density is obtained by summing the thermal contribution from free-electrons and heavy-particles, and the kinetic contribution of the gas as a whole, $\rho E = \rho e_{\text{h}} + \rho e_{\text{e}} + \rho \mathbf{v} \cdot \mathbf{v}/2 $, where the vector $\mathbf{v}$ denotes the local flow velocity. The mass density is evaluated based on the species number densities as $\rho = \sum_{s  \in \mathcal{S}} n_s \, m_s$, where the symbol $m_s$ denotes the mass of the species $s \in \mathcal{S}$. The total enthalpy density follows by adding the pressure to the total energy density, $\rho H = \rho E + p$. 

The study of the interaction between matter and radiation requires, in general, accounting for the momentum and energy content of radiation \citep{Oxenius_book}. The former plays an important role, for instance, in stellar interiors where up to $50\%$ of the pressure may be due to radiation \citep{Rogers_ApJ1996}. However, as shown in the work by \cite{Fadeyev_AA_1998,Fadeyev_AA_2000}, the radiant pressure and energy density are negligible compared to the corresponding gas values for the conditions adopted in this work.  
\subsection{Kinetics}\label{sec:StSKin}
The present Section describes the NLTE kinetic model for atomic hydrogen plasmas used in this work. This StS model provides the basis for the reduced-order models developed in Sec. \ref{sec:RedME} and accounts for both collisional and radiative processes. The following subsections provide a description of their modeling. It is worth mentioning that, in the literature, other StS models for hydrogen plasmas exist such as the one developed by \cite{Dammando_2010} and \cite{Colonna_ChemPhys_2012}.  
\subsubsection{Collisional processes} 
Collisional processes account for inelastic/elastic transitions due to collisions between heavy-particles and free-electrons and comprise:
\begin{enumerate}
   \item[i)] excitation (\textsc{e}) and ionization (\textsc{i}) by electron impact:
       \be  \label{eq:Ie}
         \text{H}(i) + \elec \rates{}{} \text{H}(j) + \elec, \quad i < j, \quad  \text{H}(i) + \elec \rates{}{} \text{H}^+ + \elec + \elec, \quad i, \, j \in \mathcal{I},
       \ee
   \item[ii)] elastic (\textsc{el}) energy exchange between heavy-particles and free-electrons.
\end{enumerate}
Excitation and ionization by heavy-particle impact have been neglected since, as shown in Sec. \ref{sec:Res}, the number of free-electrons produced by photo-ionization in the precursor guarantees that the collisional kinetics is dominated by free-electrons in the internal relaxation region. 

The endothermic rate coefficients for collisional excitation and ionization \eqnref{eq:Ie} are computed based on the assumption of stationary heavy-particles and Maxwellian free-electrons (see Sec. \ref{sec:StSAss}):
\begin{IEEEeqnarray}{rCl}
\kf{i}{j}{\textsc{e}}(\Te) & = & \sqrt{\fr{8 \kb \Te}{\pi \me }} \fr{1}{(\kb \Te)^2} \int\limits_{E_{ij}}^{+\infty} \!\! \sigf{i}{j}{\textsc{e}}(\varepsilon) \, \varepsilon \,  \exp\left( -\fr{\varepsilon}{\kb \Te}\right) d \, \varepsilon, \quad i < j, \label{eq:kExe} \\
\kf{i}{}{\textsc{i}}(\Te) & = & \sqrt{\fr{8 \kb \Te}{\pi \me }} \fr{1}{(\kb \Te)^2} \int\limits_{E_{+i}}^{+\infty} \!\! \sigf{i}{}{\textsc{i}}(\varepsilon) \, \varepsilon \,  \exp\left( -\fr{\varepsilon}{\kb \Te}\right) d \, \varepsilon, \label{eq:kIe}
\end{IEEEeqnarray}
$i, \, j \in \mathcal{I}$. Quantities $\sigf{i}{j}{\textsc{e}}$ and $\sigf{i}{}{\textsc{i}}$ denote, respectively, the total cross-section for excitation and ionization, with the corresponding threshold energies being defined as $\Ei{ij} = \Ei{j} - \Ei{i}$ and $\Ei{+i} = \Ep - \Ei{i}$. The symbol $\varepsilon$ stands for the free-electron energy. The exothermic rate coefficients for de-excitation and three-body recombination are computed based on those for excitation and ionization, respectively, by means of micro-reversibility \citep{Oxenius_book}:
\be \label{eq:kdExe}
\fr{\kf{j}{i}{\textsc{e}}(\Te)}{\kf{i}{j}{\textsc{e}}(\Te)} = \fr{g_i}{g_j}\exp\left(\fr{\Ei{ij}}{\kb \Te} \right), \quad i < j, \quad
\fr{\kf{i}{}{\textsc{r}}(\Te)}{\kf{i}{}{\textsc{i}}(\Te)}  = \fr{g_i \, \Qstari{\textsc{H}}{\Te}{\text{t}}}{g_{\text{e}} \, \Qstari{\text{e}}{\Te}{\text{t}} \, \gp \,  \Qstari{\textsc{H}^+}{\Te}{\text{t}}}\exp\left(\fr{\Ei{+i}}{\kb \Te} \right), 
\ee
$i, \, j \in \mathcal{I}$, where the translational partition functions per unit volume are  $\Qstari{c}{\Te}{\text{t}} = (2 \pi m_c \kb \Te)^{3/2}/\hpc$, $c \in \mathcal{C}$, where $\hp$ is Planck's constant. Quantity $g_{\text{e}} = 2$ is the electron degeneracy accounting for its spin.

The free-electron temperature $\Te$ is the only temperature appearing in the rate coefficients \eqnref{eq:kExe}-\eqnref{eq:kdExe}, due to the hypothesis of stationary heavy-particles. This can be shown in a rigorous way by taking the zeroth-order moment (\emph{i.e.} mass) of the Boltzmann equation \citep{Oxenius_book}.

In the present work, the cross-section for electron impact excitation and ionization processes \eqnref{eq:Ie} have been modeled according to Drawin's semi-classical formula \citep{Drawin_1963}, with the absorption oscillator strengths taken from the NIST atomic database \citep{Kramida_2010,NIST_ASD}.

For elastic collisions among free-electrons and heavy-particles, the effective collision frequency for energy transfer has been computed based on Kinetic Theory as \citep{Petschek_AnnPhys_1957,Desloge_POF_1962}:
\be \label{eq:coll_freq}
\nu^{\textsc{el}}_{\text{e}-\text{h}} = \fr{8}{3} \sum_{c \in \mathcal{C}_{\text{h}}} \left( \fr{m_{\text{e}}}{m_c} \right) \ni{c} \, \Omega^{(1,1)}_{\text{e}c}.
\ee  
Quantities $\Omega^{(1,1)}_{\text{e}c}$ denote the collision integrals for electron-heavy interactions in the first-order Laguerre-Sonine polynomial expansion \citep{Devoto_1966,Devoto_1967,Ferziger_book}. The collision integrals $\smash{\Omega^{(1,1)}_{\text{e}c}}$ are often referred to as \emph{diffusion cross-sections} since they appear in the electron-heavy binary diffusion coefficients for an ionized gas. For interactions between free-electrons and hydrogen, the collision integral $\smash{\Omega^{(1,1)}_{\text{e}\textsc{H}}}$ has been evaluated based on the work by \cite{Bruno_POP_2010}. For interactions between free-electrons and protons, a screened Coulomb potential has been used to evaluate $\smash{\Omega^{(1,1)}_{\text{e}\textsc{H}^+}}$ \citep{Devoto_1966,Devoto_1967,Spitzer_book,Bruno_POP_2010}. 

The species production terms due to excitation and ionization are computed based on zeroth-order reaction rate theory \citep{Oxenius_book}:  
\begin{IEEEeqnarray}{rCl}
\omega^{\text{col}}_{\text{e}} & = & \sum_{i \in \mathcal{I}} \nelec \, [n_i \, \kf{i}{}{\textsc{i}}(\Te)  - \nelec \, \np \, \kf{i}{}{\textsc{r}}(\Te)], \quad \omega^{\text{col}}_+ = \omega^{\text{col}}_{\text{e}}, \\
\omega^{\text{col}}_i  & = & - \nelec \, [n_i \, \kf{i}{}{\textsc{i}}(\Te)  - \nelec \, \np \kf{i}{}{\textsc{r}}(\Te)] - \sum_{\substack{j \in \mathcal{I}\\ j \ne i}} \nelec \, [n_i \, \kf{i}{j}{\textsc{e}}(\Te)  - n_j \, \kf{j}{i}{\textsc{e}}(\Te)], \label{eq:omegai_EIon}
\end{IEEEeqnarray}
$i \in \mathcal{I}$, where the acronym ``col" (\emph{i.e.} collisional) has been introduced to distinguish from the mass production terms due to radiative processes treated in Sec. \ref{sec:StSRad}. 

The volumetric time rate of change of free-electron energy due to collisional processes can be written as $\Omega^{\text{col}}_{\text{e}} = \Omega^{\textsc{el}}_{\text{e}} + \Omega^{\textsc{e}}_{\text{e}} + \Omega^{\textsc{I}}_{\text{e}}$, where the individual contributions due to elastic collisions, excitation and ionization are:  
\begin{IEEEeqnarray}{rCl}
\Omega^{\textsc{el}}_{\text{e}} & = & 3 \, \nelec \, \kb \, (\Te - \Th) \, \nu^{\textsc{el}}_{\text{e}-\text{h}}, \label{eq:Omecol_El} \\
\Omega^{\textsc{e}}_{\text{e}} & = & \sum_{\substack{i, \, j \in \mathcal{I} \\ j > i}} \nelec \, \Ei{ji} \, [n_i \, \kf{i}{j}{\textsc{e}}(\Te)  - n_j \, \kf{j}{i}{\textsc{e}}(\Te)], \quad \Omega^{\textsc{i}}_{\text{e}} = \sum_{i \in \mathcal{I}} \nelec \, \Ei{+i} \, [n_i \, \kf{i}{}{\textsc{i}}(\Te)  - \nelec \, \np \, \kf{i}{}{\textsc{r}}(\Te)]. \label{eq:Omecol_EI}
\end{IEEEeqnarray}
The inelastic energy transfer terms \eqnref{eq:Omecol_EI} are obtained by taking the second-order moment (\emph{i.e.} energy) of the Boltzmann equation for free-electrons. The formulas obtained reflect the physical fact that, under the assumption of stationary heavy-particles, the only active source/sink of free-electron energy in inelastic and ionizing collisions is the change of heavy-particle formation energies. The elastic term \eqnref{eq:Omecol_El} is derived via a similar procedure as explained in the papers by \cite{Petschek_AnnPhys_1957} and \cite{Desloge_POF_1962}, and the book by \cite{Mitchner_book}.   
\subsubsection{Radiative processes}\label{sec:StSRad} 
The modeling of radiation in plasmas requires accounting for transitions characterized by absorption, emission and scattering of light. Since the main focus of this work is the complexity reduction of the NLTE kinetics, scattering is not taken into account. 

The radiative processes leading to emission and absorption of light can be subdivided into three groups: bound-bound (\textsc{bb}), bound-free/free-bound (\textsc{bf}/\textsc{fb}) and free-free (\textsc{ff}) \citep{Mihalas_StAtm_1978,Oxenius_book}. The radiative processes considered in this work are:
\begin{enumerate}
  \item[i)] spontaneous emission/absorption and induced emission (\textsc{bb}):
    \be 
      \text{H}(j) \rates{}{} \text{H}(i) + \hp \nu_{ij}, \quad \text{H}(j) + \hp \nu_{ij} \ratesone{} \text{H}(i) +  2\hp \nu_{ij}, \quad i < j, \quad i, \, j \in \mathcal{I},
    \ee
  \item[ii)] photo-ionization and spontaneous/induced radiative recombination (\textsc{bf}/\textsc{fb}):
     \be 
      \text{H}(i) + \hp \nu \rates{}{} \text{H}^+ + \elec, \quad i \in \mathcal{I},
    \ee
  \item[iii)] spontaneous/induced and inverse Bremsstrahlung for protons (\textsc{ff}).
\end{enumerate}
\paragraph{Bound-bound transitions} The monochromatic emission and absorption coefficients due to \textsc{bb} radiation are \citep{Oxenius_book}:
\be \label{eq:emisabs_bb}
\elam{\textsc{bb}} = \sum_{\substack{i, \, j \in \mathcal{I} \\ j > i}} \fr{\hp c}{4 \pi \lambda_{ij}} \, \ni{j} \, A_{ji} \, \philam{ji}, \quad \klam{\textsc{bb}} = \sum_{\substack{i, \, j \in \mathcal{I} \\ j > i}} \fr{\hp {\lambda}_{ij}}{4 \pi} \, [n_i \, \Bij \, \psilam{ij} - n_j \, \Bji \, \philam{ji}],
\ee
where the wavelength associated with the transition $j \rightarrow i$ is  $\lambda_{ij} = \hp c/\Ei{ji}$, with $c$ being the speed of light and the energy difference $\Ei{ji}$ having the same definition as in Eqs. \eqnref{eq:kExe} and \eqnref{eq:kdExe} (\emph{i.e} $\Ei{ji} = \Ei{j} - \Ei{i}$). The symbols $\Aji$, $\Bji$ and $\Bij$ (with $i < j$) denote, respectively, the Einstein coefficients for spontaneous emission, induced emission and absorption. The Einstein coefficients are not independent of each other and satisfy the Einstein-Milne relations \citep{Mihalas_StAtm_1978}: 
\be 
\Bji = \fr{\lambda^3_{ij}}{2 \hp c} \Aji, \quad g_j \Bji = g_i \Bij, \quad i < j,
\ee
$i, \, j \in \mathcal{I}$. The functions $\philam{ji}$ and $\psilam{ij}$ in Eq. \eqnref{eq:emisabs_bb} are, respectively, the monochromatic line emission and absorption profiles. In this work, it has been assumed that the emission and absorption profiles coincide (\emph{i.e.} complete redistribution) and are described using a Voigt function by accounting for natural, Doppler and collisional broadening \citep{Mihalas_StAtm_1978}. Doppler broadening has been evaluated at the heavy-particle temperature $\Th$ as explained in \cite{Mihalas_StAtm_1978}. Collisional broadening accounts for collisions between charged particles (\emph{i.e.} Stark broadening) and has been computed at the free-electron temperature $\Te$ based on the work of \cite{Cowley_1971}. The effects of resonance and pressure broadening and Doppler shifts due to macroscopic gas motion have not been taken into account. In order to speed up the calculations, the numerical evaluation of the Voigt function (\emph{i.e.} convolution between Gaussian and Lorentzian profiles) has been accomplished using the curve fit by \cite{Whiting_JQSRT_1968}. 

Bound-bound transitions lead to a change in the occupation numbers of hydrogen bound-states, for which the related mass production terms are \citep{Oxenius_book}:
\be \label{eq:omegai_bb}
\omega^{\textsc{bb}}_i = \sum_{\substack{j \in \mathcal{I} \\ j > i}} \, [\ni{j} \, \Ajim - (\ni{i} \, \Bijm - \ni{j} \, \Bjim) \, \Phi_{ij}] - \sum_{\substack{j \in \mathcal{I} \\ j < i}} \, [\ni{i} \, \Aijm - (\ni{j} \, \Bjim - \ni{i} \, \Bijm) \, \Phi_{ij}],
\ee  
$i \in \mathcal{I}$, where the mass production (superscipt m) Einstein coefficients, defined by $\Ajim = \Aji$ (with $i < j$) and $\Bijm = \Bji \, \lambda^2_{ij}/c$, have been introduced for convenience. Quantity $\Phi_{ij}$ is defined by the lineshape integral:
\be
\Phi_{ij} = \int\limits_{0}^{+\infty} \!\! \Jlam \, \philam{ij} \, d\lambda, \quad i \neq j,
\ee
$i, \, j \in \mathcal{I}$, where the symbol $\Jlam$ denotes the average monochromatic intensity. The former is obtained by the integration of the (directionally dependent) monochromatic intensity $I_{\lambda}$ over all directions, $\Jlam = 1/4\pi \oint I_{\lambda} d\Omega$, with $\Omega$ being the solid angle \citep{Mihalas_StAtm_1978}. The changes in the occupation numbers of hydrogen bound-states are accompanied by a transfer of energy between the material gas and the radiation field. The volumetric time rate of loss of matter energy due to \textsc{bb} transitions is obtained by integrating quantity $\elam{\textsc{bb}} - I_{\lambda} \, \klam{\textsc{bb}}$ over all wavelengths and directions \citep{Oxenius_book,Mihalas_RadHyd_1984}. Performing the required integrations, one obtains:
\be \label{eq:Omegarad_bb}
\Omega^{\textsc{bb}} = \int\limits_{0}^{+\infty} \!\!\! \oint (\elam{\textsc{bb}} - I_{\lambda} \, \klam{\textsc{bb}}) \, d\lambda \, d\Omega = \sum_{\substack{i, \, j \in \mathcal{I} \\ j > i}} \, [\ni{j} \, \Ajie - (\ni{i} \, \Bije - \ni{j} \, \Bjie) \, \Phi_{ji}],
\ee
where the energy transfer (e) Einstein coefficients are defined as $\Ajie = \Ei{ji} \, \Aji$ (with $i < j$) and $\Bjie = \hp \lambda_{ij} \Bji$.
\paragraph{Bound-free/free-bound transitions} The monochromatic emission and absorption coefficients due to \textsc{bf} and \textsc{fb} radiation are \citep{Oxenius_book}:
\begin{IEEEeqnarray}{rCl}
\elam{\textsc{fb}} & = & \fr{\hpf \, c^2 \, \ni{\text{e}} \, \np}{\lambda^5 (2 \pi \mi{\text{e}} \kb \Te )^{3/2}} \sum_{i \in \mathcal{I}} \sigma^{\textsc{pi}}_{i}(\lambda) \fr{\gi{i}}{\gp} \exp\left(\fr{\Ei{+i}}{\kb \Te} - \fr{\hp c}{\kb \Te \lambda} \right), \label{eq:emiss_fb} \\
\klam{\textsc{bf}} & = & \sum_{i \in \mathcal{I}} \sigma^{\textsc{pi}}_{i}(\lambda) \left[\ni{i} -\fr{1}{2} \fr{\hpc \, \nelec \, \np}{(2 \pi \mi{\text{e}} \kb \Te )^{3/2}} \fr{\gi{i}}{\gp} \exp\left(\fr{\Ei{+i}}{\kb \Te} - \fr{\hp c}{\kb \Te \lambda} \right) \right] \label{eq:abs_bf},
\end{IEEEeqnarray}
where the ionization threshold is defined, as in Eqs. \eqnref{eq:kIe} and \eqnref{eq:kdExe}, via the relation $\Ei{+i} = \Ei{+} - \Ei{i}$. Quantity $\sigma^{\textsc{pi}}$ denotes the total photo-ionization cross-section. In the present work, the former has been evaluated based on Kramer's formula \citep{Mihalas_StAtm_1978,Zeldovich_book_1967}.

The mass production terms for free-electrons, protons, and hydrogen bound-states due to photo-ionization and radiative recombination can be expressed as:
\begin{IEEEeqnarray}{rCl}
\omega^{\textsc{bf}/\textsc{fb}}_{\text{e}} & = & \sum_{i \in \mathcal{I}} [\ni{i} \, \kf{i}{}{\textsc{pi}}(\Jlam) - \nelec \, \np \, \kf{i}{}{\textsc{rr}}(\Te, \Jlam) ], \quad \omega^{\textsc{bf}/\textsc{fb}}_+ = \omega^{\textsc{bf}/\textsc{fb}}_{\text{e}}, \label{eq:omegae_PI_RR} \\
\omega^{\textsc{bf}/\textsc{fb}}_i & = &  - [\ni{i} \, \kf{i}{}{\textsc{pi}}(\Jlam) - \nelec \, \np \, \kf{i}{}{\textsc{rr}}(\Te, \Jlam)], 
\end{IEEEeqnarray}
$i \in \mathcal{I}$, where the rate coefficients for photo-ionization and radiative recombination (accounting for both spontaneous (s) and induced (i) contributions) are:
\begin{IEEEeqnarray}{rCl}
\kf{i}{}{\textsc{pi}}(\Jlam) & = & \fr{4 \pi}{\hp c} \int\limits_{0}^{\lambda_{+i}} \! \sigma^{\textsc{pi}}_{i}(\lambda) \, \Jlam \, \lambda \, d\lambda, \quad \kf{i}{}{\textsc{rr}}(\Te, \Jlam) = \kf{i}{}{\textsc{rr}-\text{s}}(\Te) + \kf{i}{}{\textsc{rr}-\text{i}}(\Te, \Jlam), \\
\kf{i}{}{\textsc{rr}-\text{s}}(\Te) &= & \sqrt{\fr{2}{\pi}} \fr{\gi{i}}{\gp} \fr{ \hpc c }{(\mi{\text{e}} \kb \Te )^{3/2}}  \exp\left(\fr{\Ei{+i}}{\kb \Te}\right) \int\limits_{0}^{\lambda_{+i}} \! \fr{\sigma^{\textsc{pi}}_{i}(\lambda)}{\lambda^4} \, \exp\left(- \fr{\hp c}{\kb \Te \lambda} \right) d\lambda, \label{eq:kPI}  \\
\kf{i}{}{\textsc{rr}-\text{i}}(\Te, \Jlam) & = & \sqrt{\fr{1}{2\pi}} \fr{\gi{i}}{\gp} \fr{\hps}{(\mi{\text{e}} \kb \Te )^{3/2}c} \exp\left(\fr{\Ei{+i}}{\kb \Te}\right) \int\limits_{0}^{\lambda_{+i}} \! \sigma^{\textsc{pi}}_{i}(\lambda) \, \lambda \, \Jlam \, \exp\left(- \fr{\hp c}{\kb \Te \lambda} \right) d\lambda, \label{eq:kRR}
\end{IEEEeqnarray}
$i \in \mathcal{I}$, with the threshold wavelength for photo-ionization/radiative-recombination being $\lambda_{+i} = \hp c/\Ei{+i}$.

By analogy with the procedure outlined above for \textsc{bb} radiation, the volumetric time rate of loss of matter energy due to \textsc{bf}/\textsc{fb} radiation is obtained through the integration of the quantity $\elam{\textsc{fb}} - I_{\lambda} \, \klam{\textsc{bf}}$  over all wavelengths and directions:
\begin{IEEEeqnarray}{rCl}
\Omega^{\textsc{bf}/\textsc{fb}} & = & \sum_{i \in \mathcal{I}} \Bigg[ \sqrt{\fr{1}{2\pi}} \fr{\gi{i}}{\gp} \fr{\hpc \, \nelec \, \np }{(\mi{\text{e}} \kb \Te )^{3/2}}  \exp\left(\fr{\Ei{+i}}{\kb \Te}\right) \int\limits_{0}^{\lambda_{+i}} \! \sigma^{\textsc{pi}}_{i}(\lambda) \, \Jlam \, \exp\left(- \fr{\hp c}{\kb \Te \lambda} \right) d\lambda - 4\pi \ni{i} \! \int\limits_{0}^{\lambda_{+i}} \! \sigma^{\textsc{pi}}_{i}(\lambda) \, \Jlam \, d\lambda \Bigg] +  \label{eq:OmegaBF_en} \\
& & \sum_{i \in \mathcal{I}} \sqrt{\fr{2}{\pi}} \fr{\gi{i}}{\gp} \fr{\hpf c^2 \, \nelec \, \np}{(\mi{\text{e}} \kb \Te )^{3/2}}  \exp\left(\fr{\Ei{+i}}{\kb \Te}\right) \int\limits_{0}^{\lambda_{+i}} \! \fr{\sigma^{\textsc{pi}}_{i}(\lambda)}{\lambda^5} \, \exp\left(- \fr{\hp  c}{\kb \Te \lambda} \right) d\lambda. \nonumber 
\end{IEEEeqnarray}
It should be noted that the volumetric energy loss term \eqnref{eq:OmegaBF_en} refers to the whole gas, which includes free-electrons and heavy-particles. The adoption of an additional energy equation for free-electrons (see Sect. \ref{sec:StSEqs}) requires the evaluation of the corresponding of quantity $\Omega^{\textsc{bf}/\textsc{fb}}$ for the free-electron gas alone. This is accomplished in a straightforward manner by taking the second-order moment (\emph{i.e.} energy) of the collision operator for photo-ionization/radiative recombination of the Boltzmann equation for free-electrons \citep{Oxenius_book}. Under the assumption of Maxwellian free-electrons at temperature $\Te$ and stationary heavy-particles, the final result is:
\be \label{eq:OmegaBF_en_el}
\Omega^{\textsc{bf}/\textsc{fb}}_{\text{e}} = \Omega^{\textsc{bf}/\textsc{fb}} + \sum_{i \in \mathcal{I}} \Ei{+i} \, [\ni{i} \, \kf{i}{}{\textsc{pi}}(\Jlam) - \nelec \, \np \, \kf{i}{}{\textsc{rr}}(\Te, \Jlam) ].
\ee      
\paragraph{Free-free transitions} The monochromatic emission and absorption coefficients due to \textsc{ff} radiation produced by encounters between free-electrons and protons are \citep{Oxenius_book,Zeldovich_book_1967}:
\be \label{eq:emisFF}
\elam{\textsc{ff}} = \fr{8}{3} \sqrt{\fr{2\pi}{3\mi{\text{e}} \kb \Te}} \fr{ n_{\text{e}} \, \np \, q^6_{\text{e}}}{\lambda^2 \, (4\pi \epsilon_0)^3 \mi{\text{e}} c^2}  \exp\left(- \fr{\hp c}{\kb \Te \lambda}\right), \quad \klam{\textsc{ff}} = \fr{4}{3} \sqrt{\fr{2\pi}{3\mi{\text{e}} \kb \Te}} \fr{\lambda^3 \, n_{\text{e}} \, \np \, q^6_{\text{e}}}{(4\pi \epsilon_0)^3 \mi{\text{e}} \hp c^4} \left[1- \exp\left(- \fr{\hp c}{\kb \Te \lambda}\right)\right],
\ee
where quantities $\epsilon_0$ and $q_{\text{e}}$ denote, respectively, the vacuum permittivity and the electron charge.

The volumetric time rate of loss of matter energy due to \textsc{ff} radiation is obtained, as done before, by integrating quantity $\elam{\textsc{ff}} - I_{\lambda} \, \klam{\textsc{ff}}$  over all wavelengths and directions:
\be \label{eq:OmegaFF_en}
\Omega^{\textsc{ff}} = \fr{32}{3} \pi \sqrt{\fr{2 \pi}{3 \mi{\text{e}} \kb \Te}} \fr{n_{\text{e}} \, \np \, q^6_{\text{e}}}{(4 \pi \epsilon_0)^3 \mi{\text{e}} \hp}\bigg\{\kb \Te - \fr{1}{2c} \int\limits_{0}^{+\infty} \lambda^3 \left[1 - \exp\left(- \fr{\hp c}{\kb \Te \lambda} \right)  \right] \Jlam \, d\lambda \bigg\}. 
\ee

By collecting the results through Eqs. \eqnref{eq:emisabs_bb}-\eqnref{eq:OmegaFF_en}, it is possible to write down the mass production terms due to radiative transitions for free-electrons, protons and hydrogen bound-states as $\smash{\omega^{\text{rad}}_{\text{e}} = \omega^{\textsc{bf}/\textsc{fb}}_{\text{e}}}$, $\smash{\omega^{\text{rad}}_+ = \omega^{\textsc{bf}/\textsc{fb}}_+}$ and $\smash{\omega^{\text{rad}}_i = \omega^{\textsc{bb}}_i + \omega^{\textsc{bf}/\textsc{fb}}_i}$, respectively. The corresponding energy loss rates for the whole gas and free-electrons alone are $\smash{\Omega^{\text{rad}} = \Omega^{\textsc{bb}} + \Omega^{\textsc{bf}/\textsc{fb}} + \Omega^{\textsc{ff}}}$ and $\smash{\Omega^{\text{rad}}_{\text{e}} = \Omega^{\textsc{bf}/\textsc{fb}}_{\text{e}} + \Omega^{\textsc{ff}}}$, respectively.
\subsection{Governing equations}\label{sec:StSEqs}
The steady flow across a normal shock-wave of a two-temperature radiating plasma is governed by the species continuity equations, the global momentum and energy equations, and the free-electron energy equation. In the absence of transport phenomena, the former set of equations reads \citep{Murty_JQSRT_1971,Zeldovich_book_1967}:
\be \label{eq:gov_eq}
\fr{\pa}{\pa x} 
\left(
\begin{array}{c}
\ni{s} u\\
p + \rho u^2 \\
\rho H u \\
\rho e_{\text{e}} u \\
\end{array}
\right) = 
\left(
\begin{array}{c}
\omegai{s} \\
0 \\
-\Omega^{\text{rad}} \\
- p_{\text{e}} \fr{\pa u}{\pa x} - (\Omega^{\text{col}}_{\text{e}} +  \Omega^{\text{rad}}_{\text{e}}) \\
\end{array}
\right),
\ee
$s \in \mathcal{S}$, where the mass production terms are given by the sum of the collisional and radiative contributions, $\omega_s = \omega^{\text{col}}_{s} + \omega^{\text{rad}}_{s}$. Quantity $u$ stands for the flow velocity measured in the shock wave reference frame. 

It is known that electron heat conduction, neglected in Eq. \eqnref{eq:gov_eq}, may play an important role in shaping the temperature profile in the precursor (\emph{i.e.} conduction precursor) \citep{Shafranov_1957,Jukes_JFM_1957,Imshennik_1962,Jaffrin_POF_1964,Fadeyev_AA_2002}. However, for the standing shocks studied in this work, electron heat conduction is little influenced by the dynamics of excited electronic states.\footnote{The calculation of electron transport properties and fluxes is often accomplished by accounting only for the effects of elastic collisions \citep{Devoto_1966,Devoto_1967}. When assuming that the electron-heavy collision integrals do not depend on the particular electronic state (as done in this work; see Eq. \eqnref{eq:coll_freq}), it can be shown that the electron transport formulas do not show an explicit dependence on the population of excited electronic states.} In view of this, electron heat conduction is expected to play a minor role compared to radiation and chemistry on the accuracy of a reduced-order NLTE model.

For a radiating gas, the flow governing equations \eqnref{eq:gov_eq} must be coupled with the radiative transfer equation (\textsc{rte}) \citep{Oxenius_book,Mihalas_RadHyd_1984}. The former can be thought of as the kinetic equation for a photon gas and describes the evolution in space and time of a radiation field due to emission, absorption and scattering of light. In the case of a plane parallel non-scattering medium under steady-state conditions, the \textsc{rte} reads:
\be \label{eq:rte1}
\mu \fr{\pa I_{\lambda \mu}}{\pa x} = \elam{} - \klam{} I_{\lambda \mu},
\ee 
where quantity $\mu \in [-1,1]$ stands for the cosine of the angle between the line of sight and the $x$ axis. The (total) emission and absorption coefficients in the \textsc{rte} \eqnref{eq:rte1} are obtained by summing the individual contributions due to \textsc{bb}, \textsc{bf}/\textsc{fb} and \textsc{ff} transitions as $\elam{} = \elam{\textsc{bb}} + \elam{\textsc{fb}} + \elam{\textsc{ff}}$ and $\klam{} = \klam{\textsc{bb}} + \klam{\textsc{bf}} + \klam{\textsc{ff}}$, respectively. In radiative transfer problems, it is often convenient to transform the \textsc{rte} \eqnref{eq:rte1} to a second-order differential equation as proposed by \cite{Feautrier_1964}:
\be \label{eq:rte}
\mu^2 \fr{\pa^2 P_{\lambda \mu} }{\pa \tau^2_{\lambda}} = P_{\lambda \mu} - S_{\lambda},
\ee
where quantity $P_{\lambda \mu}$ is defined as $P_{\lambda \mu} = (I_{\lambda \mu} + I_{\lambda -\mu})/2$, and the monochromatic source function is given by the ratio of emission and absorption coefficients, $S_{\lambda} = \elam{}/\klam{}$. The symbol $d\tau_{\lambda}$ denotes the infinitesimal monochromatic optical thickness increment, $d\tau_{\lambda} = \klam{} \, dx$. For the \textsc{rte} \eqnref{eq:rte}, the range of the angular variable $\mu$ is restricted to $[0,1]$ \citep{Mihalas_StAtm_1978}. When written in terms of the newly introduced unknown $P_{\lambda \mu}$, the average monochromatic intensity becomes simply:
\be
\Jlam = \fr{1}{4\pi} \oint \Ilam \, d\Omega = \int\limits_{0}^1 P_{\lambda \mu} \, d\mu.
\ee
\section{Reduced-order modeling}\label{sec:Red}
In order to simplify the complexity of the NLTE kinetic mechanism, the energy levels (\emph{i.e.} species) of the StS model described in Sec. \ref{sec:StS} are lumped into groups. The governing equations for the reduced-oder model are obtained using a moment method after prescribing a distribution within each group. The general procedure is explained below.
\subsection{Level grouping: Maximum Entropy model}\label{sec:RedME}
Following the work by \cite{Yen_2015}, the logarithm of the normalized population within a given group $k$ is written as a polynomial in the internal energy $\Ei{i}$:
\be \label{eq:me_model}
\ln \left( \fr{\ni{i}}{\gi{i}} \right) = \tilde{\alpha}_k + \tilde{\beta}_k \Ei{i} + \tilde{\gamma}_k E^2_i + \textsc{h.o.t},
\ee  
$i \in \mathcal{I}_k$, $k \in \mathcal{K}$, where the sets $\mathcal{I}_k$ and $\mathcal{K}$ denote, respectively, the energy levels within group $k$ and the group indices. In the present work, terms of second and higher-order in energy are neglected. Under these circumstances, one needs to determine only the quantities $\tilde{\alpha}_k$ and $\tilde{\beta}_k$. These are related to the group populations, $\nti{k}$, and average energies, $\Eti{k}$, by the following moment constraints on particle number and energy \citep{Yen_2015}:
\be \label{eq:me_model_const} 
\nti{k} = \sum_{i \in \mathcal{I}_k} \ni{i}, \quad  \nti{k} \Eti{k} = \sum_{i \in \mathcal{I}_k} \ni{i} \Ei{i}, 
\ee     
$k \in \mathcal{K}$. Substituting Eq. \eqnref{eq:me_model} into the moment constraints \eqnref{eq:me_model_const} gives $\tilde{\alpha}_k$ and $\tilde{\beta}_k$ in terms of the group populations and energies:
\be \label{eq:me_model_const2}
\tilde{\alpha}_k = \fr{\nti{k}}{\sum_{i \in \mathcal{I}_k} \gi{i} \, \exp(\tilde{\beta}_k \Ei{i})} , \quad \tilde{\beta}_k = \fr{\sum_{i \in \mathcal{I}_k} \gi{i} \, \Ei{i} \, \exp(\tilde{\beta}_k \Ei{i})}{\sum_{i \in \mathcal{I}_k} \gi{i} \exp(\tilde{\beta}_k \Ei{i})},
\ee 
$k \in \mathcal{K}$. After introducing, for the sake of convenience, group temperatures $T_k = -1/\kb \tilde{\beta}_k$ and partition functions $\Qstarti{k}{T_k}{} = \sum_{i \in \mathcal{I}_k} \gi{i} \exp(\tilde{\beta}_k \Ei{i})$, it is possible to re-write Eq. \eqnref{eq:me_model_const2} as:
\be \label{eq:me_model_const3}
\tilde{\alpha}_k =\fr{\nti{k}}{\tilde{Z}_k (\Ti{k})}, \quad \tilde{\beta}_k = \fr{1}{ \Qstarti{k}{T_k}{}} \sum_{i \in \mathcal{I}_k} \gi{i} \, \Ei{i} \, \exp\left( -\fr{\Ei{i}}{\kb T_k}\right), 
\ee
$k \in \mathcal{K}$. Substituting Eq. \eqnref{eq:me_model_const3} into Eq. \eqnref{eq:me_model} leads finally to:
\be \label{eq:me_model_distl}
\fr{\ni{i}}{\gi{i}} = \fr{\nti{k}}{\Qstarti{k}{T_k}{}} \exp\left( -\fr{\Ei{i}}{\kb T_k}\right),
\ee
$i \in \mathcal{I}_k$, $k \in \mathcal{K}$. Equation \eqnref{eq:me_model_distl} shows that retaining terms up to first-order in energy in Eq. \eqnref{eq:me_model_const} is equivalent to assuming a Maxwell-Boltzmann distribution within each group. The Maxwell-Boltzmann distribution is the Local Thermodynamic Equilibrium (LTE) distribution for which the entropy is maximum \citep{DeGroot_Mazur_book}. This is the reason that reduced-oder models developed based on Eq. \eqnref{eq:me_model} are called Maximum Entropy (ME) models \citep{Yen_2015}. The model corresponding to Eq. \eqnref{eq:me_model_distl} is the Maximum Entropy Linear (MEL) model as only linear terms are retained in the energy polynomial \eqnref{eq:rte}. The MEL model reduces to the Maximum Entropy Uniform (MEU) model when taking the limit of infinite group temperatures (\emph{i.e.} $\tilde{\beta}_k = 0$) or, equivalently, when retaining only the zeroth-order energy term in Eq. \eqnref{eq:me_model}. In this case, Eq. \eqnref{eq:me_model_distl} reduces to:
\be \label{eq:me_model_distu}
\fr{\ni{i}}{\gi{i}} = \fr{\nti{k}}{\tilde{Z}_k},
\ee            
$i \in \mathcal{I}_k$, $k \in \mathcal{K}$, where the group partition function is now the sum of the statistical weights of the energy levels within the group, $\tilde{Z}_k = \sum_{i \in \mathcal{I}_k} \gi{i}$. It is worth mentioning that the MEU distribution \eqnref{eq:me_model_distu} does not allow retrieving equilibrium (\emph{i.e.} Maxwell-Boltzmann distribution).  
\subsection{Moment equations}\label{sec:RedMom}
The governing equations for the ME model are obtained by taking the moments with respect to the energy $\Ei{i}$ of the species continuity equations \citep{Yen_2015}. As explained in Sec. \eqnref{sec:RedME}, in the present work only the zeroth and first-order moments are needed:
\be \label{eq:mom_eqs}
\fr{\pa}{\pa x} \left(\sum_{i \in \mathcal{I}_k} \ni{i} u \right) = \sum_{i \in \mathcal{I}_k} \omegai{i}, \quad  \fr{\pa}{\pa x} \left(\sum_{i \in \mathcal{I}_k} \ni{i} \Ei{i} u \right) = \sum_{i \in \mathcal{I}_k} \omegai{i} \Ei{i},
\ee
$k \in \mathcal{K}$. Using the moment constraints \eqnref{eq:me_model_const}, Eq. \eqnref{eq:mom_eqs} becomes:
\be \label{eq:mom_eqs2}
\fr{\pa}{\pa x} \left(
\begin{array}{c}
\nti{k} u \\
\nti{k} \Eti{k} u \\
\end{array}
\right)
=
\left(
\begin{array}{c}
\omegati{k} \\
\Omegati{k} \\
\end{array}
\right),
\ee
$k \in \mathcal{K}$, where the group mass production and energy transfer terms are defined as $\omegati{k} = \sum_{i \in \mathcal{I}_k} \omegai{i}$ and $\Omegati{k} = \sum_{i \in \mathcal{I}_k} \omegai{i} \Ei{i}$, respectively. In the case of the MEU model, the group number densities $\nti{k}$ are the only unknowns. Thus, only the first of Eq. \eqnref{eq:mom_eqs2} is needed. The complete set of flow governing equations for the MEU/MEL models are obtained based on those for the StS model \eqnref{eq:gov_eq} by replacing the species continuity equations for the hydrogen bound-states with the moment equations \eqnref{eq:mom_eqs2}. The related expressions for thermodynamic properties, mass/energy production terms, and emission/absorption coefficients are given in App. \ref{sec:app}.
\\
\\
The MEU and MEL models have already been successfully applied to the study of collisional excitation, dissociation, and ionization in atomic and molecular gases \citep{Panesi_Lani_CR_2013,Munafo_PRE_2014,Yen_2015,Munafo_POF_2015}. In that work, the determination of the states contained in a given group can be based, for instance, on an even subdivision of the internal energy ladder. This is justified by the fact that the rate coefficients for inelastic collisional processes (\emph{e.g.} electron impact ionization) are larger for states with similar energy. The inclusion of radiative transitions (in particular \textsc{bb} radiation) completely changes the picture for states which are close in energy but are strongly coupled through radiative transitions. Such states would be inaccurately modeled by being placed within the same energy group.     
\section{Computational method}\label{sec:Num}
The self-consistent solution of the governing equations for the material gas \eqnref{eq:gov_eq} and the radiation field \eqnref{eq:rte} is, in general, challenging due to the non-local nature of radiation, which introduces a global coupling between the solution at all points. In mathematical terms, the inclusion of radiation transforms the non-radiating shock flow problem, which is an initial value problem, to a mixed initial-boundary value problem. In view of this, one has to resort to an iterative approach for numerical solutions. Various techniques have been proposed to solve radiation hydrodynamics problems \citep{Mihalas_RadHyd_1984}. For the present study, the method of global iterations developed in the series of papers by \cite{Fadeyev_AA_1998,Fadeyev_AA_2000,Fadeyev_AA_2001} and \cite{Fadeyev_AA_2002} is employed. The former is essentially a lambda iteration method, where the flow governing equations at iteration $n$ are solved using the radiative rates from iteration $n-1$. To speed up the calculations the solution is usually restarted from a previously computed one with a slightly different free-stream velocity. A similar approach has been developed independently by \cite{Panesi_AIAA_2011} to study ionization phenomena in air for atmospheric entry flows. It is known that the lambda iteration may have very poor convergence when accounting for thick continua (\emph{e.g.} Lyman continuum) and lines \citep{Mihalas_StAtm_1978}. To overcome these deficiencies, methods such as the complete linearization by \cite{Auer_Mihalas_ApJ_1969I,Auer_Mihalas_ApJ_1969II,Auer_Mihalas_ApJ_1969III} or the accelerated lambda iteration by \cite{Ribicky_AA_1991,Ribicky_AA_1992} have been proposed. These techniques outperform the conventional lambda iteration. However, they come at the price of a more complex and lengthy implementation (especially for the complete linearization). Since the purpose of this work is to reduce the complexity of the NLTE kinetic mechanism, it was decided to adopt a simpler method such as the one by \cite{Fadeyev_AA_1998}. Moreover, as shown by the results in Sec. \ref{sec:Res}, the Balmer and Paschen lines are mostly in emission throughout the shock layer, which is a favorable condition when using a method resembling lambda iteration.      
\subsection{Spatial, wavelength and angular grids}
For convenience, the shock, which is treated as a discontinuous surface, is placed at $x = 0$. The left and right boundaries are placed at $x = - x_{\text{L}}$ and $x = x_{\text{R}}$, respectively. The lengths $x_{\text{L}}$ and $x_{\text{R}}$ are set to values of the order of \num{e2}-\SI{e3}{m} in order to include the whole extent of the precursor and radiative relaxation regions, respectively. In order to properly resolve the smaller scales of the internal relaxation region ($\simeq$ \num{e-2}-\SI{e1}{m}), an exponential stretching is applied to reduce the grid size around the shock location. 

The wavelength domain is discretized as suggested by \cite{Fadeyev_AA_1998,Fadeyev_AA_2000}. When accounting only for continuum radiation the procedure goes as follows. After prescribing the minimum and maximum wavelengths ($\lambda_{\text{min}}$ and $\lambda_{\text{max}}$, respectively), the interval $\smash{[\lambda_{\text{min}},\lambda_{\text{max}}]}$ is divided into sub-intervals determined by the photo-ionization thresholds $\lambda_{+i}$. Each of these sub-intervals is then discretized using Gauss-Legendre quadrature points. This procedure is slightly modified to account for line radiation by adding additional sub-intervals for each atomic line. This is motivated by the rapid variation of emission and absorption coefficients (and source functions as well) over the lines. For this reason, the wavelength domain close to an atomic line $l$ is discretized by adopting a Gauss-Legendre or uniform grid over the interval $[\lambda_l - \delta \lambda_l,\lambda_l + \delta \lambda_l]$, where the width $\delta \lambda_l$ is set to \SI{5}{}-\SI{10}{\angstrom}. The remaining sub-intervals are discretized as done for continuum radiation.

The angular variable $\mu$ is also discretized by using Gauss-Legendre quadrature points.       
\subsection{Numerical solution of the governing equations}
The flow governing equations \eqnref{eq:gov_eq} are solved using a space marching approach in both the pre-shock and the post-shock regions. This requires the specification of initial conditions. For the pre-shock region, LTE conditions are assumed during the first iteration. However, the gas in the far precursor is not in LTE due to non-equilibrium excitation caused by absorption of resonant radiation in atomic line wings \citep{Murty_JQSRT_1968,Foley_POF_1973}. To take this into account, the occupation numbers of free-electrons, protons and hydrogen bound-states are computed (starting from the second iteration) by solving the statistical equilibrium equations \citep{Mihalas_StAtm_1978}. These equations are obtained by setting to zero the convective term in the species continuity equations (\emph{i.e.} $\omegai{s} = 0$) and are solved iteratively by a Newton-Raphson procedure. Once the shock location is reached, the integration of Eq. \eqnref{eq:gov_eq} is stopped and the Rankine-Hugoniot jump relations \citep{Zeldovich_book_1967} are applied to determine the kinematic and thermo-chemical state of the gas just behind the shock. This provides the initial solution for the post-shock region. The jump relations are solved under the assumption of \emph{frozen} kinetics and by neglecting the effects of radiant energy fluxes \citep{Marshak_POF_1958}. Free-electrons are assumed isothermal within the shock. Alternatively, one could consider a slightly more accurate method by treating the compression of free-electrons as iso-entropic \citep{Zeldovich_book_1967,Fadeyev_AA_1998}. Preliminary calculations indicated, however, that this second approach does not lead to appreciable improvements compared to the first one (isothermal compression). In the present work, the flow governing equations \eqnref{eq:gov_eq} are numerically integrated by using a fifth-order Backward Differentiation Formula method \citep{Gear_book} implemented in the \textsc{lsode} library for stiff initial value problems \citep{lsode_1993}. For convenience, the numerical integration is performed by using the mass fractions, velocity, and temperatures as solution variables \citep{Panesi_JTHT_2009,Panesi_JTHT_2011,Munafo_PRE_2014,Munafo_POF_2015}. This is motivated by the particularly simple form assumed by Eq. \eqnref{eq:gov_eq} when re-written in terms of these variables. 

The \textsc{rte} \eqnref{eq:rte} in Feautrier form is discretized by using a conventional second-order finite difference method under the assumption of no incoming radiation from both boundaries \citep{Auer_ApJ_1967,Mihalas_StAtm_1978}. For a non-scattering medium where Doppler shifts due to bulk motions are neglected, the above procedure leads to a set of uncoupled tridiagonal systems of equations (one for each discrete angle-wavelength point). The former are solved using the elimination scheme proposed by \cite{Ribicky_AA_1991} for the sake of better numerical conditioning.  
\section{Results}\label{sec:Res}
The present Section describes the results obtained by applying the StS and ME models to radiative shock waves propagating through a partially ionized atomic hydrogen plasma. The discussion is organized in two parts. The first, Sec. \ref{sec:Res_StS}, illustrates the general features of the testcase under investigation. In the second, Sec. \ref{sec:Res_ME}, the StS predictions are systematically compared with those of the reduced-order MEU and MEL models to assess the accuracy. The comparison is performed on both gas and radiation quantities.

The free-stream temperature and pressure are set to \SI{5000}{K} and \SI{5}{Pa}, respectively. The former corresponds to a number and mass density of \SI{7.24e19}{m^{-3}} and \SI{1.21e-7}{kg/m^{-3}}. The velocity at the left boundary is varied between a minimum of \SI{40}{km/s} and a maximum of \SI{70}{km/s} in steps of \SI{5}{km/s}. The adopted free-stream conditions are typical of atmospheres of pulsating stars (\emph{e.g.} Cepheid variables) and are similar to the ones used by \cite{Fadeyev_AA_1998,Fadeyev_AA_2000}. 

In all the calculations, the precursor and radiative relaxation lengths are set to, respectively, \SI{2000}{m} and \SI{500}{m}. The minimum and maximum wavelengths are taken at \SI{600}{\angstrom} and \SI{20000}{\angstrom}, respectively. In this work, the number of atomic lines considered is \num{9} namely the $\alpha$, $\beta$ and $\gamma$ lines of the Lyman, Balmer, and Paschen series \citep{Mihalas_StAtm_1978}.  

Preliminary convergence studies have been performed on the spatial, wavelength and angular grids. For the spatial grid, calculations indicated that using \num{1690} nodes, with a maximum grid size of \SI{5}{m} at the boundaries and a minimum of \SI{1e-3}{m} close to the shock, lead to accurate predictions. For the wavelength domain, it has been observed that placing more than \num{16}-\num{32} Gauss-Legendre nodes in each photo-ionization sub-interval did not result in appreciable improvements when accounting only for continuum radiation. This outcome is not surprising and is due to the smoothness of continuum emission and absorption coefficients as functions of wavelength/frequency. The inclusion of line radiation required the adoption of a higher number of nodes. In the present work, up to \num{300} wavelength nodes (distributed uniformly) have been used for each atomic line. Finally, for the angular grid, results have demonstrated that \num{32} Gauss-Legendre were sufficient to resolve the directional dependence of the radiation field.          
\subsection{General features of the StS predictions}\label{sec:Res_StS}
\begin{figure}[htb!]
\centering
\includegraphics[bb=0 0 539 254,clip,scale=0.82,keepaspectratio]{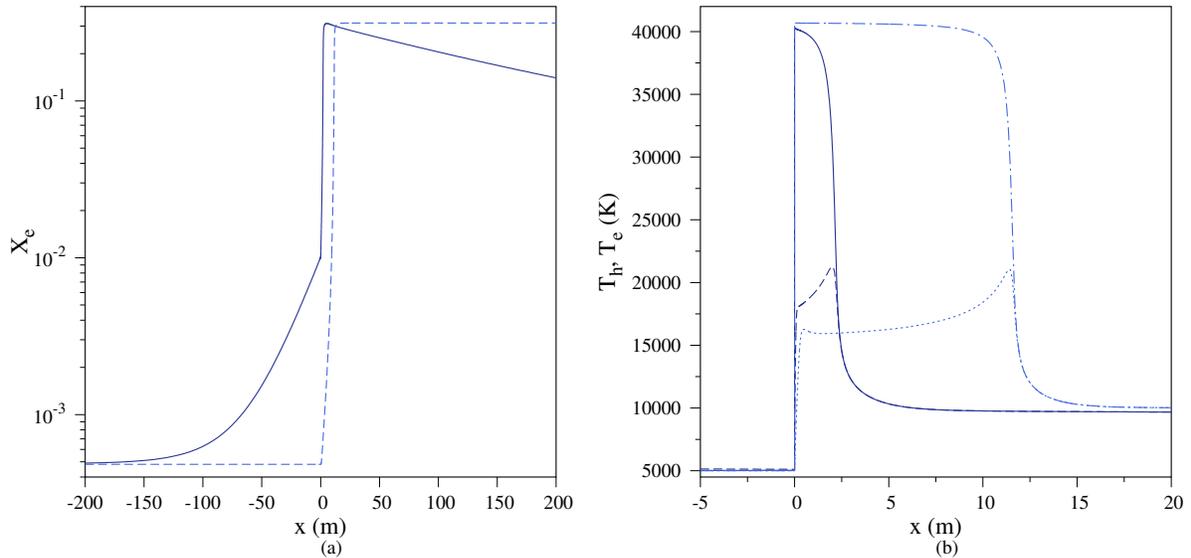}
\caption{Electron mole fraction (a) and temperatures (b) switching on/off radiative transitions. In (a) solid line with radiation, dashed line w/o radiation. In (b) solid line $\Ti{h}$ with radiation, dashed line $\Ti{e}$ with radiation, dotted-dashed line $\Ti{h}$ w/o radiation, dotted line $\Ti{e}$ w/o radiation (StS model; $\uinf = \SI{40}{km/s} $).}\label{fig:Xe_ThTe_with_wo_Rad}
\end{figure}

Figure \ref{fig:Xe_ThTe_with_wo_Rad} shows the evolution of the temperatures and electron mole fraction across the internal relaxation and the near precursor regions for a shock propagating at \SI{40}{km/s}. The absorption of Lyman continuum radiation photo-ionizes the gas in the free-stream and leads to an increase in the electron concentration, which is about \SI{1}{\percent} at the shock location. The spatial extent of the near precursor region is two-orders of magnitude larger than that of the internal relaxation region. The temperature evolution within the latter shows the typical structure observed in atomic plasmas \citep{Zeldovich_book_1967}. An initial sudden increase of the free-electron temperature is followed by a region where global ionization is indirectly controlled by the rate of elastic energy transfer between heavy-particle and free-electrons. In this zone, elastic collisions act as a source of energy for free-electrons by replenishing the energy lost in excitation and ionization processes. The production of free-electrons through electron-atom collisions stimulates the further ionization of the gas. This creates an \emph{avalanche} mechanism that is terminated by three-body recombination of protons and electrons balancing the effect of ionization. This process causes the rapid drop in the heavy-particle temperature seen in Fig. \ref{fig:Xe_ThTe_with_wo_Rad}(b). The ionized gas produced by the shock then enters the radiative relaxation region, where the electron concentration decreases due to radiative recombination (see Fig. \ref{fig:Xe_ThTe_with_wo_Rad}(a)). Figure \ref{fig:Xe_ThTe_with_wo_Rad} also shows the temperature and electron mole fraction evolution when taking into account only collisional processes. This has been done to show the effects of radiation on the shock structure. Neglecting radiative transitions (in particular photo-ionization), leads to a larger internal relaxation region, due to the smaller concentration of free-electrons at the shock location. In the present case, the length of the above zone is almost five times larger. Once the ionization is completed through the avalanche mechanism, the degree of ionization of the gas does not change and maintains its post-shock LTE value. The same holds true for the heavy-particle and free-electron temperatures, whose values are consistently higher compared to the radiative case.          
\begin{figure}[htb!]
\centering
\includegraphics[bb= 0 0 534 254,clip,scale=0.82,keepaspectratio]{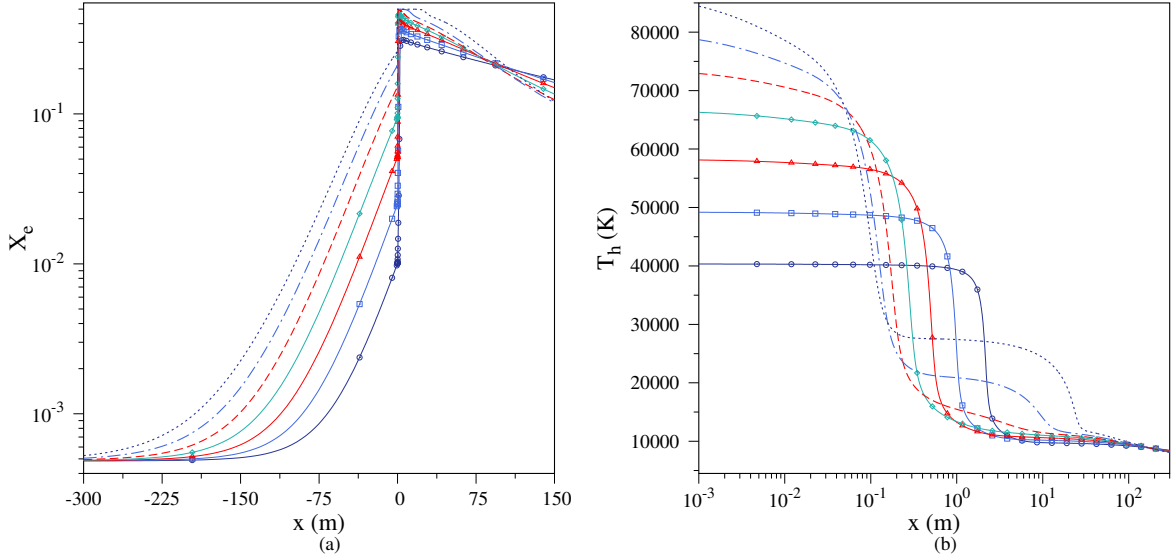}
\caption{Electron mole fraction (a) and heavy-particle temperature (b) for different free-stream velocities: line with circle $\uinf = \SI{40}{km/s}$, line with square $\uinf = \SI{45}{km/s}$, line with triangle $\uinf = \SI{50}{km/s}$, line with diamond $\uinf = \SI{55}{km/s}$, dashed line $\uinf = \SI{60}{km/s}$, dot-dashed line $\uinf = \SI{65}{km/s}$, dotted line $\uinf = \SI{70}{km/s}$ (StS model).}\label{fig:Xe_Th_vel}
\end{figure}

\begin{figure}[htb!]
\centering
\includegraphics[bb=0 0 609 254,clip,scale=0.82,keepaspectratio]{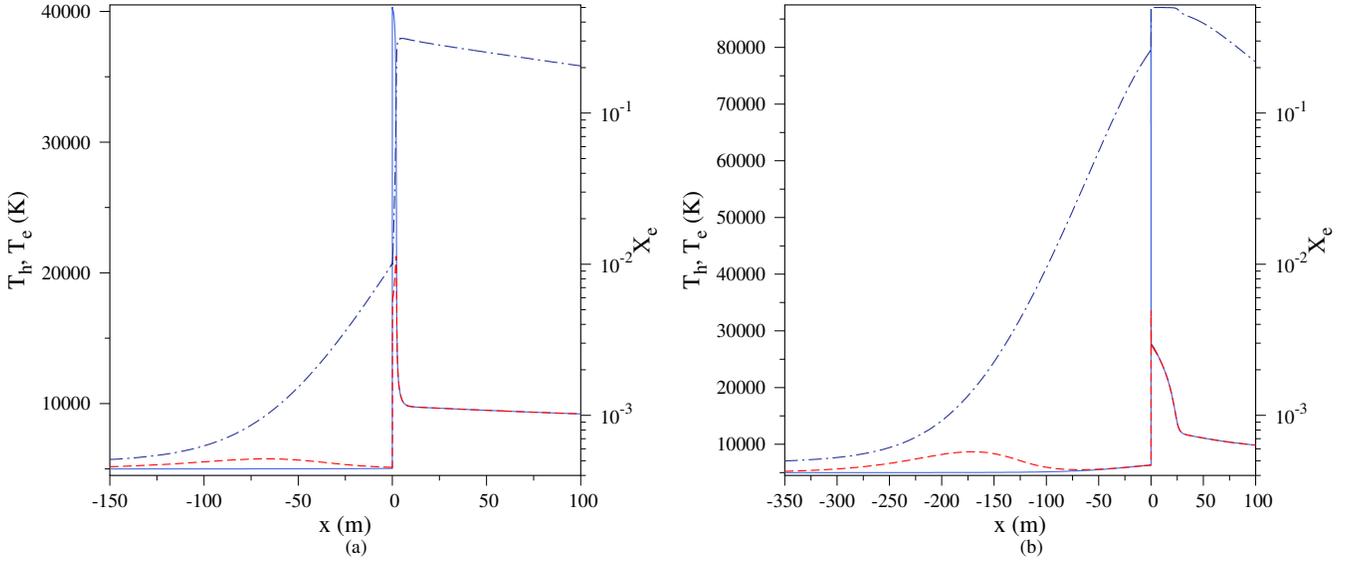}
\caption{Temperatures and electron mole fraction for (a) $\uinf = \SI{40}{km/s}$ and (b) $\uinf = \SI{70}{km/s}$: solid line $\Ti{h}$, dashed line $\Ti{e}$, dot-dashed line $X_{\text{e}}$ (StS model).}\label{fig:ThTeXe_40_70}
\end{figure}

Increasing the shock speed leads to more energetic photons from the radiative cooling region. This causes, in turn, a higher degree of ionization in the precursor as shown in Fig. \ref{fig:Xe_Th_vel}. In the same picture, the evolution of the heavy-particle translational temperature at different speeds (using a logarithmic scale for the $x$ axis) is shown. The increase in the degree of ionization due to larger shock speeds has the effect of shrinking the length of the internal relaxation region by more than an order of magnitude. For shock speeds higher than \SI{60}{km/s}, the hydrogen plasma is fully ionized. Under these circumstances, when the gas enters the radiative cooling region, the electron mole fraction remains almost constant at the beginning, and then decreases due to radiative recombination. The extent of the region of (almost) constant degree of ionization increases with the shock speed and is reflected in the \emph{plateau} observed for the heavy-particle temperature in Fig. \ref{fig:Xe_Th_vel}(b). At large shock speeds, the ionization avalanche is followed by a fast cooling zone which is slowed by the onset of radiative recombination. This is demonstrated in Fig. \ref{fig:ThTeXe_40_70} which superimposes the electron mole fraction and temperature evolution at \SI{40}{km/s} and \SI{70}{km/s}.             

The fast temperature drop at \SI{70}{km/s} (Fig. \ref{fig:ThTeXe_40_70}(b)) is due to the absence of neutral absorbers, which makes the plasma completely transparent to continuum radiation (see Eq. \eqnref{eq:OmegaBF_en}). The temperature decrease favors radiative recombination as the rate coefficient for this process is larger at low temperatures (see Eqs. \eqnref{eq:kPI}-\eqnref{eq:kRR}). Once the amount of ground-state hydrogen atoms is large enough, the radiation emitted in the Lyman continuum is partially re-absorbed, causing the temperature inflection point observed in Fig. \ref{fig:ThTeXe_40_70}(b). At lower speeds (see Fig. \ref{fig:ThTeXe_40_70}(a)), the above temperature drop is not observed as the shock strength is not sufficient to fully ionize the incoming gas. The trends shown in Figs. \ref{fig:Xe_Th_vel}-\ref{fig:ThTeXe_40_70} are similar to those reported in the work by \cite{Fadeyev_AA_1998,Fadeyev_AA_2000}.

The evolution of the free-electron temperature in Fig. \ref{fig:ThTeXe_40_70} shows a non-monotone behavior in the precursor. The initial rise is due to energy deposition by photo-ionization (see Eq. \eqnref{eq:OmegaBF_en_el}). During this stage, elastic collisions between charged particles are unable to bring heavy-particles and free-electrons into equilibrium due to the low degree of ionization. This trend continues till the further increase of ionization causes elastic energy transfer to dominate over photo-ionization. As a result, the free-electron temperature reaches a maximum and then relaxes towards the heavy-particle temperature. The observed precursor behavior of the free-electron temperature is consistent with the findings of \cite{Foley_POF_1973} who investigated radiative shock waves in helium and argon plasmas. The only significant difference from the above reference is that, in the present work, free-electrons are essentially in thermal equilibrium with heavy-particles at the shock location. This is most probably due to the more efficient energy transfer in electron-atom collisions caused by the lower atomic weight of hydrogen (which is, respectively, four and forty times lighter than helium and argon). This fact also has a strong influence on the length of the internal relaxation region (where elastic collisions play a crucial role), as recognized in the early theoretical and experimental work by \cite{Belozerov_JFM_1969}.             

\begin{figure}[htb!]
\centering
\includegraphics[bb=0 0 527 519,clip,scale=0.82,keepaspectratio]{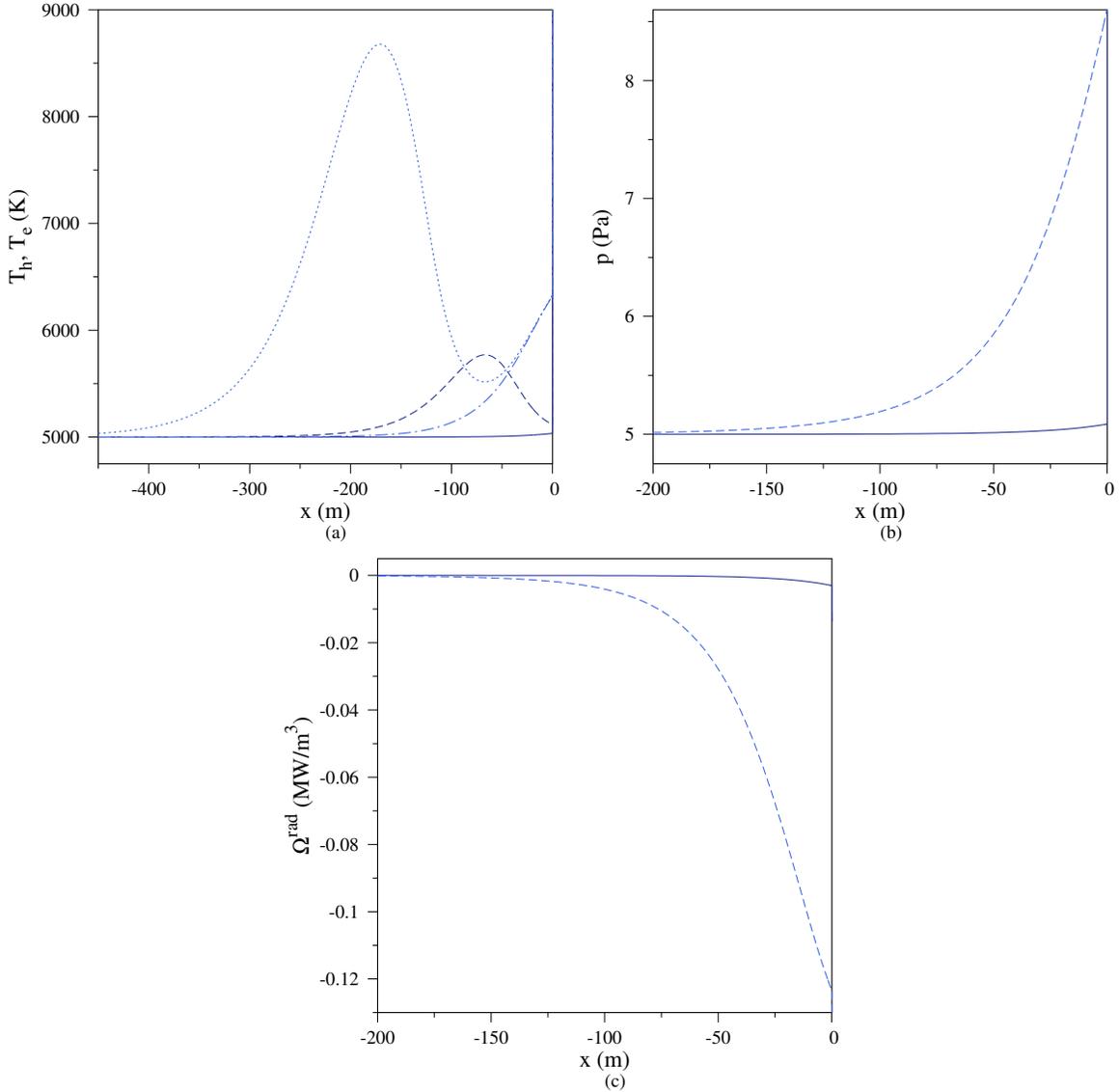}
\caption{Evolution of (a) the heavy-particle and free-electron temperatures, (b) the gas pressure and (c) the volumetric radiative loss term in the precursor for different shock speeds. In (a) solid line $\Th$ for $\uinf = \SI{40}{km/s}$, dashed line $\Te$ for $\uinf = \SI{40}{km/s}$, dot-dashed line $\Th$ for $\uinf = \SI{70}{km/s}$, dotted line $\Te$ for $\uinf = \SI{70}{km/s}$. In (b) and (c) solid line $\uinf = \SI{40}{km/s}$, dashed line $\uinf = \SI{70}{km/s}$ (StS model).}\label{fig:ThTe_p_zoom_prec_40_70}
\end{figure}

The precursor heating for the whole gas becomes substantial only at large shock speeds, as indicated by the rise of the heavy-particle temperature and pressure in Figs. \ref{fig:ThTe_p_zoom_prec_40_70}(a)-(b), and by the volumetric radiant loss term plotted in Fig. \ref{fig:ThTe_p_zoom_prec_40_70}(c). It is interesting to notice that at large speeds (\emph{i.e.} \SI{70}{km/s}), the free-electron temperature reaches a local minimum, and then rises again before reaching the shock. The observed behavior is due to the combined effect of the energy absorbed by radiation, which tends to increase the thermal energy of heavy-particles, and electron-proton elastic collisions which tend to keep the heavy-particle and free-electron temperatures together. The absorption of radiation in the precursor, in addition to altering the temperature and concentration profiles with respect to their free-stream values, also modifies the post-shock conditions due to the decrease and increase, respectively, of the Mach number and the pressure of the incoming gas. This is shown in detail in Table \ref{table:shock_jump}, which shows the pre and post-shock conditions with and without radiation. At large speeds, the absorption of radiation reduces the post-shock temperature by more than \SI{20}{\percent}. As opposed to pressure and temperature, the mass density is quite insensible to precursor radiation (see Table \ref{table:shock_jump}). This is a consequence of the (near) constancy of the flow velocity (not provided in Table \ref{table:shock_jump}), which can be explained by recalling that the absorbed radiation goes mainly into thermal energy of the material gas. For a standing shock, global mass conservation requires that mass flux $\rho u$ must be constant, so that the density cannot vary appreciably when the flow velocity is changed by a small amount.   

\begin{figure}[htb!]
\centering
\includegraphics[bb=0 0 536 254,clip,scale=0.82,keepaspectratio]{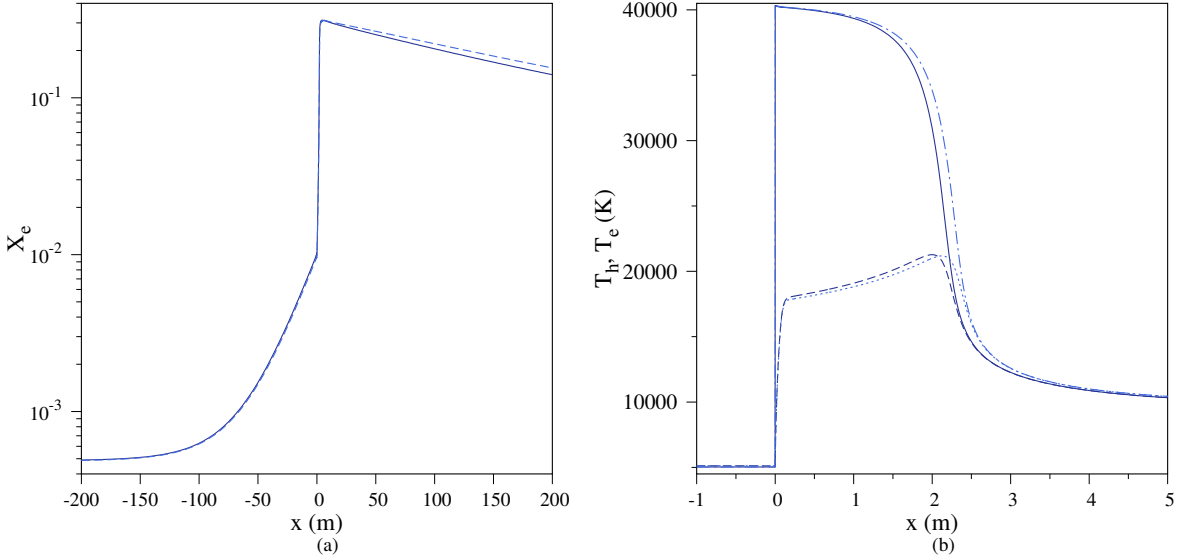}
\caption{Electron mole fraction (a) and temperatures (b)  switching bound-bound radiation on/off. In (a) solid line with bound-bound radiation, dashed line w/o bound-bound radiation. In (b) solid line $\Ti{h}$ with bound-bound radiation, dashed line $\Ti{e}$ with bound-bound radiation, dot-dashed line $\Ti{h}$ w/o bound-bound radiation, dotted line $\Ti{e}$ w/o bound-bound radiation ($\uinf = \SI{40}{km/s}$; StS model).}\label{fig:ThTeXe_40_line}
\end{figure}

\begin{figure}[ht!]
\centering
\includegraphics[bb=0 0 544 254,clip,scale=0.82,keepaspectratio]{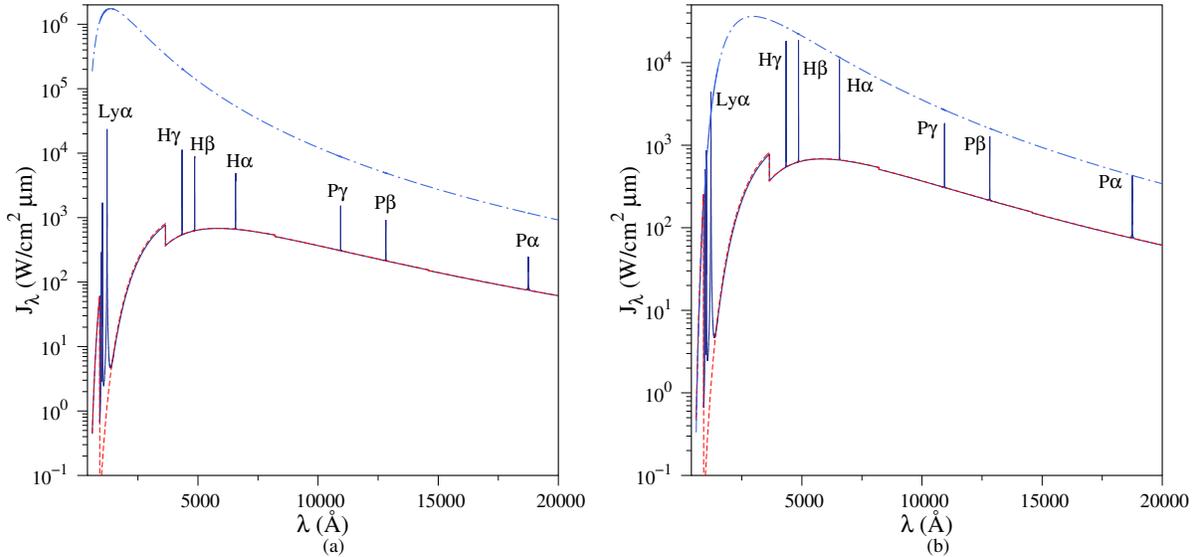}
\caption{Average monochromatic intensity at the locations (a) $x = \SI{2}{m}$ (internal relaxation region) and (b) $x = \SI{10}{m}$ (radiative cooling region) behind the shock switching on/off bound-bound radiation: solid line with bound-bound radiation, dashed line w/o bound-bound radiation. The dot-dashed line represents the Planck function at the free-electron temperature, $B_{\lambda}(\Ti{e})$, when including bound-bound radiation ($\uinf = \SI{40}{km/s}$; StS model).}\label{fig:Jlam_40_with_wo_line}
\end{figure}

Before moving to the comparison with the reduced-order ME models discussed in Sec. \ref{sec:Res_ME}, it is worth to briefly analyze the effect of including/excluding bound-bound transitions from the calculations. Figure \ref{fig:ThTeXe_40_line} shows a sample of this investigation for the temperatures and electron mole fraction profiles obtained for a shock propagating at \SI{40}{km/s}. The inclusion of line radiation has the effect of shrinking the internal relaxation region and speeding up radiative recombination. The observed behavior originates from emission in the Balmer and Paschen lines, which are optically thin in the post-shock region, as shown in Fig. \ref{fig:Jlam_40_with_wo_line}, which shows the average monochromatic intensity at $x = \SI{2}{m}$ and $x = \SI{10}{m}$. The effect of an optically thin line is to provide an efficient channel to depopulate the higher bound electronic states. This is accompanied by a loss of energy which, in turn, induces the faster temperature drop observed in Fig. \ref{fig:ThTeXe_40_line}(b). At the same time, in the radiative cooling region, the depopulation of high-lying states due to line emission indirectly favors radiative recombination because it makes quantity $(\ni{i} \, \kf{i}{}{\textsc{pi}} - \nelec \, \np \, \kf{i}{}{\textsc{rr}})$ more negative for these states (see Eq. \eqnref{eq:omegae_PI_RR}).  

\begin{table}[h!]
 \begin{center}
 \caption{Pressure, temperature and density at the pre and post-shock locations switching on/off radiation (StS model). The pre-shock pressure, temperature and density are not provided in case w/o radiation as these quantities are given by their free-stream LTE values ($\pinf = \SI{5}{Pa}$, $\Tinf = \SI{5000}{K}$, $\rhoinf = \SI{1.21e-7}{kg/m^{-3}}$).\label{table:shock_jump}}
\begin{tabular}{l|lll|llllll}
\hline\hline
 & \multicolumn{3}{l|}{w/o radiation} & \multicolumn{6}{l}{with radiation} \\
 & \multicolumn{3}{l|}{post-shock}   & \multicolumn{3}{l}{pre-shock} & \multicolumn{3}{l}{post-shock}  \\  
\hline
$\uinf \, [\si{km/s}]$ &  $p \, [\si{Pa}]$ & $T \, [\si{K}]$ & $\rho \, [\si{kg/m^3}]$ & $p \, [\si{Pa}]$ & $T \, [\si{K}]$ & $\rho \, [\si{kg/m^3}]$ & $p \, [\si{Pa}]$ & $T \, [\si{K}]$ & $\rho \, [\si{kg/m^3}]$  \\
\hline
\num{40}  & \num{144.086} & \num{40683.75}  & \num{4.29e-07} & \num{5.087} & \num{5034.509} & \num{1.21e-7} & \num{142.79}  & \num{40331.96} & \num{4.28e-07} \\
\num{45}  & \num{182.69}  & \num{50347.42}  & \num{4.40e-7}  & \num{5.22}  & \num{5082.56}  & \num{1.21e-7} & \num{178.46}  & \num{49209.53} & \num{4.38e-07} \\
\num{50}  & \num{225.83}  & \num{6114.46}   & \num{4.48e-7}  & \num{5.46}  & \num{5171.12}  & \num{1.21e-7} & \num{214.91}  & \num{58231.51} & \num{4.45e-7} \\
\num{55}  & \num{273.52}  & \num{73076.09}  & \num{4.54e-07} & \num{5.89}  & \num{5320.73}  & \num{1.21e-7} & \num{249.15}  & \num{66616.85} & \num{4.45e-07} \\ 
\num{60}  & \num{325.74}  & \num{86142.43}  & \num{4.58e-07} & \num{6.56}  & \num{5548.98}  & \num{1.22e-7} & \num{278.84}  & \num{73744.47} & \num{4.52e-7} \\
\num{65}  & \num{382.51}  & \num{100343.87} & \num{4.62e-7}  & \num{7.50}  & \num{5892.28}  & \num{1.22e-7} & \num{306.054} & \num{80208.88} & \num{4.54e-7} \\
\num{70}  & \num{443.82}  & \num{115680.72} & \num{4.65e-7}  & \num{8.61}  & \num{6322.32}  & \num{1.22e-7} & \num{333.62}  & \num{86745.66} & \num{4.54e-7} \\
\hline\hline
  \end{tabular}
 \end{center}
\end{table}
\newpage
\subsection{Comparison between the StS and the ME models}\label{sec:Res_ME}
After analyzing the general features of the radiative shock waves considered in this work, predictions obtained by means of the ME models were systematically compared with the StS results. The comparison was first performed by excluding line radiation (Sec. \ref{sec:Res_ME_cont}), which was then added back into the final runs (Sec. \ref{sec:Res_ME_line_cont}
).

Before discussing the results, it is worth saying a few words on how the grouping was done in practice. As anticipated at the end of Sec. \ref{sec:RedMom}, one should avoid lumping together states which are coupled by strong radiative transitions. This is particularly true for optically thin lines (\emph{e.g.} $\text{H}\alpha$, $\text{H}\beta$). For this reason, the first and second energy groups (when two or more are used) were assigned to the $n=1$ and $n=2$ states, respectively, where $n$ stands for the principal quantum number \citep{Pauling_book_1935,Mihalas_StAtm_1978}. No internal temperatures are used for these groups in the case of the MEL model, allowing for a further reduction of the number of unknowns. The higher states ($n \geq 3$) are lumped based on Eq. \eqnref{eq:me_model_distl} (or Eq. \eqnref{eq:me_model_distu} for the MEU model). In what follows, the notations MEU($k$) and MEL($k$), are used to indicate the reduced-order model used and the number of groups, $k$.
\subsection{Continuum radiation only}\label{sec:Res_ME_cont}
\begin{figure}[ht!]
\centering
\includegraphics[bb=0 0 537 257,clip,scale=0.82,keepaspectratio]{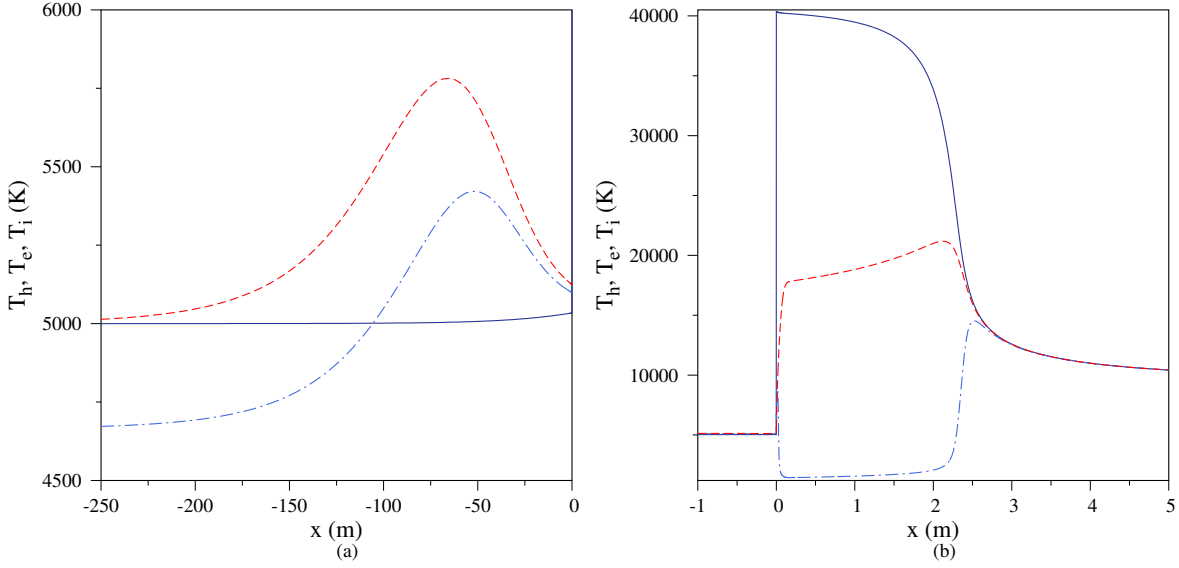}
\caption{Temperature evolution across (a) the precursor and (b) the internal relaxation region for the MEL(3) model: solid line $\Th$, dashed line $\Te$, dot-dashed line $\Ti{i}$ ($\uinf = \SI{40}{km/s}$; no bound-bound radiation).}\label{fig:ThTeTi_MEL3}
\end{figure}

Figure \ref{fig:ThTeTi_MEL3} shows the temperature evolution for a shock propagating at \SI{40}{km/s} when using the MEL(3) model. The (only) internal temperature, indicated by the dot-dashed line, refers to the states with $n \geq 3$. In the precursor, the high-lying levels are not in thermal equilibrium with free-electrons since $\Ti{i} \neq \Te$. This behavior is due to the concurrent action of photo-ionization from high-lying states and de-excitation collisions \citep{Foley_POF_1973}. The internal temperature replicates the observed non-monotone trend of that of the free-electrons. When the incoming gas reaches the shock, the high-lying states are almost in thermal equilibrium with free-electrons.

\begin{figure}[ht!]
\centering
\includegraphics[bb=0 0 536 254,clip,scale=0.82,keepaspectratio]{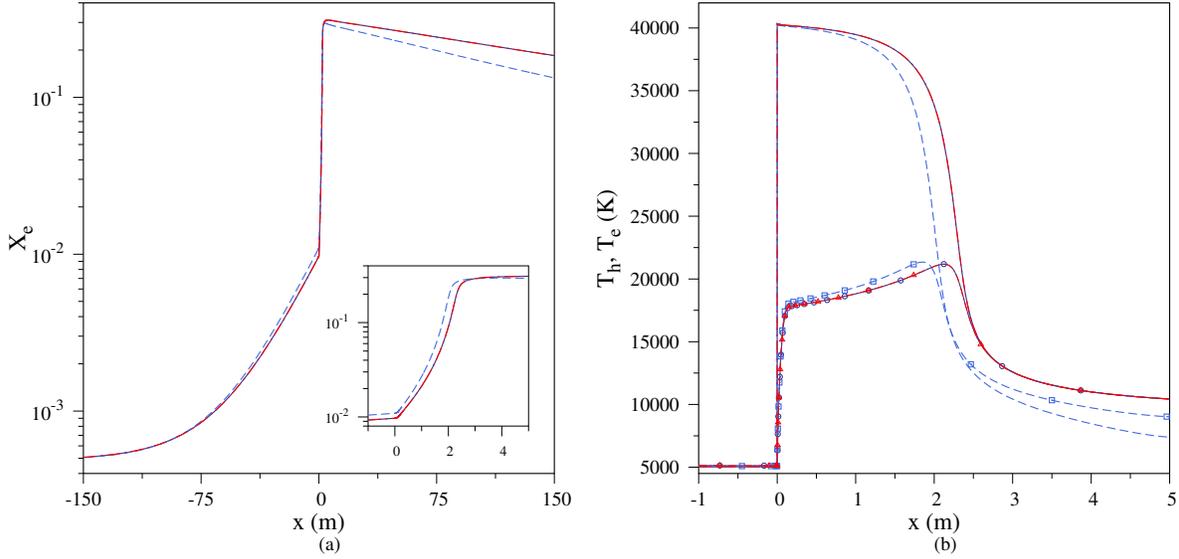}
\caption{Comparison between the StS, MEU(3) and MEL(3) model predictions for (a) the electron mole fraction and (b) the heavy-particle and free-electron temperatures. In (a) solid line StS model, dashed line MEU(3) model, dot-dashed line MEL(3) model. In (b) solid line $\Ti{h}$ StS model, dashed line $\Ti{h}$ MEU(3) model, dot-dashed line $\Ti{h}$ MEL(3) model, line with circle $\Ti{e}$ StS model, dashed line with square $\Ti{e}$ MEU(3) model, dot-dashed line with triangle $\Ti{e}$ MEL(3) model ($\uinf = \SI{40}{km/s}$; no bound-bound radiation).}\label{fig:XeThTe_cont_MEL_MEU_StS}
\end{figure}

The results of Fig. \ref{fig:ThTeTi_MEL3} were compared with the StS predictions in Fig. \ref{fig:XeThTe_cont_MEL_MEU_StS} for the electron mole fraction and the heavy-particle and free-electron temperatures. The MEU(3) solution is also reported. The MEL(3) model is in excellent agreement with the StS results. The two solutions are essentially indistinguishable. These results indicate that using three energy groups plus one internal temperature is sufficient to achieve an accurate prediction of radiative shock waves in hydrogen plasmas. In practical terms, this means that the number of unknowns is reduced by two orders of magnitude, which makes the MEL model very attractive for Computational Fluid Dynamics (CFD) applications. On the other hand, The MEU(3) predictions are in strong disagreement with the StS results. 

\begin{figure}[ht!]
\centering
\includegraphics[bb=0 0 538 246,clip,scale=0.82,keepaspectratio]{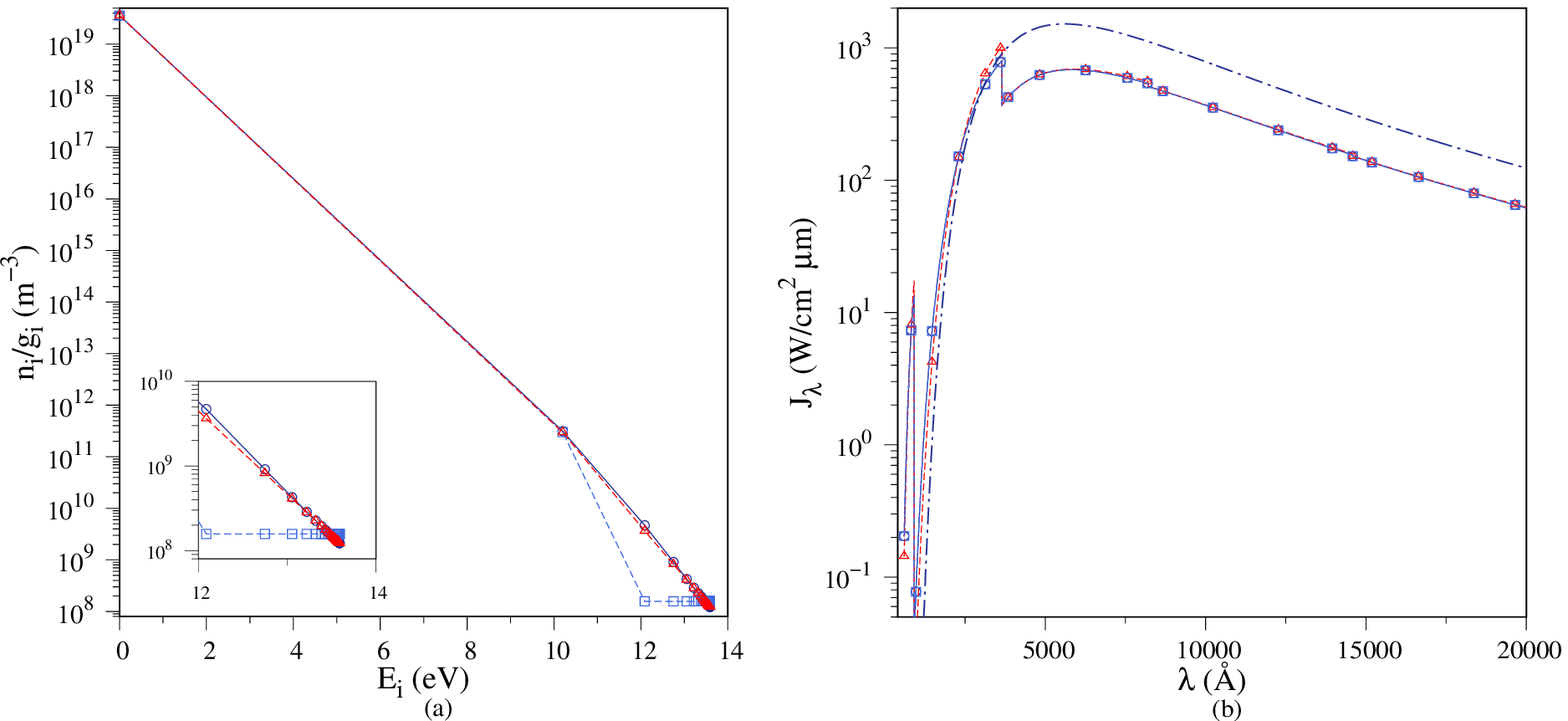}
\caption{Comparison between the StS, MEU(3) and MEL(3) model predictions for (a) the population of H bound levels and (b) the average monochromatic intensity at the location $x = \SI{-10}{m}$ ahead of the shock (near precursor region): line with circle StS model, line with square MEU(3) model, line with triangle MEL(3) model. The dot-dashed line in (b) represents the Planck function at the free-electron temperature, $B_{\lambda}(\Ti{e})$, for the StS model ($\uinf = \SI{40}{km/s}$; no bound-bound radiation).}\label{fig:pop_Jlam_prec_x_m10_cont}
\end{figure}

\begin{figure}[ht!]
\centering
\includegraphics[bb=0 0 538 246,clip,scale=0.82,keepaspectratio]{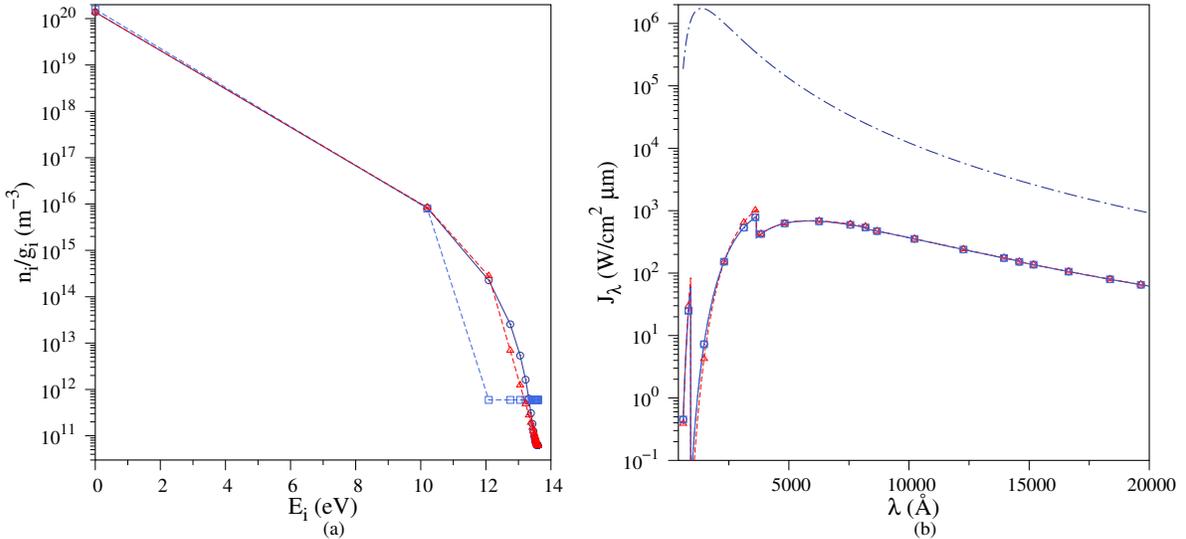}
\caption{Comparison between the StS, MEU(3) and MEL(3) model predictions for (a) the population of H bound levels and (b) the average monochromatic intensity at the location $x = \SI{2}{m}$ behind the shock (internal relaxation region): line with circle StS model, line with triangle MEU(3) model, line with square MEL(3) model. The dot-dashed line in (b) represents the Planck function at the free-electron temperature, $B_{\lambda}(\Ti{e})$, for the StS model ($\uinf = \SI{40}{km/s}$; no bound-bound radiation).}\label{fig:pop_Jlam_prec_x_p2_cont}
\end{figure}

\begin{figure}[ht!]
\centering
\includegraphics[bb=0 0 538 246,clip,scale=0.81,keepaspectratio]{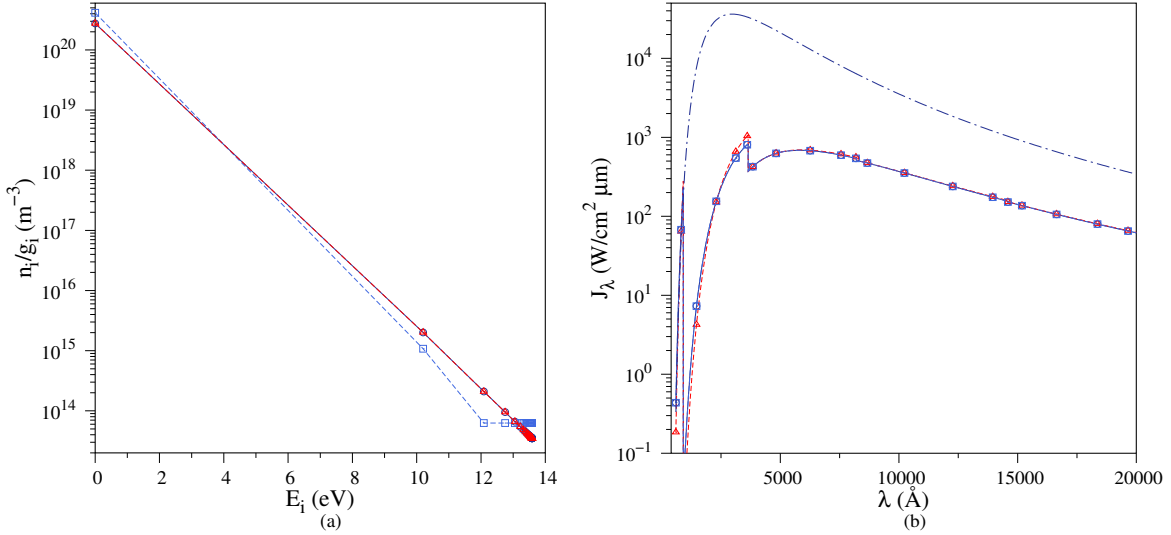}
\caption{Comparison between the StS, MEU(3) and MEL(3) model predictions for (a) the population of H bound levels and (b) the average monochromatic intensity at the location $x = \SI{20}{m}$ behind the shock (radiative cooling region): line with circle StS model, line with square MEU(3) model, line with triangle MEL(3) model. The dot-dashed line in (b) represents the Planck function at the free-electron temperature, $B_{\lambda}(\Ti{e})$, for the StS model ($\uinf = \SI{40}{km/s}$; no bound-bound radiation).}\label{fig:pop_Jlam_prec_x_p20_cont}
\end{figure}

To gain more insight from the comparison of Fig. \ref{fig:XeThTe_cont_MEL_MEU_StS}, the (normalized) populations of the internal energy states have been extracted at the locations $x = - 10, \, 2$ and \SI{20}{m} (see Figs. \ref{fig:pop_Jlam_prec_x_m10_cont}-\ref{fig:pop_Jlam_prec_x_p20_cont}). The first location (Fig. \ref{fig:pop_Jlam_prec_x_m10_cont}) refers to the near precursor, the second (Fig. \ref{fig:pop_Jlam_prec_x_p2_cont}) to the internal relaxation region, and the third (Fig. \ref{fig:pop_Jlam_prec_x_p20_cont}) to the radiative cooling region. In the same figures, the corresponding average monochromatic intensity is also provided. For the MEL and MEU models, the population distributions have been re-constructed based on the group number densities and internal temperatures using Eqs. \eqnref{eq:me_model_distl} and \eqnref{eq:me_model_distu}, respectively. The results show that the superior description of the MEL model lies in its ability to (almost) replicate the StS behavior of the high-lying states. This fact is of great importance, especially for the ionization within the internal relaxation region. The MEU model gives a poor description of the population dynamics due to the assumption of \emph{infinite} internal temperatures. As already stated in Sec. \ref{sec:RedME}, this hypothesis does not allow retrieval of equilibrium (\emph{i.e.} Boltzmann distribution) and has a strong impact on the temperature evolution in the radiative relaxation region.              

In this zone, where the degree of ionization decreases due to radiative recombination, the collisional rate among the gas particles is large enough to maintain equilibrium at the local free-electron temperature (see Fig. \ref{fig:pop_Jlam_prec_x_p20_cont}(a)). The same holds true for the radiation field only in the Lymann continuum, as shown by the observed departure from the local Planck function, $B_{\lambda}(\Te)$, in the Balmer and Paschen continuum (see Fig. \ref{fig:pop_Jlam_prec_x_p20_cont}(b)). The use of a uniform grouping prevents obtaining a Boltzmann distribution and leads to a higher population of the high-lying states and an under-predicted population for the $n = 2$ and $n = 3$ states (see Fig. \ref{fig:pop_Jlam_prec_x_p20_cont}(a)). This produces lower temperatures compared to the MEL(3) and StS solutions. The observed difference is not negligible and is also enhanced by the large the statistical weight of the high-lying states.\footnote{The non-relativistic solution of Schr\"{o}dinger's equation for an isolated hydrogen atom predicts that the energy and the statistical weight of a bound energy state with principal quantum number $n$ are, respectively, $\Ei{n} = - I_{\text{H}}/n^2$ and $g_n = 2 n^2$, where $I_{\text{H}}$ is the hydrogen ionization potential \citep{Pauling_book_1935}. Hence, the statistical weight grows rapidly with increasing energy.} The poor description of the level dynamics by the MEU(3) model has also an adverse effect on the predicted radiative signature of the plasma. This is demonstrated by the comparison in terms of the average monochromatic intensity given in Figs. \ref{fig:pop_Jlam_prec_x_m10_cont}(b), \ref{fig:pop_Jlam_prec_x_p2_cont}(b) and \ref{fig:pop_Jlam_prec_x_p20_cont}(b). At all locations, the difference is higher at the Lymann and Balmer edges. The above findings on the MEU(3) and MEL(3) models have been also obtained when analyzing cases at larger shock speeds. For the sake of brevity, these results have not been added to the manuscript. 

Before including the effects of line radiation on the reduced-order models (see Sec. \ref{sec:Res_ME_line_cont}), it was decided to investigate the sensitivity of the solution to the number of groups. The study included both the MEU and MEL models. 

\begin{figure}[ht!]
\centering
\includegraphics[bb=0 0 536 778,clip,scale=0.82,keepaspectratio]{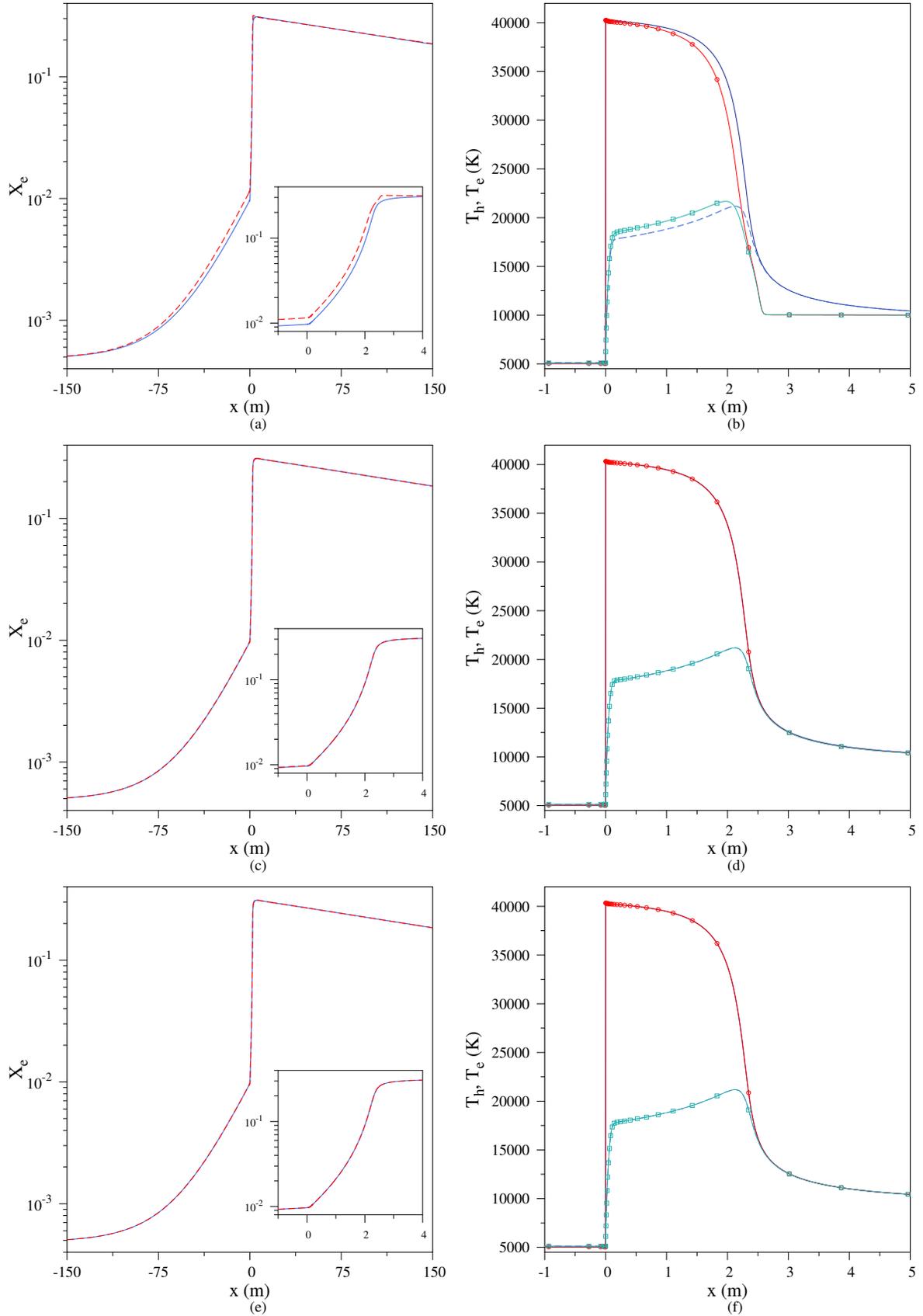}
\caption{Comparison between the StS and the MEL(1), (a)-(b), MEL(2), (c)-(d), and the MEL(3), (e)-(f), model predictions for the electron mole fraction, and the heavy-particle and free-electron temperatures. In (a), (c) and (e) solid line StS model, dashed line MEL($i$) model. In (b), (d) and (f) solid line $\Ti{h}$ StS model, dashed line $\Ti{e}$ StS model, line with circle $\Ti{h}$ MEL($i$) model, line with square $\Ti{e}$ MEL($i$) model ($\uinf = \SI{40}{km/s}$; no bound-bound radiation).}\label{fig:Xe_ThTe_cont_MEL_StS}
\end{figure}

\begin{figure}[ht!]
\centering
\includegraphics[bb=97 56 354 294,clip,scale=0.9,keepaspectratio]{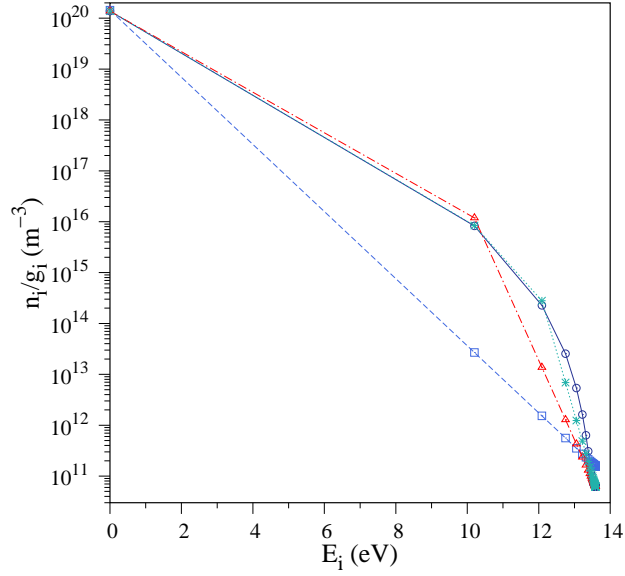}
\caption{Comparison between the StS and the MEL($i$) model predictions for the population of H bound levels at the location $x = \SI{2}{m}$ behind the shock (internal relaxation region): line with circle StS model, line with square MEL(1) model, line with triangle MEL(2) model, line with star MEL(3) model ($\uinf = \SI{40}{km/s}$; no bound-bound radiation).}\label{fig:pop_StS_MEL_x_p2m_post_shock_cont}
\end{figure}

Figure \ref{fig:Xe_ThTe_cont_MEL_StS} shows the free-electron mole fraction and all temperatures for the MEL(1), MEL(2) and MEL(3) models. When all the hydrogen bound states are lumped into one group (\emph{i.e.} MEL(1) model), it is not possible to account for the under-population experienced by the high-lying states in the precursor and within the internal relaxation region. As a matter of fact, the MEL(1) model forces the high-lying states to be in thermal equilibrium with free-electrons, resulting in a larger ionization rate (see Figs. \ref{fig:Xe_ThTe_cont_MEL_StS}(a)-\ref{fig:Xe_ThTe_cont_MEL_StS}(b)). The inclusion of one additional group (\emph{i.e.} MEL(2) model) for the states with $n \geq 2$ greatly improves the solution and leads essentially to the same result obtained with the MEL(3) model. The MEL(3) gives, however, a superior description of the population distribution as demonstrated by the comparison between the three models in Fig. \ref{fig:pop_StS_MEL_x_p2m_post_shock_cont}. It is worth noting that the lower accuracy for the population of the $n=2$ and $n=3$ states for the MEL(2) model plays a negligible role on gas quantities such as chemical composition and temperatures. This is no longer true when line radiation is taken into account (see Sec. \ref{sec:Res_ME_line_cont}). One may conclude that using only two energy groups is already enough to obtain an accurate prediction when neglecting line radiation.    

\begin{figure}[ht!]
\centering
\includegraphics[bb=0 0 527 254,clip,scale=0.82,keepaspectratio]{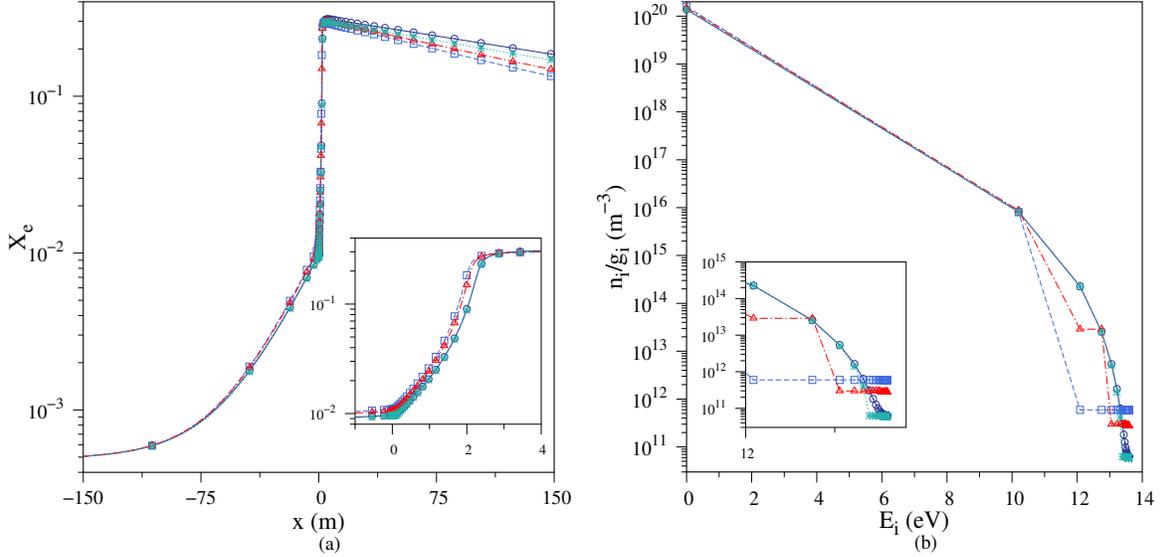}
\caption{Comparison between the StS and the MEU($i$) model predictions for (a) the electron mole fraction and (b) the population of H bound levels at the location $x = \SI{2}{m}$ behind the shock: line with circle StS model, line with square MEU(3) model, line with triangle MEU(7) model, line with star MEU(11) model ($\uinf = \SI{40}{km/s}$; no bound-bound radiation).}\label{fig:Xe_pop_StS_MEU}
\end{figure}

Figure \ref{fig:Xe_pop_StS_MEU} illustrates the results of the sensitivity study for the MEU model. In analogy with the MEL model, increasing the number of groups improves the solution. However, in the present case, this is achieved with a larger number of groups (11), which makes the MEU model less attractive for CFD applications. The use of eleven groups allows to accurately predict both temperatures and species concentrations in the precursor and in the internal relaxation region. However, some discrepancies with the StS results still appear in the radiative cooling region. As demonstrated earlier, these are caused by the intrinsic defect of the MEU model, namely, the impossibility of retrieving a Boltzmann distribution for the bound states. In view of these results, the MEU model has not been considered in the calculations performed including the effects of line radiation as discussed in Sec. \ref{sec:Res_ME_line_cont}. 

\subsection{Line and continuum radiation}\label{sec:Res_ME_line_cont}
\begin{figure}[ht!]
\centering
\includegraphics[bb=0 0 536 516,clip,scale=0.82,keepaspectratio]{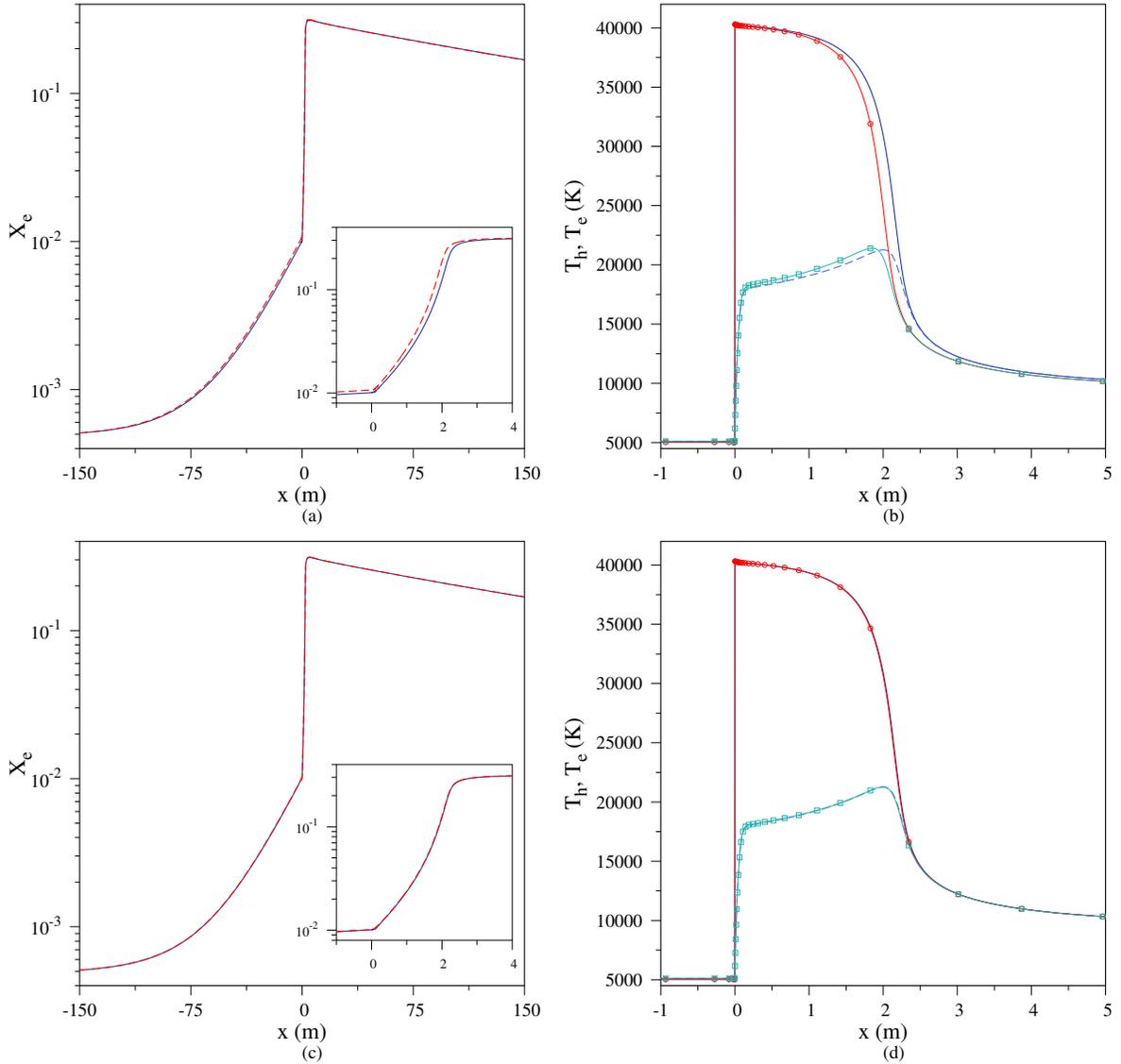}
\caption{Comparison between the StS and the MEL(2), (a)-(b), and the MEL(3), (c)-(d), model predictions for the electron mole fraction, and the heavy-particle and free-electron temperatures. In (a) and (c) solid line StS model, dashed line MEL($i$) model. In (b) and (d) solid line $\Ti{h}$ StS model, dashed line $\Ti{e}$ StS model, line with circle $\Ti{h}$ MEL($i$) model, line with square $\Ti{e}$ MEL($i$) model ($\uinf = \SI{40}{km/s}$).}\label{fig:Xe_ThTe_line_StS_MEL}
\end{figure}

\begin{figure}[ht!]
\centering
\includegraphics[bb=0 0 525 246,clip,scale=0.82,keepaspectratio]{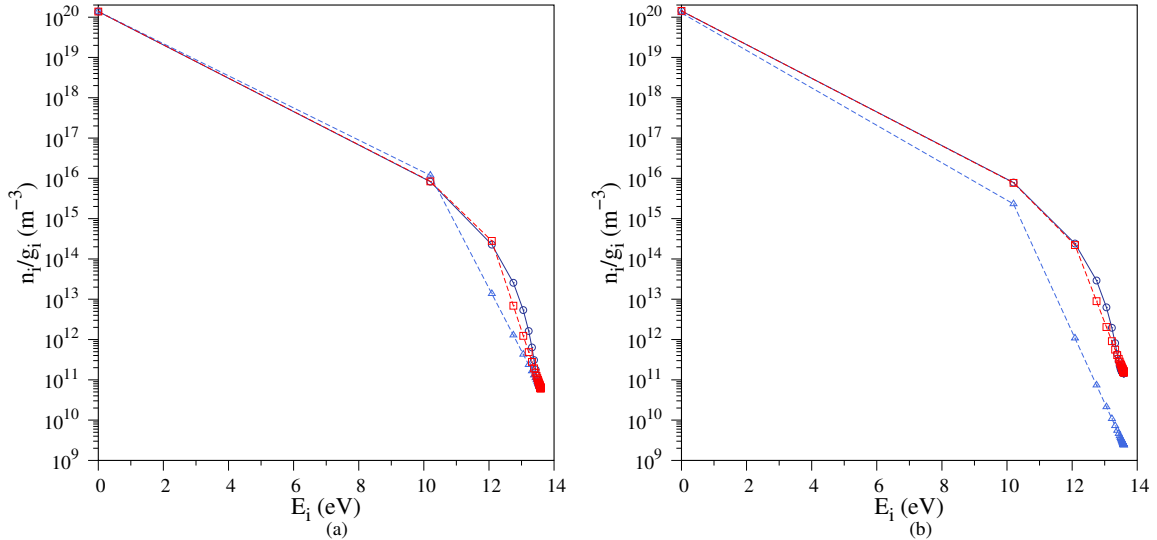}
\caption{Comparison between the StS, MEL(2) and MEL(3) model predictions for the population of H bound levels at the location $x = \SI{2}{m}$ behind the shock (internal relaxation region): line with circle StS model, line with square MEL(2) model, line with triangle MEL(3). In (a) w/o bound-bound radiation, in (b) with bound-bound radiation ($\uinf = \SI{40}{km/s}$).}\label{fig:pop_StS_MEL_x_p2m_line_cont}
\end{figure}

The sensitivity analysis at the end of Sec. \ref{sec:Res_ME_line_cont} has shown that two energy groups are sufficient to obtain an accurate solution when accounting only for continuum radiation. To test whether this finding is retrieved when atomic lines are added, the StS and MEL(2) solutions have been compared in  Fig. \ref{fig:Xe_ThTe_line_StS_MEL} for the electron mole fraction and heavy-particle and free-electron temperatures. In the same picture the MEL(3) solution is also shown. Overall, the MEL(2) solution now shows a sensible departure from the StS prediction, the difference being larger in the internal relaxation region (see \ref{fig:Xe_ThTe_line_StS_MEL}(a)-(b)). In the precursor, the MEL(2) model slightly over-estimates photo-ionization, while in the radiative cooling region the disagreement with the StS solution is barely noticeable. The MEL(3) model is again in excellent agreement with the StS prediction (see Figs. \ref{fig:Xe_ThTe_line_StS_MEL}(c)-(d)). The performance degradation of the MEL(2) model is not surprising and is the result of grouping together the $n=2$ and $n=3$ states. This annihilates the effect of radiative decay through the optically thin $\text{H}\alpha$ line, which, as shown in Fig. \ref{fig:ThTeXe_40_line}, has a non negligible impact on the internal relaxation and radiative cooling regions. On the other hand, in the MEL(3) model, the $n=2$ and $n=3$ states are grouped separately, thus allowing for a more accurate prediction of bound-bound radiative losses. The above qualitative arguments are confirmed when comparing the StS and MEL(2-3) population distributions as done in Fig. \ref{fig:pop_StS_MEL_x_p2m_line_cont}. The above picture refers to the location $x = \SI{2}{m}$ behind the shock (internal relaxation region), and, for the sake of completeness, also shows results obtained when neglecting line radiation. It is readily seen that, when the $n = 2$ and $n = 3$ states are lumped together, the accuracy of the reconstructed distribution strongly degrades. In particular, the population of the above states (and also those close to the ionization limit) is under-predicted by at least one order of magnitude. It is worth noticing that the observed departure between the StS and MEL(3) predictions for the population of the $n=3$ and $n=4$ states (see Fig. \ref{fig:pop_StS_MEL_x_p2m_line_cont}(b)) has essentially no effect on the solution accuracy. This can be explained by recalling that, in general, the Einstein coefficient for the $\text{H}\alpha$ line is one order of magnitude larger than of the $\text{H}\beta$ and $\text{H}\gamma$ lines \citep{Mihalas_StAtm_1978}.  
\begin{figure}[ht!]
\centering
\includegraphics[bb=0 0 554 522,clip,scale=0.82,keepaspectratio]{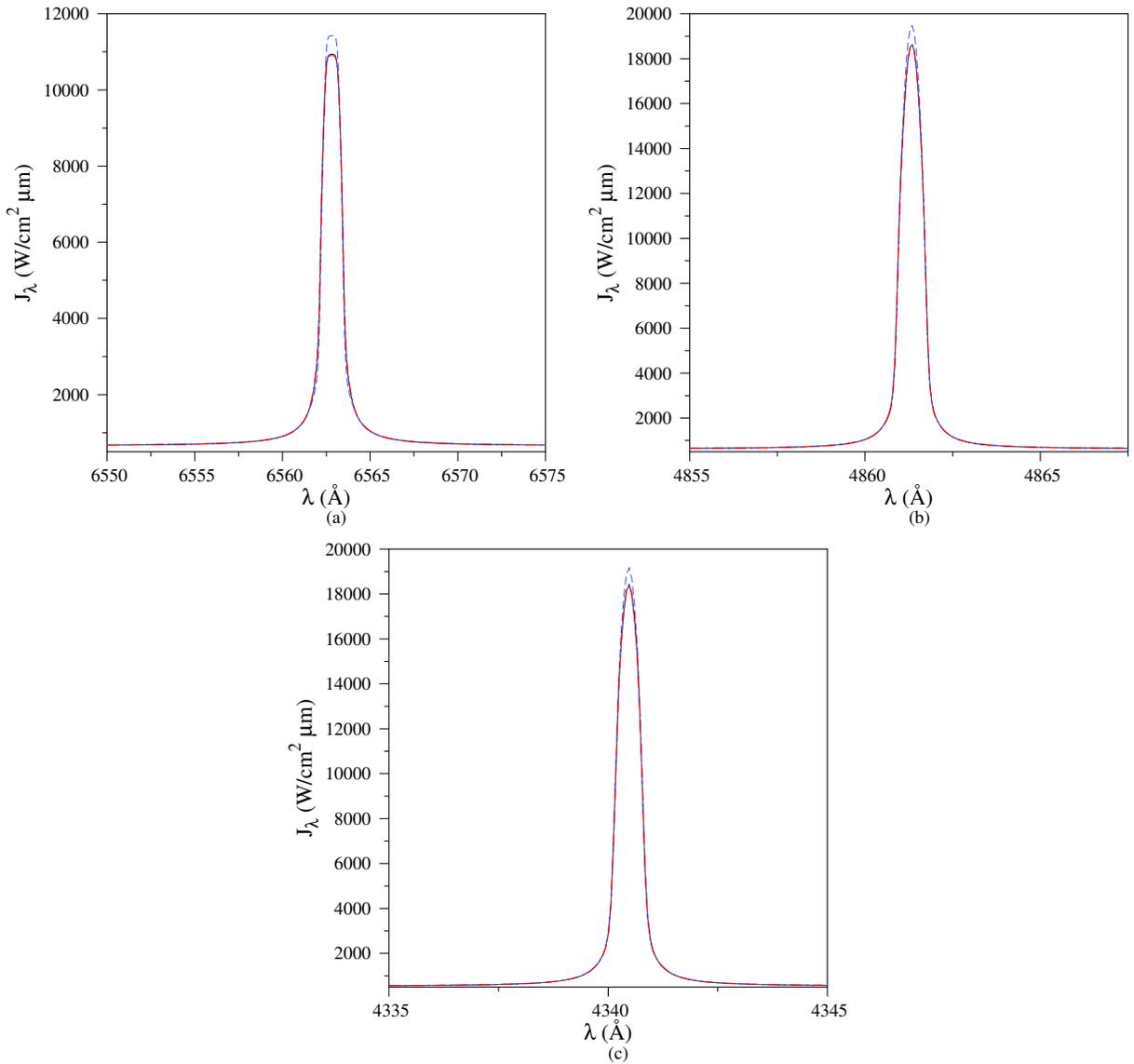}
\caption{Comparison between the StS, MEL(2) and MEU(3) model predictions for the average monochromatic intensity near the (a) H${\alpha}$, (b) H${\beta}$ and (c) H${\gamma}$ atomic lines at the location $x = \SI{10}{m}$ behind the shock (radiative cooling region): solid line StS model, dashed line MEL(2) model, dot-dashed line MEL(3) model ($\uinf = \SI{40}{km/s}$).}\label{fig:Habg_StS_MEL_x_p10m_post_shock_line}
\end{figure}
\begin{figure}[ht!]
\centering
\includegraphics[bb=0 0 529 516,clip,scale=0.82,keepaspectratio]{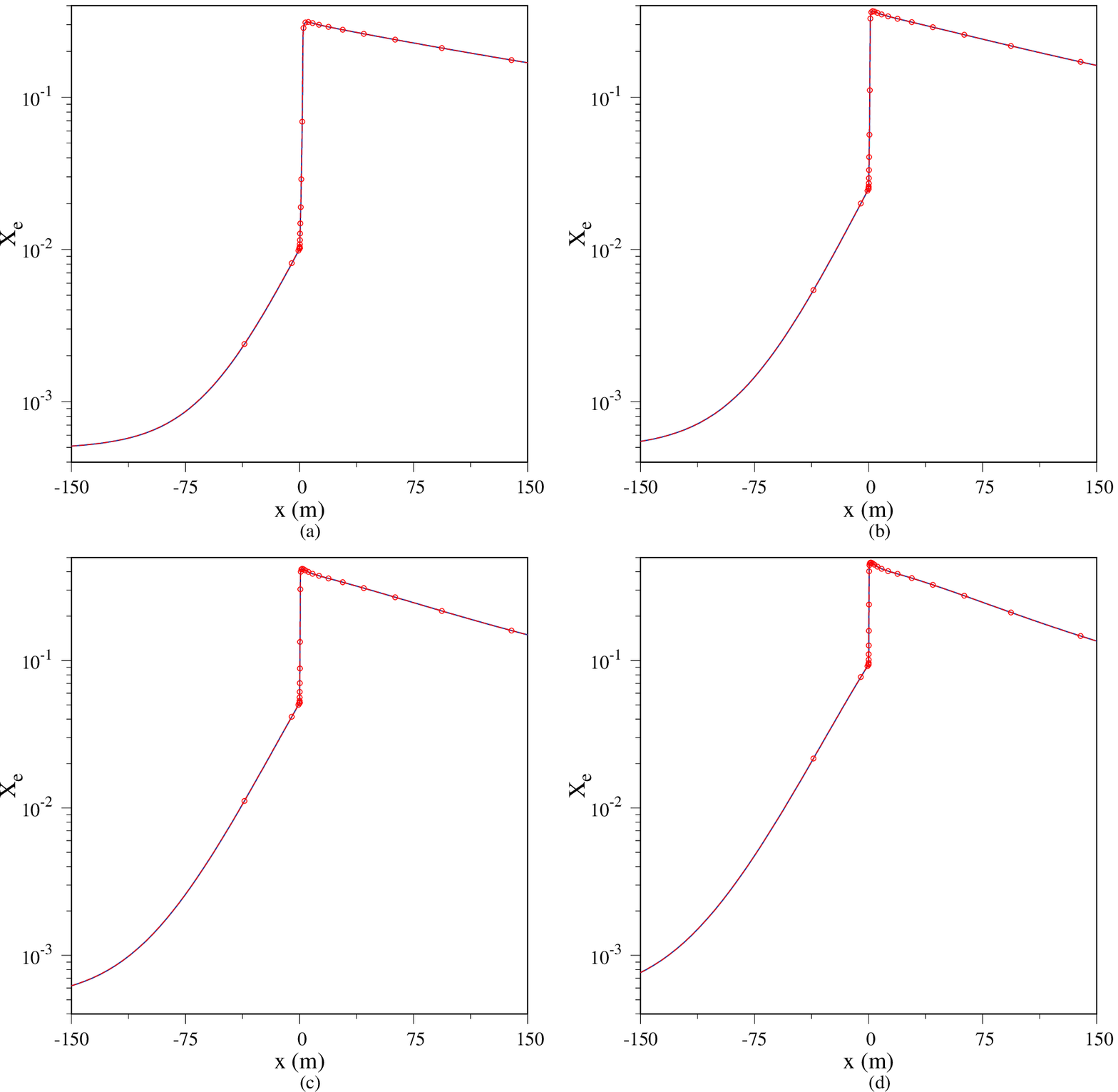}
\caption{Comparison between the StS and the MEL(3) model predictions for the electron mole fraction for (a) $\uinf = \SI{40}{km/s}$, (b) $\uinf = \SI{45}{km/s}$, (c) $\uinf = \SI{50}{km/s}$, (d) $\uinf = \SI{55}{km/s}$: solid line StS model, line with circle MEL(3) model.}\label{fig:Xe_StS_MEL_40_55}
\end{figure}

Figure \ref{fig:Habg_StS_MEL_x_p10m_post_shock_line} compares the average monochromatic intensity around the $\text{H}\alpha$, $\text{H}\beta$ and $\text{H}\gamma$ lines predicted by the StS and MEL(2-3) models at the location $x = \SI{10}{m}$ (radiative cooling region). Despite the good agreement observed for gas quantities (such as the electron mole fraction shown in Fig. \ref{fig:Xe_ThTe_line_StS_MEL}), the MEL(2) model systematically overestimates the peak value of the intensity of all lines. The MEL(3) model is, on the other hand, in excellent agreement with the StS results. 

The accuracy of the MEL(3) model has been further confirmed by repeating the calculations treated in this section at larger shock speeds. This is demonstrated in Fig. \ref{fig:Xe_StS_MEL_40_55} which compares the StS and MEL(3) electron mole fraction profiles for increasing shock speeds. In analogy with what is observed for the \SI{40}{km/s} case, the two solutions overlap.  
\section{Conclusions}\label{sec:Concl}
A State-to-State kinetic model for NLTE astrophysical hydrogen plasmas has been constructed by collecting the most up-to-date \emph{ab-initio} and/or experimental data. Based on the Maximum Entropy (ME) principle, the complexity of the StS kinetic model has been reduced by lumping the bound energy states of hydrogen in energy groups. Two different grouping strategies have been considered in this work: uniform (MEU model) and Maxwell-Boltzmann (MEL model). The reduced set of governing equations for the material gas have been obtained based on a moment method. The accuracy of the MEU/MEL models has been tested by means of a systematic comparison with the StS predictions. Applications considered the flow across radiative shock waves for conditions typical of pulsating stars. The results have shown that, with the use of only two to three energy groups, the MEL model is in excellent agreement with the StS predictions. To be more specific, two energy groups are already enough to achieve an accurate solution when neglecting line radiation. The inclusion of atomic lines requires the adoption of one additional group to account for the optically thin losses in the Balmer and Paschen lines. This makes the MEL very attractive for potential CFD applications. The MEU model, on the other hand, is less accurate than the MEL model and requires the adoption of a larger number of groups to achieve a fair agreement with the StS model. The persistent disagreement (even with a large number of groups) is due to the assumption of a uniform distribution which prevents retrieving a Boltzmann distribution in the radiative cooling region.                     
\section*{Acknowledgements}
The research of A. M. was supported by the NASA Award grant NNX 14AN44G. The research of M. P. was supported by the NASA grant NNX15AQ57A. The views and conclusions contained herein are those of the authors  and should not  be interpreted  as necessarily representing the official policies or endorsements, either expressed or implied, of NASA.
\newpage
\appendix
\section{Thermodynamics, kinetics and radiation of the reduced-order models}\label{sec:app}
This Appendix provides the thermodynamic and kinetic constitutive relations for the reduced-order NLTE models developed in Sec. \ref{sec:Red}. These relations are obtained by substituting the group distributions \eqnref{eq:me_model_distl}/\eqnref{eq:me_model_distu} in the expressions for thermodynamic properties, source terms and emission/absorption coefficients for the StS model. Only the MEL model is considered. The corresponding formulas for the MEU model can be readily obtained taking the limit of infinite internal temperatures (as already explained in Sec. \ref{sec:RedME}). For the sake of a more concise notation, the following group averaging operators are introduced:
\be 
\avg{\alpha_i}{k} = \fr{1}{\Qstarti{k}{T_k}{}} \sum_{i \in \mathcal{I}_k} \alpha_i \, \gi{i} \, \exp\left( -\fr{\Ei{i}}{\kb T_k}\right), \quad \avgt{\beta_{ij}}{kl} = \sum_{j \in \mathcal{I}_l} \avg{\beta_{ij}}{k},
\ee 
$k, \, l \in \mathcal{K}$, where the symbols $\alpha_i$ and $\beta_{ij}$ stand for the quantities being averaged. To avoid ambiguity in the evaluation of the averages $\avgt{\beta_{ij}}{}$, the dummy index of the internal sum is set equal to the first lower-script (\emph{i.e.} $i$ in this case).   
\subsection{Thermodynamics}
The gas pressure is always given by Dalton's law, $p = \ni{\text{e}} \kb \Te + \ni{\text{h}} \kb \Th$, where the heavy-particle number density is obtained by summing the group and proton contributions, $\ni{\text{h}} =  \sum_{k \in \mathcal{K}} \nti{k} + n_+$. In view of Eq.  \eqnref{eq:me_model_distl}, the heavy-particle thermal energy density \eqnref{eq:rhoe_he} becomes:
\be 
\rho e_{\text{h}} = \fr{3}{2} \ph + \sum_{k \in \mathcal{K}} \nti{k} \avg{\Ei{i}}{k} + \np \Ep.
\ee
\subsection{Collisional kinetics}
The group mass production terms due to collisional excitation and ionization are obtained via the substitution of Eq. \eqnref{eq:me_model_distl} in \eqnref{eq:omegai_EIon} and then summing the result obtained over all the levels within each group (\emph{i.e.} $\omegati{k} = \sum_{i \in \mathcal{I}_k} \omegai{i} $ ). After some algebraic manipulation, one obtains:
\begin{IEEEeqnarray}{rCl}
\omega^{\text{col}}_{\text{e}} & = & \sum_{k \in \mathcal{K}} \nelec \, [\nti{k} \, \ktf{k}{}{\textsc{i}}(\Te,T_k)  - \nelec \, \np \, \ktf{k}{}{\textsc{r}}(\Te)], \quad \omega^{\text{col}}_+ = \omega^{\text{col}}_{\text{e}}, \\
{\tilde{\omega}}^{\text{col}}_k  & = & - \nelec \, [\nti{k} \, \ktf{k}{}{\textsc{i}}(\Te, T_k)  - \nelec \, \np \ktf{k}{}{\textsc{r}}(\Te)] - \sum_{\substack{l \in \mathcal{K}\\ l \ne k}} \nelec \, [\nti{k} \, \ktf{k}{l}{\textsc{e}}(\Te, T_k)  - \nti{l} \, \ktf{l}{k}{\textsc{e}}(\Te, T_l)], 
\end{IEEEeqnarray}  
$k \in \mathcal{K}$, where the group endothermic rate coefficients for excitation and ionization, and those for the related exothermic de-excitation and three-body recombination processes are: 
\begin{IEEEeqnarray}{rCl}
\ktf{k}{}{\textsc{i}}(\Te,T_k)  & = & \avg{\kf{i}{}{\textsc{i}}(\Te)}{k}, \quad\quad\,\,\,\, \ktf{k}{}{\textsc{r}}(\Te) = \ktf{k}{}{\textsc{i}}(\Te,\Te) \fr{\Qstari{\textsc{H}}{\Te}{\text{t}} \, \Qstarti{k}{\Te}{}}{g_{\text{e}} \, \Qstari{\text{e}}{\Te}{\text{t}} \, \gp \,  \Qstari{\textsc{H}^+}{\Te}{\text{t}}}\exp\left(\fr{\Ei{+}}{\kb \Te}  \right), \label{eq:kig} \\
\ktf{k}{l}{\textsc{e}}(\Te, T_k) & = & \avgt{\kf{i}{j}{\textsc{e}}(\Te)}{kl}, \quad \ktf{l}{k}{\textsc{e}}(\Te, T_l) = \avgt{\kf{j}{i}{\textsc{e}}(\Te)}{lk}, \label{eq:kdexg} 
\end{IEEEeqnarray}
$k, \, l \in \mathcal{K}$, where $k < l$ in Eq. \eqnref{eq:kdexg}. It is worth noticing that the group rate coefficients for ionization and three-body recombination \eqnref{eq:kig} are related by a detailed balance relation among the groups. The same holds true for excitation/de-excitation only in the case of thermal equilibrium with free-electrons (\emph{i.e.} $T_k = \Te, \, \forall \, k \in \mathcal{K}$). 

By applying the same procedure as above, one arrives at the following relations for the group energy transfer source terms due to ionization and excitation:
\be 
\tilde{\Omega}^{\textsc{i}}_k = - \sum_{k \in \mathcal{K}} \! \nelec \,  [\nti{k} \, \Gtf{k}{}{\textsc{i}}(\Te, T_k)  - \nelec \, \np  \Gtf{k}{}{\textsc{r}}(\Te)], \quad 
\tilde{\Omega}^{\textsc{e}}_k = - \sum_{k, \, l \in \mathcal{K}} \! \nelec \, [\nti{k} \, \Gtf{k}{l}{\textsc{e}_f}(\Te, T_k)  - \nti{l} \, \Gtf{l}{k}{\textsc{e}_b}(\Te, T_l)],  
\ee  
$k \in \mathcal{K}$, where the related transfer rates due to ionization, recombination and excitation/de-excitation are:
\begin{IEEEeqnarray}{rCl}
\Gtf{k}{}{\textsc{i}}(\Te, T_k)  & = & \avg{\Ei{i} \, \kf{i}{}{\textsc{i}}(\Te)}{k}, \quad\quad\,\,\,\, \Gtf{k}{}{\textsc{r}}(\Te) = \Gtf{k}{}{\textsc{i}}(\Te, \Te) \, \fr{\Qstari{\textsc{H}}{\Te}{\text{t}} \, \Qstarti{k}{\Te}{}}{g_{\text{e}} \, \Qstari{\text{e}}{\Te}{\text{t}} \, \gp \,  \Qstari{\textsc{H}^+}{\Te}{\text{t}}}\exp\left(\fr{\Ei{+}}{\kb \Te}  \right), \\
\Gtf{k}{l}{\textsc{e}_f}(\Te, T_k) & = & \avgt{\Ei{i} \, \kf{i}{j}{\textsc{e}}(\Te)}{kl}, \quad \Gtf{l}{k}{\textsc{e}_b}(\Te, T_l) = \avgt{\Ei{i} \, \kf{j}{i}{\textsc{e}}(\Te)}{lk}, \label{eq:Edexg} 
\end{IEEEeqnarray}
$k, \, l \in \mathcal{K}$. The use of the Boltzmann grouping relation \eqnref{eq:me_model_distl} in Eq. \eqnref{eq:Omecol_EI} allows re-writing the ionization and excitation energy transfer terms for the free-electron gas as:
\be \label{eq:omegak_Ex3}
\tilde{\Omega}^{\textsc{i}}_{\text{e}} = - \sum_{k \in \mathcal{K}} \tilde{\Omega}^{\textsc{i}}_k - \Ei{+} \, \omega^{\textsc{i}}_{+}, \quad \tilde{\Omega}^{\textsc{i}}_{\text{e}} = - \sum_{k \in \mathcal{K}} \tilde{\Omega}^{\textsc{e}}_k,
\ee  
where $\omega^{\textsc{i}}_{+} = \omega^{\text{col}}_{+}$.
\subsection{Radiative kinetics}
\paragraph{Bound-bound transitions} The substitution of Eq. \eqnref{eq:me_model_distl} in Eq.  \eqnref{eq:emisabs_bb} allows re-writing the monochromatic emission and absorption coefficients due to \textsc{bb} radiation as: 
\be 
\elam{\textsc{bb}} = \sum_{\substack{k, \, l \in \mathcal{K} \\ l \geq k}} \fr{\hp c}{4 \pi} \, \nti{l} \, \tilde{A}_{lk}(\lambda, T_l), \quad \klam{\textsc{bb}} = \sum_{\substack{k, \, l \in \mathcal{K} \\ l \geq k}} \fr{\hp}{4 \pi} \, [\nti{k} \, \tilde{B}_{kl}(\lambda, T_k) - \nti{l} \, \tilde{B}_{lk}(\lambda, T_l)],
\ee
where the group wavelength and temperature-dependent Einstein coefficients are defined as:
\be
\tilde{A}_{lk}(\lambda, T_l) = \avgt{ \Aji \, \philam{ji} }{lk} , \quad \tilde{B}_{lk}(\lambda, T_l) = \avgt{\lambda_{ji} \, \Bji \, \philam{ji}}{lk}, \quad \tilde{B}_{kl}(\lambda, T_k) = \avgt{\lambda_{ij} \, \Bij \, \philam{ji}}{kl}, \, k \leq l 
\ee
$k, \, l \in \mathcal{K}$, where use has been done of the obvious symmetry relations satisfied by the line emission/absorption profile and wavelength (\emph{i.e.} $\philam{ij} = \philam{ji}$ and $\lambda_{ij} = \lambda_{ji}$, respectively). The application of the above procedure to Eqs. \eqnref{eq:omegai_bb} and \eqnref{eq:Omegarad_bb} gives, respectively, the group mass production rates and the volumetric rate of energy loss of matter energy due to \textsc{bb} radiation:
 \begin{IEEEeqnarray}{rCl} 
{\tilde{\omega}}^{\textsc{bb}}_k  & = & \sum_{\substack{l \in \mathcal{K} \\ l > k}} \, \{\nti{l} \, \tilde{A}^{\text{m}}_{lk}(T_l) - [\nti{k} \, \tilde{B}^{\text{m}}_{kl}(T_k) - \nti{l} \, \tilde{B}^{\text{m}}_{lk}(T_l)]\} - \sum_{\substack{l \in \mathcal{K} \\ l < k}} \, \{\nti{k} \, \tilde{A}^{\text{e}}_{kl}(T_k) - [\nti{l} \, \tilde{B}^{\text{e}}_{lk}(T_l) - \nti{k} \, \tilde{B}^{\text{e}}_{kl}(T_k)]\} , \quad k \in \mathcal{K}, \\
\Omega^{\textsc{bb}} & = & \sum_{\substack{k, \, l \in \mathcal{K} \\ l \geq k}} \, \{\nti{l} \, \tilde{A}^{\text{e}}_{lk}(T_l) - [\nti{k} \, \tilde{B}^{\text{e}}_{kl}(T_k) - \nti{l} \, \tilde{B}^{\text{e}}_{lk}(T_l)]\},
\end{IEEEeqnarray}
where the group mass production and energy transfer Einstein coefficients are given by the following Boltzmann averages: 
\begin{IEEEeqnarray}{rCl}
\tilde{A}^{\text{m}}_{lk}(T_l) & = & \avgt{A^{\text{m}}_{ji}}{lk}, \quad \tilde{A}^{\text{e}}_{lk}(T_l) = \avgt{A^{\text{e}}_{ji}}{lk}, \\
\tilde{B}^{\text{m}}_{lk}(T_l) & = & \avgt{B^{\text{m}}_{ji}}{lk}, \quad \tilde{B}^{\text{e}}_{lk}(T_l) = \avgt{B^{\text{e}}_{ji}}{lk}, \\
\tilde{B}^{\text{m}}_{kl}(T_k) & = & \avgt{B^{\text{m}}_{ij}}{kl}, \quad \tilde{B}^{\text{e}}_{kl}(T_k) = \avgt{B^{\text{e}}_{ij}}{kl},
\end{IEEEeqnarray}
$k, \, l \in \mathcal{K}$, where $k \leq l$.
\paragraph{Bound-free/free-bound transitions} The substitution of Eq. \eqnref{eq:me_model_distl} in Eqs. \eqnref{eq:emiss_fb}-\eqnref{eq:abs_bf} allows re-writing the monochromatic emission and absorption coefficients due to \textsc{bf} and \textsc{fb} radiation as:
\begin{IEEEeqnarray}{rCl}
\elam{\textsc{fb}} & = & \fr{\hpf \, c^2 \, \ni{\text{e}} \, \np}{\lambda^5 (2 \pi \mi{\text{e}} \kb \Te )^{3/2} \gp} \exp\left(\fr{\Ei{+}}{\kb \Te} - \fr{\hp c}{\kb \Te \lambda} \right) \sum_{k \in \mathcal{K}} \Qstarti{k}{\Te}{} \, {\tilde{\sigma}}^{\textsc{pi}}_{k}(\lambda, \Te), \\
\klam{\textsc{bf}} & = & \sum_{k \in \mathcal{K}} {\tilde{\sigma}}^{\textsc{pi}}_{k}(\lambda, T_k) \left[\nti{k} -\fr{1}{2} \fr{\hpc \, \nelec \, \np}{(2 \pi \mi{\text{e}} \kb \Te )^{3/2}} \fr{\Qstarti{k}{\Te}{}}{\gp} \fr{{\tilde{\sigma}}^{\textsc{pi}}_{k}(\lambda, \Te)}{{\tilde{\sigma}}^{\textsc{pi}}_{k}(\lambda, T_k)} \exp\left(\fr{\Ei{+}}{\kb \Te} - \fr{\hp c}{\kb \Te \lambda} \right) \right],
\end{IEEEeqnarray}
where quantity ${\tilde{\sigma}}^{\textsc{pi}}_{k}$ stands for the temperature dependent group photo-ionization cross-section. The former is computed via the Boltzmann average ${\tilde{\sigma}}^{\textsc{pi}}_{k} = \avg{\sigma^{\textsc{pi}}_{i}}{k}$. The notation ${\tilde{\sigma}}^{\textsc{pi}}_{k} (\lambda, \Te)$ indicates that the Boltzmann average has to be evaluated at the free-electron temperature.

The mass production terms are:
\begin{IEEEeqnarray}{rCl}
\omega^{\textsc{bf}/\textsc{fb}}_{\text{e}} & = & \sum_{k \in \mathcal{K}} [\nti{k} \, \ktf{k}{}{\textsc{pi}}(\Jlam, T_k) - \nelec \, \np \, \ktf{k}{}{\textsc{rr}}(\Te, \Jlam) ], \quad \omega^{\textsc{bf}/\textsc{fb}}_+ = \omega^{\textsc{bf}/\textsc{fb}}_{\text{e}}, \\
{\tilde{\omega}}^{\textsc{bf}/\textsc{fb}}_k & = &  - [\nti{k} \, \ktf{k}{}{\textsc{pi}}(\Jlam, T_k) - \nelec \, \np \, \ktf{k}{}{\textsc{rr}}(\Te, \Jlam)], 
\end{IEEEeqnarray}
$k \in \mathcal{K}$, where the group rate coefficients for photo-ionization and radiative recombination read $\ktf{k}{}{\textsc{pi}} = \avg{\kf{i}{}{\textsc{pi}}}{k}$ and $\ktf{k}{}{\textsc{rr}} = \sum_{i \in \mathcal{I}_k} \kf{i}{}{\textsc{rr}}$, respectively. By setting the elementary photo-ionization cross-sections to zero for wavelengths above the photo-ionization limit (\emph{i.e.} $\sigma^{\textsc{pi}}_{i} = 0$ for $\lambda > \lambda_{+i}$), it is possible to extend the wavelength integrals to infinity and exchange the order between summation and integration when computing the rate coefficients $\ktf{k}{}{\textsc{pi}}$ and $\ktf{k}{}{\textsc{rr}}$. Performing these operations one obtains: 
\begin{IEEEeqnarray}{rCl}
\ktf{k}{}{\textsc{pi}}(\Jlam, T_k) & = & \fr{4 \pi}{\hp c} \int\limits_{0}^{+\infty} \! {\tilde{\sigma}}^{\textsc{pi}}_{k}(\lambda, T_k) \, \Jlam \, \lambda \, d\lambda, \quad \ktf{k}{}{\textsc{rr}}(\Te, \Jlam) = \ktf{k}{}{\textsc{rr}-\text{s}}(\Te) + \ktf{k}{}{\textsc{rr}-\text{i}}(\Te, \Jlam), \label{eq:kPIg} \\
\ktf{k}{}{\textsc{rr}-\text{s}}(\Te) &= & \sqrt{\fr{2}{\pi}} \fr{\Qstarti{k}{\Te}{}}{\gp} \fr{ \hpc c }{(\mi{\text{e}} \kb \Te )^{3/2}}  \exp\left(\fr{\Ei{+}}{\kb \Te}\right) \int\limits_{0}^{+\infty} \! \fr{{\tilde{\sigma}}^{\textsc{pi}}_{k}(\lambda, \Te)}{\lambda^4} \, \exp\left(- \fr{\hp c}{\kb \Te \lambda} \right) d\lambda,\\
\ktf{k}{}{\textsc{rr}-\text{k}}(\Te, \Jlam) & = & \sqrt{\fr{1}{2\pi}} \fr{\Qstarti{k}{\Te}{}}{\gp} \fr{\hps}{(\mi{\text{e}} \kb \Te )^{3/2}c} \exp\left(\fr{\Ei{+}}{\kb \Te}\right) \int\limits_{0}^{+\infty} \! {\tilde{\sigma}}^{\textsc{pi}}_{k}(\lambda, \Te) \, \lambda \, \Jlam \, \exp\left(- \fr{\hp c}{\kb \Te \lambda} \right) d\lambda, \label{eq:kRRg}
\end{IEEEeqnarray} 
$k \in \mathcal{K}$. Equations \eqnref{eq:kPIg}-\eqnref{eq:kRRg} show that the group rate coefficients for photo-ionization and radiative recombination are formally identical to those for the StS model \eqnref{eq:kPI}-\eqnref{eq:kRR} and that the former can be obtained from the latter by replacing the statistical weights with the group partition functions, and the elementary cross-sections with the group cross-sections. 

The group energy transfer terms due to photo-ionization and radiative recombination are:
\be
\tilde{\Omega}^{\textsc{bf/fb}}_k = - [\nti{k} \, \Gtf{k}{}{\textsc{pi}}(\Jlam, T_k) - \nelec \, \np \, \Gtf{k}{}{\textsc{rr}}(\Jlam, \Te)],
\ee
$k \in \mathcal{K}$, where energy transfer rates due to photo-ionization and radiative recombination are $\Gtf{k}{}{\textsc{pi}} = \avg{\Ei{i} \, \kf{i}{}{\textsc{pi}}}{k}$ and $\Gtf{k}{}{\textsc{rr}} = \sum_{i \in \mathcal{I}_k} \Ei{i} \, \kf{i}{}{\textsc{rr}}$, respectively.

By repeating the above procedure, the net volumetric energy loss rates for the material gas \eqnref{eq:OmegaBF_en} and free-electrons \eqnref{eq:OmegaBF_en_el} become:
\begin{IEEEeqnarray}{rCl}
\Omega^{\textsc{bf}/\textsc{fb}} & = & \sum_{k \in \mathcal{K}} \Bigg[ \sqrt{\fr{1}{2\pi}} \fr{\Qstarti{k}{\Te}{}}{\gp} \fr{\hpc \, \nelec \, \np }{(\mi{\text{e}} \kb \Te )^{3/2}}  \exp\left(\fr{\Ei{+}}{\kb \Te}\right) \int\limits_{0}^{+\infty} \! {\tilde{\sigma}}^{\textsc{pi}}_{k}(\lambda, \Te) \, \Jlam \, \exp\left(- \fr{\hp c}{\kb \Te \lambda} \right) d\lambda - 4\pi \, \nti{k} \!\!\! \int\limits_{0}^{+\infty} \! {\tilde{\sigma}}^{\textsc{pi}}_{k}(\lambda, T_k) \, \Jlam \, d\lambda \Bigg] + \nonumber \\
& & \sum_{k \in \mathcal{K}} \sqrt{\fr{2}{\pi}} \fr{\Qstarti{k}{\Te}{}}{\gp} \fr{\hpf c^2 \, \nelec \, \np}{(\mi{\text{e}} \kb \Te )^{3/2}}  \exp\left(\fr{\Ei{+}}{\kb \Te}\right) \int\limits_{0}^{+\infty} \! \fr{{\tilde{\sigma}}^{\textsc{pi}}_{k}(\lambda, \Te)}{\lambda^5} \, \exp\left(- \fr{\hp  c}{\kb \Te \lambda} \right) d\lambda, \\
\Omega^{\textsc{bf}/\textsc{fb}}_{\text{e}} & = & \Omega^{\textsc{bf}/\textsc{fb}} + \sum_{k \in \mathcal{K}} [\nti{k} \, \Gtf{k}{}{\textsc{pi}}(\Jlam, T_k) - \nelec \, \np \, \Gtf{k}{}{\textsc{rr}}(\Jlam, \Te) ] + \Ei{+} \, \omega^{\textsc{bf}/\textsc{fb}}_+.
\end{IEEEeqnarray}  
\paragraph{Free-free transitions} The emission and absorption coefficients, and the net volumetric energy loss rate due to \textsc{ff} transitions are not affected by the grouping as only charged-charged interactions are taken into account in this work. Hence, Eqs. \eqnref{eq:emisFF}-\eqnref{eq:OmegaFF_en} are used in the same way as for the StS model. 

\begin{thebibliography}{76}%
\makeatletter
\providecommand \@ifxundefined [1]{%
 \@ifx{#1\undefined}
}%
\providecommand \@ifnum [1]{%
 \ifnum #1\expandafter \@firstoftwo
 \else \expandafter \@secondoftwo
 \fi
}%
\providecommand \@ifx [1]{%
 \ifx #1\expandafter \@firstoftwo
 \else \expandafter \@secondoftwo
 \fi
}%
\providecommand \natexlab [1]{#1}%
\providecommand \enquote  [1]{``#1''}%
\providecommand \bibnamefont  [1]{#1}%
\providecommand \bibfnamefont [1]{#1}%
\providecommand \citenamefont [1]{#1}%
\providecommand \href@noop [0]{\@secondoftwo}%
\providecommand \href [0]{\begingroup \@sanitize@url \@href}%
\providecommand \@href[1]{\@@startlink{#1}\@@href}%
\providecommand \@@href[1]{\endgroup#1\@@endlink}%
\providecommand \@sanitize@url [0]{\catcode `\\12\catcode `\$12\catcode
  `\&12\catcode `\#12\catcode `\^12\catcode `\_12\catcode `\%12\relax}%
\providecommand \@@startlink[1]{}%
\providecommand \@@endlink[0]{}%
\providecommand \url  [0]{\begingroup\@sanitize@url \@url }%
\providecommand \@url [1]{\endgroup\@href {#1}{\urlprefix }}%
\providecommand \urlprefix  [0]{URL }%
\providecommand \Eprint [0]{\href }%
\providecommand \doibase [0]{http://dx.doi.org/}%
\providecommand \selectlanguage [0]{\@gobble}%
\providecommand \bibinfo  [0]{\@secondoftwo}%
\providecommand \bibfield  [0]{\@secondoftwo}%
\providecommand \translation [1]{[#1]}%
\providecommand \BibitemOpen [0]{}%
\providecommand \bibitemStop [0]{}%
\providecommand \bibitemNoStop [0]{.\EOS\space}%
\providecommand \EOS [0]{\spacefactor3000\relax}%
\providecommand \BibitemShut  [1]{\csname bibitem#1\endcsname}%
\let\auto@bib@innerbib\@empty
\bibitem [{\citenamefont {Priest}(1982)}]{Priest_book}%
  \BibitemOpen
  \bibfield  {author} {\bibinfo {author} {\bibfnamefont {E.~R.}\ \bibnamefont
  {Priest}},\ }\href@noop {} {\emph {\bibinfo {title} {Solar
  {M}agnetohydrodynamics}}}\ (\bibinfo  {publisher} {Springer},\ \bibinfo
  {address} {Berlin},\ \bibinfo {year} {1982})\BibitemShut {NoStop}%
\bibitem [{\citenamefont {Mihalas}\ and\ \citenamefont
  {Mihalas}(1984)}]{Mihalas_RadHyd_1984}%
  \BibitemOpen
  \bibfield  {author} {\bibinfo {author} {\bibfnamefont {D.}~\bibnamefont
  {Mihalas}}\ and\ \bibinfo {author} {\bibfnamefont {B.~W.}\ \bibnamefont
  {Mihalas}},\ }\href@noop {} {\emph {\bibinfo {title} {Foundations of
  {R}adiation {H}ydrodynamics}}}\ (\bibinfo  {publisher} {Oxford {U}niversity
  {P}ress},\ \bibinfo {year} {1984})\BibitemShut {NoStop}%
\bibitem [{\citenamefont {Mihalas}(1978)}]{Mihalas_StAtm_1978}%
  \BibitemOpen
  \bibfield  {author} {\bibinfo {author} {\bibfnamefont {D.}~\bibnamefont
  {Mihalas}},\ }\href@noop {} {\emph {\bibinfo {title} {Stellar
  {A}tmospheres}}}\ (\bibinfo  {publisher} {W. H. Freeman},\ \bibinfo {year}
  {1978})\ \bibinfo {note} {2nd edition}\BibitemShut {NoStop}%
\bibitem [{\citenamefont {Oxenius}(1986)}]{Oxenius_book}%
  \BibitemOpen
  \bibfield  {author} {\bibinfo {author} {\bibfnamefont {J.}~\bibnamefont
  {Oxenius}},\ }\href@noop {} {\emph {\bibinfo {title} {Kinetic {T}heory of
  {P}articles and {P}hotons}}},\ \bibinfo {series} {{S}pringer {S}eries in
  {E}lectronics and {P}hotonics}, Vol.~\bibinfo {volume} {20}\ (\bibinfo
  {publisher} {Springer-Verlag Berlin Heidelberg},\ \bibinfo {address}
  {Berlin},\ \bibinfo {year} {1986})\BibitemShut {NoStop}%
\bibitem [{\citenamefont {Capitelli}\ \emph {et~al.}(2000)\citenamefont
  {Capitelli}, \citenamefont {Ferreira}, \citenamefont {Gordiets},\ and\
  \citenamefont {Osipov}}]{Capitelli_book}%
  \BibitemOpen
  \bibfield  {author} {\bibinfo {author} {\bibfnamefont {M.}~\bibnamefont
  {Capitelli}}, \bibinfo {author} {\bibfnamefont {C.~M.}\ \bibnamefont
  {Ferreira}}, \bibinfo {author} {\bibfnamefont {B.~F.}\ \bibnamefont
  {Gordiets}}, \ and\ \bibinfo {author} {\bibfnamefont {A.~I.}\ \bibnamefont
  {Osipov}},\ }\href@noop {} {\emph {\bibinfo {title} {{P}lasma {K}inetics in
  {A}tmospheric {G}ases}}}\ (\bibinfo  {publisher} {Springer},\ \bibinfo {year}
  {2000})\BibitemShut {NoStop}%
\bibitem [{\citenamefont {Nagnibeda}\ and\ \citenamefont
  {Kustova}(2009)}]{Kustova_book}%
  \BibitemOpen
  \bibfield  {author} {\bibinfo {author} {\bibfnamefont {E.}~\bibnamefont
  {Nagnibeda}}\ and\ \bibinfo {author} {\bibfnamefont {E.}~\bibnamefont
  {Kustova}},\ }\href@noop {} {\emph {\bibinfo {title} {Non-{E}quilibrium
  {R}eacting {G}as {F}lows}}}\ (\bibinfo  {publisher} {Springer},\ \bibinfo
  {address} {Berlin},\ \bibinfo {year} {2009})\BibitemShut {NoStop}%
\bibitem [{\citenamefont {Kneer}\ and\ \citenamefont
  {Nagakawa}(1976)}]{Kneer_1976}%
  \BibitemOpen
  \bibfield  {author} {\bibinfo {author} {\bibfnamefont {F.}~\bibnamefont
  {Kneer}}\ and\ \bibinfo {author} {\bibfnamefont {Y.}~\bibnamefont
  {Nagakawa}},\ }\href@noop {} {\bibfield  {journal} {\bibinfo  {journal}
  {Astron. Astrophys.}\ }\textbf {\bibinfo {volume} {47}},\ \bibinfo {pages}
  {65} (\bibinfo {year} {1976})}\BibitemShut {NoStop}%
\bibitem [{\citenamefont {Klein}\ \emph {et~al.}(1976)\citenamefont {Klein},
  \citenamefont {Stein},\ and\ \citenamefont {Kalkofen}}]{Klein_1976}%
  \BibitemOpen
  \bibfield  {author} {\bibinfo {author} {\bibfnamefont {R.~K.}\ \bibnamefont
  {Klein}}, \bibinfo {author} {\bibfnamefont {R.~F.}\ \bibnamefont {Stein}}, \
  and\ \bibinfo {author} {\bibfnamefont {W.}~\bibnamefont {Kalkofen}},\
  }\href@noop {} {\bibfield  {journal} {\bibinfo  {journal} {Astrophys. J.}\
  }\textbf {\bibinfo {volume} {205}},\ \bibinfo {pages} {499} (\bibinfo {year}
  {1976})}\BibitemShut {NoStop}%
\bibitem [{\citenamefont {Klein}\ \emph {et~al.}(1978)\citenamefont {Klein},
  \citenamefont {Stein},\ and\ \citenamefont {Kalkofen}}]{Klein_1978}%
  \BibitemOpen
  \bibfield  {author} {\bibinfo {author} {\bibfnamefont {R.~K.}\ \bibnamefont
  {Klein}}, \bibinfo {author} {\bibfnamefont {R.~F.}\ \bibnamefont {Stein}}, \
  and\ \bibinfo {author} {\bibfnamefont {W.}~\bibnamefont {Kalkofen}},\
  }\href@noop {} {\bibfield  {journal} {\bibinfo  {journal} {Astrophys. J.}\
  }\textbf {\bibinfo {volume} {220}},\ \bibinfo {pages} {1024} (\bibinfo {year}
  {1978})}\BibitemShut {NoStop}%
\bibitem [{\citenamefont {Carlsson}\ and\ \citenamefont
  {Stein}(1992)}]{Carlsson_1992}%
  \BibitemOpen
  \bibfield  {author} {\bibinfo {author} {\bibfnamefont {M.}~\bibnamefont
  {Carlsson}}\ and\ \bibinfo {author} {\bibfnamefont {R.~F.}\ \bibnamefont
  {Stein}},\ }\href@noop {} {\bibfield  {journal} {\bibinfo  {journal}
  {Astrophys. J.}\ }\textbf {\bibinfo {volume} {397}},\ \bibinfo {pages} {L59}
  (\bibinfo {year} {1992})}\BibitemShut {NoStop}%
\bibitem [{\citenamefont {Carlsson}\ and\ \citenamefont
  {Stein}(1995)}]{Carlsson_1995}%
  \BibitemOpen
  \bibfield  {author} {\bibinfo {author} {\bibfnamefont {M.}~\bibnamefont
  {Carlsson}}\ and\ \bibinfo {author} {\bibfnamefont {R.~F.}\ \bibnamefont
  {Stein}},\ }\href@noop {} {\bibfield  {journal} {\bibinfo  {journal}
  {Astrophys. J.}\ }\textbf {\bibinfo {volume} {440}},\ \bibinfo {pages} {L29}
  (\bibinfo {year} {1995})}\BibitemShut {NoStop}%
\bibitem [{\citenamefont {Carlsson}\ and\ \citenamefont
  {Stein}(1997)}]{Carlsson_1997}%
  \BibitemOpen
  \bibfield  {author} {\bibinfo {author} {\bibfnamefont {M.}~\bibnamefont
  {Carlsson}}\ and\ \bibinfo {author} {\bibfnamefont {R.~F.}\ \bibnamefont
  {Stein}},\ }\href@noop {} {\bibfield  {journal} {\bibinfo  {journal}
  {Astrophys. J.}\ }\textbf {\bibinfo {volume} {481}},\ \bibinfo {pages} {500}
  (\bibinfo {year} {1997})}\BibitemShut {NoStop}%
\bibitem [{\citenamefont {Carlsson}\ and\ \citenamefont
  {Stein}(2002)}]{Carlsson_2002}%
  \BibitemOpen
  \bibfield  {author} {\bibinfo {author} {\bibfnamefont {M.}~\bibnamefont
  {Carlsson}}\ and\ \bibinfo {author} {\bibfnamefont {R.~F.}\ \bibnamefont
  {Stein}},\ }\href@noop {} {\bibfield  {journal} {\bibinfo  {journal}
  {Astrophys. J.}\ }\textbf {\bibinfo {volume} {572}},\ \bibinfo {pages} {626}
  (\bibinfo {year} {2002})}\BibitemShut {NoStop}%
\bibitem [{\citenamefont {Liu}\ \emph {et~al.}(2015)\citenamefont {Liu},
  \citenamefont {Panesi}, \citenamefont {Sahai},\ and\ \citenamefont
  {Vinokur}}]{Yen_2015}%
  \BibitemOpen
  \bibfield  {author} {\bibinfo {author} {\bibfnamefont {Y.}~\bibnamefont
  {Liu}}, \bibinfo {author} {\bibfnamefont {M.}~\bibnamefont {Panesi}},
  \bibinfo {author} {\bibfnamefont {A.}~\bibnamefont {Sahai}}, \ and\ \bibinfo
  {author} {\bibfnamefont {M.}~\bibnamefont {Vinokur}},\ }\href@noop {}
  {\bibfield  {journal} {\bibinfo  {journal} {J. Chem. Phys.}\ }\textbf
  {\bibinfo {volume} {142}},\ \bibinfo {pages} {134109} (\bibinfo {year}
  {2015})}\BibitemShut {NoStop}%
\bibitem [{\citenamefont {Panesi}\ and\ \citenamefont
  {Lani}(2013)}]{Panesi_Lani_CR_2013}%
  \BibitemOpen
  \bibfield  {author} {\bibinfo {author} {\bibfnamefont {M.}~\bibnamefont
  {Panesi}}\ and\ \bibinfo {author} {\bibfnamefont {A.}~\bibnamefont {Lani}},\
  }\href@noop {} {\bibfield  {journal} {\bibinfo  {journal} {Phys. Fluids}\
  }\textbf {\bibinfo {volume} {25}},\ \bibinfo {pages} {057101} (\bibinfo
  {year} {2013})}\BibitemShut {NoStop}%
\bibitem [{\citenamefont {Munaf\`{o}}\ \emph {et~al.}(2014)\citenamefont
  {Munaf\`{o}}, \citenamefont {Panesi},\ and\ \citenamefont
  {Magin}}]{Munafo_PRE_2014}%
  \BibitemOpen
  \bibfield  {author} {\bibinfo {author} {\bibfnamefont {A.}~\bibnamefont
  {Munaf\`{o}}}, \bibinfo {author} {\bibfnamefont {M.}~\bibnamefont {Panesi}},
  \ and\ \bibinfo {author} {\bibfnamefont {T.~E.}\ \bibnamefont {Magin}},\
  }\href@noop {} {\bibfield  {journal} {\bibinfo  {journal} {Phys. Rev. E}\
  }\textbf {\bibinfo {volume} {89}},\ \bibinfo {pages} {023001} (\bibinfo
  {year} {2014})}\BibitemShut {NoStop}%
\bibitem [{\citenamefont {Munaf\`{o}}\ \emph {et~al.}(2015)\citenamefont
  {Munaf\`{o}}, \citenamefont {Liu},\ and\ \citenamefont
  {Panesi}}]{Munafo_POF_2015}%
  \BibitemOpen
  \bibfield  {author} {\bibinfo {author} {\bibfnamefont {A.}~\bibnamefont
  {Munaf\`{o}}}, \bibinfo {author} {\bibfnamefont {Y.}~\bibnamefont {Liu}}, \
  and\ \bibinfo {author} {\bibfnamefont {M.}~\bibnamefont {Panesi}},\
  }\href@noop {} {\bibfield  {journal} {\bibinfo  {journal} {Phys. Fluids}\
  }\textbf {\bibinfo {volume} {27}},\ \bibinfo {pages} {127101} (\bibinfo
  {year} {2015})}\BibitemShut {NoStop}%
\bibitem [{\citenamefont {Whitney}\ and\ \citenamefont
  {Skalafuris}(1963)}]{Skalafuris_ApJ_1963}%
  \BibitemOpen
  \bibfield  {author} {\bibinfo {author} {\bibfnamefont {C.~A.}\ \bibnamefont
  {Whitney}}\ and\ \bibinfo {author} {\bibfnamefont {A.~J.}\ \bibnamefont
  {Skalafuris}},\ }\href@noop {} {\bibfield  {journal} {\bibinfo  {journal}
  {Astrophys. J.}\ }\textbf {\bibinfo {volume} {138}},\ \bibinfo {pages} {199}
  (\bibinfo {year} {1963})}\BibitemShut {NoStop}%
\bibitem [{\citenamefont {Skalafuris}(1965)}]{Skalafuris_ApJ_1965}%
  \BibitemOpen
  \bibfield  {author} {\bibinfo {author} {\bibfnamefont {A.~J.}\ \bibnamefont
  {Skalafuris}},\ }\href@noop {} {\bibfield  {journal} {\bibinfo  {journal}
  {Astrophys. J.}\ }\textbf {\bibinfo {volume} {142}},\ \bibinfo {pages} {351}
  (\bibinfo {year} {1965})}\BibitemShut {NoStop}%
\bibitem [{\citenamefont
  {Skalafuris}(1968{\natexlab{a}})}]{Skalafuris_ApJ_1968}%
  \BibitemOpen
  \bibfield  {author} {\bibinfo {author} {\bibfnamefont {A.~J.}\ \bibnamefont
  {Skalafuris}},\ }\href@noop {} {\bibfield  {journal} {\bibinfo  {journal}
  {Astrophys. Space Sci.}\ }\textbf {\bibinfo {volume} {2}},\ \bibinfo {pages}
  {258} (\bibinfo {year} {1968}{\natexlab{a}})}\BibitemShut {NoStop}%
\bibitem [{\citenamefont
  {Skalafuris}(1968{\natexlab{b}})}]{Skalafuris_ApJ_1968_II}%
  \BibitemOpen
  \bibfield  {author} {\bibinfo {author} {\bibfnamefont {A.~J.}\ \bibnamefont
  {Skalafuris}},\ }\href@noop {} {\bibfield  {journal} {\bibinfo  {journal}
  {Astrophys. Space Sci.}\ }\textbf {\bibinfo {volume} {3}},\ \bibinfo {pages}
  {234} (\bibinfo {year} {1968}{\natexlab{b}})}\BibitemShut {NoStop}%
\bibitem [{\citenamefont {Murty}(1971)}]{Murty_JQSRT_1971}%
  \BibitemOpen
  \bibfield  {author} {\bibinfo {author} {\bibfnamefont {S.~S.~R.}\
  \bibnamefont {Murty}},\ }\href@noop {} {\bibfield  {journal} {\bibinfo
  {journal} {J. Quant. Spectrosc. Radiat. Transfer}\ }\textbf {\bibinfo
  {volume} {11}},\ \bibinfo {pages} {1681} (\bibinfo {year}
  {1971})}\BibitemShut {NoStop}%
\bibitem [{\citenamefont {Clarke}\ and\ \citenamefont
  {Ferrari}(1965)}]{Clarke_Ferrari_POF_1965}%
  \BibitemOpen
  \bibfield  {author} {\bibinfo {author} {\bibfnamefont {J.~H.}\ \bibnamefont
  {Clarke}}\ and\ \bibinfo {author} {\bibfnamefont {C.}~\bibnamefont
  {Ferrari}},\ }\href@noop {} {\bibfield  {journal} {\bibinfo  {journal} {Phys.
  Fluids}\ }\textbf {\bibinfo {volume} {8}},\ \bibinfo {pages} {2121} (\bibinfo
  {year} {1965})}\BibitemShut {NoStop}%
\bibitem [{\citenamefont {Farnsworth}\ and\ \citenamefont
  {Clarke}(1971)}]{Clarke_POF_1971}%
  \BibitemOpen
  \bibfield  {author} {\bibinfo {author} {\bibfnamefont {A.~V.}\ \bibnamefont
  {Farnsworth}}\ and\ \bibinfo {author} {\bibfnamefont {J.~H.}\ \bibnamefont
  {Clarke}},\ }\href@noop {} {\bibfield  {journal} {\bibinfo  {journal} {Phys.
  Fluids}\ }\textbf {\bibinfo {volume} {14}},\ \bibinfo {pages} {1352}
  (\bibinfo {year} {1971})}\BibitemShut {NoStop}%
\bibitem [{\citenamefont {Foley}\ and\ \citenamefont
  {Clarke}(1973)}]{Foley_POF_1973}%
  \BibitemOpen
  \bibfield  {author} {\bibinfo {author} {\bibfnamefont {W.~H.}\ \bibnamefont
  {Foley}}\ and\ \bibinfo {author} {\bibfnamefont {J.~H.}\ \bibnamefont
  {Clarke}},\ }\href@noop {} {\bibfield  {journal} {\bibinfo  {journal} {Phys.
  Fluids}\ }\textbf {\bibinfo {volume} {16}},\ \bibinfo {pages} {375} (\bibinfo
  {year} {1973})}\BibitemShut {NoStop}%
\bibitem [{\citenamefont {Fadeyev}\ and\ \citenamefont
  {Gillet}(1998)}]{Fadeyev_AA_1998}%
  \BibitemOpen
  \bibfield  {author} {\bibinfo {author} {\bibfnamefont {Y.~A.}\ \bibnamefont
  {Fadeyev}}\ and\ \bibinfo {author} {\bibfnamefont {D.}~\bibnamefont
  {Gillet}},\ }\href@noop {} {\bibfield  {journal} {\bibinfo  {journal}
  {Astron. Astrophys.}\ }\textbf {\bibinfo {volume} {333}},\ \bibinfo {pages}
  {687} (\bibinfo {year} {1998})}\BibitemShut {NoStop}%
\bibitem [{\citenamefont {Fadeyev}\ and\ \citenamefont
  {Gillet}(2000)}]{Fadeyev_AA_2000}%
  \BibitemOpen
  \bibfield  {author} {\bibinfo {author} {\bibfnamefont {Y.~A.}\ \bibnamefont
  {Fadeyev}}\ and\ \bibinfo {author} {\bibfnamefont {D.}~\bibnamefont
  {Gillet}},\ }\href@noop {} {\bibfield  {journal} {\bibinfo  {journal}
  {Astron. Astrophys.}\ }\textbf {\bibinfo {volume} {354}},\ \bibinfo {pages}
  {349} (\bibinfo {year} {2000})}\BibitemShut {NoStop}%
\bibitem [{\citenamefont {Fadeyev}\ and\ \citenamefont
  {Gillet}(2001)}]{Fadeyev_AA_2001}%
  \BibitemOpen
  \bibfield  {author} {\bibinfo {author} {\bibfnamefont {Y.~A.}\ \bibnamefont
  {Fadeyev}}\ and\ \bibinfo {author} {\bibfnamefont {D.}~\bibnamefont
  {Gillet}},\ }\href@noop {} {\bibfield  {journal} {\bibinfo  {journal}
  {Astron. Astrophys.}\ }\textbf {\bibinfo {volume} {368}},\ \bibinfo {pages}
  {901} (\bibinfo {year} {2001})}\BibitemShut {NoStop}%
\bibitem [{\citenamefont {Fadeyev}\ \emph {et~al.}(2002)\citenamefont
  {Fadeyev}, \citenamefont {Coroller},\ and\ \citenamefont
  {Gillet}}]{Fadeyev_AA_2002}%
  \BibitemOpen
  \bibfield  {author} {\bibinfo {author} {\bibfnamefont {Y.~A.}\ \bibnamefont
  {Fadeyev}}, \bibinfo {author} {\bibfnamefont {H.~L.}\ \bibnamefont
  {Coroller}}, \ and\ \bibinfo {author} {\bibfnamefont {D.}~\bibnamefont
  {Gillet}},\ }\href@noop {} {\bibfield  {journal} {\bibinfo  {journal}
  {Astron. and Astrophys.}\ }\textbf {\bibinfo {volume} {392}},\ \bibinfo
  {pages} {735} (\bibinfo {year} {2002})}\BibitemShut {NoStop}%
\bibitem [{\citenamefont {Panesi}\ \emph {et~al.}(2009)\citenamefont {Panesi},
  \citenamefont {Magin}, \citenamefont {Bourdon}, \citenamefont {Bultel},\ and\
  \citenamefont {Chazot}}]{Panesi_JTHT_2009}%
  \BibitemOpen
  \bibfield  {author} {\bibinfo {author} {\bibfnamefont {M.}~\bibnamefont
  {Panesi}}, \bibinfo {author} {\bibfnamefont {T.~E.}\ \bibnamefont {Magin}},
  \bibinfo {author} {\bibfnamefont {A.}~\bibnamefont {Bourdon}}, \bibinfo
  {author} {\bibfnamefont {A.}~\bibnamefont {Bultel}}, \ and\ \bibinfo {author}
  {\bibfnamefont {O.}~\bibnamefont {Chazot}},\ }\href@noop {} {\bibfield
  {journal} {\bibinfo  {journal} {J. Thermophys. Heat Transfer}\ }\textbf
  {\bibinfo {volume} {23}},\ \bibinfo {pages} {236} (\bibinfo {year}
  {2009})}\BibitemShut {NoStop}%
\bibitem [{\citenamefont {Panesi}\ \emph {et~al.}(2011)\citenamefont {Panesi},
  \citenamefont {Magin}, \citenamefont {Bourdon}, \citenamefont {Bultel},\ and\
  \citenamefont {Chazot}}]{Panesi_JTHT_2011}%
  \BibitemOpen
  \bibfield  {author} {\bibinfo {author} {\bibfnamefont {M.}~\bibnamefont
  {Panesi}}, \bibinfo {author} {\bibfnamefont {T.~E.}\ \bibnamefont {Magin}},
  \bibinfo {author} {\bibfnamefont {A.}~\bibnamefont {Bourdon}}, \bibinfo
  {author} {\bibfnamefont {A.}~\bibnamefont {Bultel}}, \ and\ \bibinfo {author}
  {\bibfnamefont {O.}~\bibnamefont {Chazot}},\ }\href@noop {} {\bibfield
  {journal} {\bibinfo  {journal} {J. Thermophys. Heat Transfer}\ }\textbf
  {\bibinfo {volume} {25}},\ \bibinfo {pages} {361} (\bibinfo {year}
  {2011})}\BibitemShut {NoStop}%
\bibitem [{\citenamefont {Kapper}\ and\ \citenamefont
  {Cambier}(2011{\natexlab{a}})}]{Kapper_2011_1}%
  \BibitemOpen
  \bibfield  {author} {\bibinfo {author} {\bibfnamefont {M.~G.}\ \bibnamefont
  {Kapper}}\ and\ \bibinfo {author} {\bibfnamefont {J.-L.}\ \bibnamefont
  {Cambier}},\ }\href@noop {} {\bibfield  {journal} {\bibinfo  {journal} {J.
  Appl. Phys.}\ }\textbf {\bibinfo {volume} {109}},\ \bibinfo {pages} {113308}
  (\bibinfo {year} {2011}{\natexlab{a}})}\BibitemShut {NoStop}%
\bibitem [{\citenamefont {Kapper}\ and\ \citenamefont
  {Cambier}(2011{\natexlab{b}})}]{Kapper_2011_2}%
  \BibitemOpen
  \bibfield  {author} {\bibinfo {author} {\bibfnamefont {M.~G.}\ \bibnamefont
  {Kapper}}\ and\ \bibinfo {author} {\bibfnamefont {J.-L.}\ \bibnamefont
  {Cambier}},\ }\href@noop {} {\bibfield  {journal} {\bibinfo  {journal} {J.
  Appl. Phys.}\ }\textbf {\bibinfo {volume} {109}},\ \bibinfo {pages} {113309}
  (\bibinfo {year} {2011}{\natexlab{b}})}\BibitemShut {NoStop}%
\bibitem [{\citenamefont {Capitelli}\ \emph {et~al.}(2013)\citenamefont
  {Capitelli}, \citenamefont {Colonna}, \citenamefont {Pietanza},\ and\
  \citenamefont {D'Ammando}}]{Dammando_2013}%
  \BibitemOpen
  \bibfield  {author} {\bibinfo {author} {\bibfnamefont {M.}~\bibnamefont
  {Capitelli}}, \bibinfo {author} {\bibfnamefont {G.}~\bibnamefont {Colonna}},
  \bibinfo {author} {\bibfnamefont {L.~D.}\ \bibnamefont {Pietanza}}, \ and\
  \bibinfo {author} {\bibfnamefont {G.}~\bibnamefont {D'Ammando}},\ }\href@noop
  {} {\bibfield  {journal} {\bibinfo  {journal} {Spectrochim. Acta Part B}\
  }\textbf {\bibinfo {volume} {83--84}},\ \bibinfo {pages} {1} (\bibinfo {year}
  {2013})}\BibitemShut {NoStop}%
\bibitem [{\citenamefont {Libermann}\ and\ \citenamefont
  {Velikovich}(1986)}]{Liberman_Book}%
  \BibitemOpen
  \bibfield  {author} {\bibinfo {author} {\bibfnamefont {M.~A.}\ \bibnamefont
  {Libermann}}\ and\ \bibinfo {author} {\bibfnamefont {A.~L.}\ \bibnamefont
  {Velikovich}},\ }\href@noop {} {\emph {\bibinfo {title} {Physics of {S}hock
  {W}aves in {G}ases and {P}lasmas}}},\ \bibinfo {series} {{S}pringer {S}eries
  in {E}lectronics and {P}hotonics}, Vol.~\bibinfo {volume} {19}\ (\bibinfo
  {publisher} {Springer-Verlag Berlin Heidelberg},\ \bibinfo {address}
  {Berlin},\ \bibinfo {year} {1986})\BibitemShut {NoStop}%
\bibitem [{\citenamefont {Kramida}(2010)}]{Kramida_2010}%
  \BibitemOpen
  \bibfield  {author} {\bibinfo {author} {\bibfnamefont {A.~E.}\ \bibnamefont
  {Kramida}},\ }\href@noop {} {\bibfield  {journal} {\bibinfo  {journal} {At.
  Data Nucl. Data Tables}\ }\textbf {\bibinfo {volume} {96}},\ \bibinfo {pages}
  {586} (\bibinfo {year} {2010})}\BibitemShut {NoStop}%
\bibitem [{\citenamefont {Kramida}\ \emph {et~al.}(2014)\citenamefont
  {Kramida}, \citenamefont {{Yu.~Ralchenko}}, \citenamefont {Reader},\ and\
  \citenamefont {{and NIST ASD Team}}}]{NIST_ASD}%
  \BibitemOpen
  \bibfield  {author} {\bibinfo {author} {\bibfnamefont {A.}~\bibnamefont
  {Kramida}}, \bibinfo {author} {\bibnamefont {{Yu.~Ralchenko}}}, \bibinfo
  {author} {\bibfnamefont {J.}~\bibnamefont {Reader}}, \ and\ \bibinfo {author}
  {\bibnamefont {{and NIST ASD Team}}},\ }\href@noop {} {}\bibinfo
  {howpublished} {{NIST Atomic Spectra Database (ver. 5.2), [Online].
  Available: {\tt{http://physics.nist.gov/asd}} [2015, October 28]. National
  Institute of Standards and Technology, Gaithersburg, MD.}} (\bibinfo {year}
  {2014})\BibitemShut {NoStop}%
\bibitem [{\citenamefont {Herzberg}(1963)}]{Herzberg_book}%
  \BibitemOpen
  \bibfield  {author} {\bibinfo {author} {\bibfnamefont {G.}~\bibnamefont
  {Herzberg}},\ }\href@noop {} {\emph {\bibinfo {title} {{M}olecular {S}pectra
  and {M}olecular {S}tructure}}}\ (\bibinfo  {publisher} {D. Van Nostrand,
  Inc.},\ \bibinfo {address} {New York, NY},\ \bibinfo {year}
  {1963})\BibitemShut {NoStop}%
\bibitem [{\citenamefont {Bates}\ \emph
  {et~al.}(1962{\natexlab{a}})\citenamefont {Bates}, \citenamefont {Kingston},\
  and\ \citenamefont {Mc{W}hirter}}]{Bates_1962I}%
  \BibitemOpen
  \bibfield  {author} {\bibinfo {author} {\bibfnamefont {D.~R.}\ \bibnamefont
  {Bates}}, \bibinfo {author} {\bibfnamefont {A.~E.}\ \bibnamefont {Kingston}},
  \ and\ \bibinfo {author} {\bibfnamefont {R.~W.~P.}\ \bibnamefont
  {Mc{W}hirter}},\ }\href@noop {} {\bibfield  {journal} {\bibinfo  {journal}
  {Proc. R. Soc. A}\ }\textbf {\bibinfo {volume} {267}},\ \bibinfo {pages}
  {297} (\bibinfo {year} {1962}{\natexlab{a}})}\BibitemShut {NoStop}%
\bibitem [{\citenamefont {Bates}\ \emph
  {et~al.}(1962{\natexlab{b}})\citenamefont {Bates}, \citenamefont {Kingston},\
  and\ \citenamefont {Mc{W}hirter}}]{Bates_1962II}%
  \BibitemOpen
  \bibfield  {author} {\bibinfo {author} {\bibfnamefont {D.~R.}\ \bibnamefont
  {Bates}}, \bibinfo {author} {\bibfnamefont {A.~E.}\ \bibnamefont {Kingston}},
  \ and\ \bibinfo {author} {\bibfnamefont {R.~W.~P.}\ \bibnamefont
  {Mc{W}hirter}},\ }\href@noop {} {\bibfield  {journal} {\bibinfo  {journal}
  {Proc. R. Soc. A}\ }\textbf {\bibinfo {volume} {270}},\ \bibinfo {pages}
  {155} (\bibinfo {year} {1962}{\natexlab{b}})}\BibitemShut {NoStop}%
\bibitem [{\citenamefont {Rogers}(1974)}]{Rogers_PhysRevA_1974}%
  \BibitemOpen
  \bibfield  {author} {\bibinfo {author} {\bibfnamefont {F.~J.}\ \bibnamefont
  {Rogers}},\ }\href@noop {} {\bibfield  {journal} {\bibinfo  {journal} {Phys.
  Rev. A}\ }\textbf {\bibinfo {volume} {10}},\ \bibinfo {pages} {2441}
  (\bibinfo {year} {1974})}\BibitemShut {NoStop}%
\bibitem [{\citenamefont {Rogers}\ \emph {et~al.}(1996)\citenamefont {Rogers},
  \citenamefont {Swenson},\ and\ \citenamefont {Iglesias}}]{Rogers_ApJ1996}%
  \BibitemOpen
  \bibfield  {author} {\bibinfo {author} {\bibfnamefont {F.~J.}\ \bibnamefont
  {Rogers}}, \bibinfo {author} {\bibfnamefont {F.~J.}\ \bibnamefont {Swenson}},
  \ and\ \bibinfo {author} {\bibfnamefont {C.~A.}\ \bibnamefont {Iglesias}},\
  }\href@noop {} {\bibfield  {journal} {\bibinfo  {journal} {Astrophys. J.}\
  }\textbf {\bibinfo {volume} {456}},\ \bibinfo {pages} {902} (\bibinfo {year}
  {1996})}\BibitemShut {NoStop}%
\bibitem [{\citenamefont {Appleton}\ and\ \citenamefont
  {Bray}(1964)}]{App_Bray_1964}%
  \BibitemOpen
  \bibfield  {author} {\bibinfo {author} {\bibfnamefont {J.~P.}\ \bibnamefont
  {Appleton}}\ and\ \bibinfo {author} {\bibfnamefont {K.~N.~C.}\ \bibnamefont
  {Bray}},\ }\href@noop {} {\bibfield  {journal} {\bibinfo  {journal} {J. Fluid
  Mech.}\ }\textbf {\bibinfo {volume} {20}},\ \bibinfo {pages} {659} (\bibinfo
  {year} {1964})}\BibitemShut {NoStop}%
\bibitem [{\citenamefont {D'Ammando}\ \emph {et~al.}(2010)\citenamefont
  {D'Ammando}, \citenamefont {Pietanza}, \citenamefont {Colonna}, \citenamefont
  {Longo},\ and\ \citenamefont {Capitelli}}]{Dammando_2010}%
  \BibitemOpen
  \bibfield  {author} {\bibinfo {author} {\bibfnamefont {G.}~\bibnamefont
  {D'Ammando}}, \bibinfo {author} {\bibfnamefont {L.~D.}\ \bibnamefont
  {Pietanza}}, \bibinfo {author} {\bibfnamefont {G.}~\bibnamefont {Colonna}},
  \bibinfo {author} {\bibfnamefont {S.}~\bibnamefont {Longo}}, \ and\ \bibinfo
  {author} {\bibfnamefont {M.}~\bibnamefont {Capitelli}},\ }\href@noop {}
  {\bibfield  {journal} {\bibinfo  {journal} {Spectrochim. Acta Part B}\
  }\textbf {\bibinfo {volume} {65}},\ \bibinfo {pages} {120} (\bibinfo {year}
  {2010})}\BibitemShut {NoStop}%
\bibitem [{\citenamefont {Colonna}\ \emph {et~al.}(2012)\citenamefont
  {Colonna}, \citenamefont {Pietanza},\ and\ \citenamefont
  {D'Ammando}}]{Colonna_ChemPhys_2012}%
  \BibitemOpen
  \bibfield  {author} {\bibinfo {author} {\bibfnamefont {G.}~\bibnamefont
  {Colonna}}, \bibinfo {author} {\bibfnamefont {L.~D.}\ \bibnamefont
  {Pietanza}}, \ and\ \bibinfo {author} {\bibfnamefont {G.}~\bibnamefont
  {D'Ammando}},\ }\href@noop {} {\bibfield  {journal} {\bibinfo  {journal}
  {Chem. Phys.}\ }\textbf {\bibinfo {volume} {398}},\ \bibinfo {pages} {37}
  (\bibinfo {year} {2012})}\BibitemShut {NoStop}%
\bibitem [{\citenamefont {Drawin}(1963)}]{Drawin_1963}%
  \BibitemOpen
  \bibfield  {author} {\bibinfo {author} {\bibfnamefont {H.}~\bibnamefont
  {Drawin}},\ }\href@noop {} {\emph {\bibinfo {title} {Atomic cross-sections
  for inelastic electronic collisions}}},\ \bibinfo {type} {{EUR}-{CEA}-{FC}}\
  \bibinfo {number} {236}\ (\bibinfo {year} {1963})\BibitemShut {NoStop}%
\bibitem [{\citenamefont {Petschek}\ and\ \citenamefont
  {Byron}(1957)}]{Petschek_AnnPhys_1957}%
  \BibitemOpen
  \bibfield  {author} {\bibinfo {author} {\bibfnamefont {H.}~\bibnamefont
  {Petschek}}\ and\ \bibinfo {author} {\bibfnamefont {S.}~\bibnamefont
  {Byron}},\ }\href@noop {} {\bibfield  {journal} {\bibinfo  {journal} {Ann.
  Phys.}\ }\textbf {\bibinfo {volume} {1}},\ \bibinfo {pages} {270} (\bibinfo
  {year} {1957})}\BibitemShut {NoStop}%
\bibitem [{\citenamefont {Desloge}(1962)}]{Desloge_POF_1962}%
  \BibitemOpen
  \bibfield  {author} {\bibinfo {author} {\bibfnamefont {E.~A.}\ \bibnamefont
  {Desloge}},\ }\href@noop {} {\bibfield  {journal} {\bibinfo  {journal} {Phys.
  Fluids}\ }\textbf {\bibinfo {volume} {5}},\ \bibinfo {pages} {1223} (\bibinfo
  {year} {1962})}\BibitemShut {NoStop}%
\bibitem [{\citenamefont {Devoto}(1966)}]{Devoto_1966}%
  \BibitemOpen
  \bibfield  {author} {\bibinfo {author} {\bibfnamefont {R.~S.}\ \bibnamefont
  {Devoto}},\ }\href@noop {} {\bibfield  {journal} {\bibinfo  {journal} {Phys.
  Fluids}\ }\textbf {\bibinfo {volume} {9}},\ \bibinfo {pages} {1230} (\bibinfo
  {year} {1966})}\BibitemShut {NoStop}%
\bibitem [{\citenamefont {Devoto}(1967)}]{Devoto_1967}%
  \BibitemOpen
  \bibfield  {author} {\bibinfo {author} {\bibfnamefont {R.~S.}\ \bibnamefont
  {Devoto}},\ }\href@noop {} {\bibfield  {journal} {\bibinfo  {journal} {Phys.
  Fluids}\ }\textbf {\bibinfo {volume} {10}},\ \bibinfo {pages} {354} (\bibinfo
  {year} {1967})}\BibitemShut {NoStop}%
\bibitem [{\citenamefont {Ferziger}\ and\ \citenamefont
  {Kaper}(1972)}]{Ferziger_book}%
  \BibitemOpen
  \bibfield  {author} {\bibinfo {author} {\bibfnamefont {J.~H.}\ \bibnamefont
  {Ferziger}}\ and\ \bibinfo {author} {\bibfnamefont {H.~G.}\ \bibnamefont
  {Kaper}},\ }\href@noop {} {\emph {\bibinfo {title} {Mathematical Theory of
  Transport Processes in Gases}}}\ (\bibinfo  {publisher} {North-Holland Pub.
  Co.},\ \bibinfo {year} {1972})\BibitemShut {NoStop}%
\bibitem [{\citenamefont {Bruno}\ \emph {et~al.}(2010)\citenamefont {Bruno},
  \citenamefont {Catalfamo}, \citenamefont {M.~Capitelli~and}, \citenamefont
  {Diomede}, \citenamefont {Gorse}, \citenamefont {Laricchiuta}, \citenamefont
  {Longo}, \citenamefont {Giordano},\ and\ \citenamefont
  {Pirani}}]{Bruno_POP_2010}%
  \BibitemOpen
  \bibfield  {author} {\bibinfo {author} {\bibfnamefont {D.}~\bibnamefont
  {Bruno}}, \bibinfo {author} {\bibfnamefont {C.}~\bibnamefont {Catalfamo}},
  \bibinfo {author} {\bibfnamefont {O.~D.~P.}\ \bibnamefont {M.~Capitelli~and},
  \bibfnamefont {G.~Colonna~and}}, \bibinfo {author} {\bibfnamefont
  {P.}~\bibnamefont {Diomede}}, \bibinfo {author} {\bibfnamefont
  {C.}~\bibnamefont {Gorse}}, \bibinfo {author} {\bibfnamefont
  {A.}~\bibnamefont {Laricchiuta}}, \bibinfo {author} {\bibfnamefont
  {S.}~\bibnamefont {Longo}}, \bibinfo {author} {\bibfnamefont
  {D.}~\bibnamefont {Giordano}}, \ and\ \bibinfo {author} {\bibfnamefont
  {F.}~\bibnamefont {Pirani}},\ }\href@noop {} {\bibfield  {journal} {\bibinfo
  {journal} {Phys. Plasmas}\ }\textbf {\bibinfo {volume} {17}},\ \bibinfo
  {pages} {112315} (\bibinfo {year} {2010})}\BibitemShut {NoStop}%
\bibitem [{\citenamefont {Spitzer}(2006)}]{Spitzer_book}%
  \BibitemOpen
  \bibfield  {author} {\bibinfo {author} {\bibfnamefont {L.}~\bibnamefont
  {Spitzer}},\ }\href@noop {} {\emph {\bibinfo {title} {Physics of {F}ully
  {I}onized {G}ases}}},\ Dover Books on Physics\ (\bibinfo  {publisher} {Dover
  Publications},\ \bibinfo {address} {Mineola, NY},\ \bibinfo {year}
  {2006})\BibitemShut {NoStop}%
\bibitem [{\citenamefont {Mitchner}\ and\ \citenamefont
  {Kruger}(1973)}]{Mitchner_book}%
  \BibitemOpen
  \bibfield  {author} {\bibinfo {author} {\bibfnamefont {M.}~\bibnamefont
  {Mitchner}}\ and\ \bibinfo {author} {\bibfnamefont {C.~H.}\ \bibnamefont
  {Kruger}},\ }\href@noop {} {\emph {\bibinfo {title} {Partially Ionized
  Gases}}}\ (\bibinfo  {publisher} {John Wiley \& Sons},\ \bibinfo {year}
  {1973})\BibitemShut {NoStop}%
\bibitem [{\citenamefont {Cowley}(1971)}]{Cowley_1971}%
  \BibitemOpen
  \bibfield  {author} {\bibinfo {author} {\bibfnamefont {C.~R.}\ \bibnamefont
  {Cowley}},\ }\href@noop {} {\bibfield  {journal} {\bibinfo  {journal} {The
  Observatory}\ }\textbf {\bibinfo {volume} {91}},\ \bibinfo {pages} {139}
  (\bibinfo {year} {1971})}\BibitemShut {NoStop}%
\bibitem [{\citenamefont {Whiting}(1968)}]{Whiting_JQSRT_1968}%
  \BibitemOpen
  \bibfield  {author} {\bibinfo {author} {\bibfnamefont {E.~E.}\ \bibnamefont
  {Whiting}},\ }\href@noop {} {\bibfield  {journal} {\bibinfo  {journal} {J.
  Quant. Spectrosc. Radiat. Transfer}\ }\textbf {\bibinfo {volume} {8}},\
  \bibinfo {pages} {1379} (\bibinfo {year} {1968})}\BibitemShut {NoStop}%
\bibitem [{\citenamefont {Zel'dovich}\ and\ \citenamefont
  {Raizer}(1967)}]{Zeldovich_book_1967}%
  \BibitemOpen
  \bibfield  {author} {\bibinfo {author} {\bibfnamefont {Y.~B.}\ \bibnamefont
  {Zel'dovich}}\ and\ \bibinfo {author} {\bibfnamefont {{\relax Yu}.~P.}\
  \bibnamefont {Raizer}},\ }\href@noop {} {\emph {\bibinfo {title} {Physics of
  {S}hock {W}aves and {H}igh-{T}emperature {H}ydrodynamic {P}henomena}}}\
  (\bibinfo  {publisher} {{A}cademic {P}ress {I}nc.},\ \bibinfo {address} {New
  York, NY},\ \bibinfo {year} {1967})\BibitemShut {NoStop}%
\bibitem [{\citenamefont {Shafranov}(1957)}]{Shafranov_1957}%
  \BibitemOpen
  \bibfield  {author} {\bibinfo {author} {\bibfnamefont {V.}~\bibnamefont
  {Shafranov}},\ }\href@noop {} {\bibfield  {journal} {\bibinfo  {journal}
  {Sov. Phys. JETP}\ }\textbf {\bibinfo {volume} {5}},\ \bibinfo {pages} {1183}
  (\bibinfo {year} {1957})}\BibitemShut {NoStop}%
\bibitem [{\citenamefont {Jukes}(1957)}]{Jukes_JFM_1957}%
  \BibitemOpen
  \bibfield  {author} {\bibinfo {author} {\bibfnamefont {J.~D.}\ \bibnamefont
  {Jukes}},\ }\href@noop {} {\bibfield  {journal} {\bibinfo  {journal} {J.
  Fluid Mech.}\ }\textbf {\bibinfo {volume} {3}},\ \bibinfo {pages} {275}
  (\bibinfo {year} {1957})}\BibitemShut {NoStop}%
\bibitem [{\citenamefont {Imshennik}(1962)}]{Imshennik_1962}%
  \BibitemOpen
  \bibfield  {author} {\bibinfo {author} {\bibfnamefont {V.~S.}\ \bibnamefont
  {Imshennik}},\ }\href@noop {} {\bibfield  {journal} {\bibinfo  {journal}
  {Sov. Phys. JETP}\ }\textbf {\bibinfo {volume} {15}},\ \bibinfo {pages} {167}
  (\bibinfo {year} {1962})}\BibitemShut {NoStop}%
\bibitem [{\citenamefont {Jaffrin}\ and\ \citenamefont
  {Probstein}(1964)}]{Jaffrin_POF_1964}%
  \BibitemOpen
  \bibfield  {author} {\bibinfo {author} {\bibfnamefont {M.~Y.}\ \bibnamefont
  {Jaffrin}}\ and\ \bibinfo {author} {\bibfnamefont {R.~F.}\ \bibnamefont
  {Probstein}},\ }\href@noop {} {\bibfield  {journal} {\bibinfo  {journal}
  {Phys. Fluids}\ }\textbf {\bibinfo {volume} {7}},\ \bibinfo {pages} {1658}
  (\bibinfo {year} {1964})}\BibitemShut {NoStop}%
\bibitem [{\citenamefont {Feautrier}(1964)}]{Feautrier_1964}%
  \BibitemOpen
  \bibfield  {author} {\bibinfo {author} {\bibfnamefont {P.}~\bibnamefont
  {Feautrier}},\ }\href@noop {} {\bibfield  {journal} {\bibinfo  {journal} {Cr.
  Acad. Sci.}\ }\textbf {\bibinfo {volume} {258}},\ \bibinfo {pages} {3189}
  (\bibinfo {year} {1964})},\ \bibinfo {note} {in French}\BibitemShut {NoStop}%
\bibitem [{\citenamefont {Groot}\ and\ \citenamefont
  {Mazur}(2011)}]{DeGroot_Mazur_book}%
  \BibitemOpen
  \bibfield  {author} {\bibinfo {author} {\bibfnamefont {S.~R.~D.}\
  \bibnamefont {Groot}}\ and\ \bibinfo {author} {\bibfnamefont
  {P.}~\bibnamefont {Mazur}},\ }\href@noop {} {\emph {\bibinfo {title}
  {Nonequilibrium thermodynamics}}},\ Dover Books on Physics\ (\bibinfo
  {publisher} {Dover Publications},\ \bibinfo {address} {Mineola, NY},\
  \bibinfo {year} {2011})\BibitemShut {NoStop}%
\bibitem [{\citenamefont {Panesi}\ and\ \citenamefont
  {Huo}(2011)}]{Panesi_AIAA_2011}%
  \BibitemOpen
  \bibfield  {author} {\bibinfo {author} {\bibfnamefont {M.}~\bibnamefont
  {Panesi}}\ and\ \bibinfo {author} {\bibfnamefont {W.}~\bibnamefont {Huo}},\
  }\href@noop {} {\emph {\bibinfo {title} {Non-equilibrium ionization phenomena
  behind shock waves}}},\ \bibinfo {type} {AIAA Paper}\ \bibinfo {number}
  {2011--3629}\ (\bibinfo {year} {2011})\ \bibinfo {note} {11th AIAA/ASME Joint
  Thermophysics and Heat Transfer Conference, Honolulu, HW}\BibitemShut
  {NoStop}%
\bibitem [{\citenamefont {Auer}\ and\ \citenamefont
  {Mihalas}(1969{\natexlab{a}})}]{Auer_Mihalas_ApJ_1969I}%
  \BibitemOpen
  \bibfield  {author} {\bibinfo {author} {\bibfnamefont {L.~H.}\ \bibnamefont
  {Auer}}\ and\ \bibinfo {author} {\bibfnamefont {D.}~\bibnamefont {Mihalas}},\
  }\href@noop {} {\bibfield  {journal} {\bibinfo  {journal} {Astrophys. J.}\
  }\textbf {\bibinfo {volume} {156}},\ \bibinfo {pages} {157} (\bibinfo {year}
  {1969}{\natexlab{a}})}\BibitemShut {NoStop}%
\bibitem [{\citenamefont {Auer}\ and\ \citenamefont
  {Mihalas}(1969{\natexlab{b}})}]{Auer_Mihalas_ApJ_1969II}%
  \BibitemOpen
  \bibfield  {author} {\bibinfo {author} {\bibfnamefont {L.~H.}\ \bibnamefont
  {Auer}}\ and\ \bibinfo {author} {\bibfnamefont {D.}~\bibnamefont {Mihalas}},\
  }\href@noop {} {\bibfield  {journal} {\bibinfo  {journal} {Astrophys. J.}\
  }\textbf {\bibinfo {volume} {156}},\ \bibinfo {pages} {681} (\bibinfo {year}
  {1969}{\natexlab{b}})}\BibitemShut {NoStop}%
\bibitem [{\citenamefont {Auer}\ and\ \citenamefont
  {Mihalas}(1969{\natexlab{c}})}]{Auer_Mihalas_ApJ_1969III}%
  \BibitemOpen
  \bibfield  {author} {\bibinfo {author} {\bibfnamefont {L.~H.}\ \bibnamefont
  {Auer}}\ and\ \bibinfo {author} {\bibfnamefont {D.}~\bibnamefont {Mihalas}},\
  }\href@noop {} {\bibfield  {journal} {\bibinfo  {journal} {Astrophys. J.}\
  }\textbf {\bibinfo {volume} {158}},\ \bibinfo {pages} {641} (\bibinfo {year}
  {1969}{\natexlab{c}})}\BibitemShut {NoStop}%
\bibitem [{\citenamefont {Ribicky}\ and\ \citenamefont
  {Hummer}(1991)}]{Ribicky_AA_1991}%
  \BibitemOpen
  \bibfield  {author} {\bibinfo {author} {\bibfnamefont {G.~B.}\ \bibnamefont
  {Ribicky}}\ and\ \bibinfo {author} {\bibfnamefont {D.~G.}\ \bibnamefont
  {Hummer}},\ }\href@noop {} {\bibfield  {journal} {\bibinfo  {journal}
  {Astron. Astrophys.}\ }\textbf {\bibinfo {volume} {245}},\ \bibinfo {pages}
  {171} (\bibinfo {year} {1991})}\BibitemShut {NoStop}%
\bibitem [{\citenamefont {Ribicky}\ and\ \citenamefont
  {Hummer}(1992)}]{Ribicky_AA_1992}%
  \BibitemOpen
  \bibfield  {author} {\bibinfo {author} {\bibfnamefont {G.~B.}\ \bibnamefont
  {Ribicky}}\ and\ \bibinfo {author} {\bibfnamefont {D.~G.}\ \bibnamefont
  {Hummer}},\ }\href@noop {} {\bibfield  {journal} {\bibinfo  {journal}
  {Astron. Astrophys.}\ }\textbf {\bibinfo {volume} {262}},\ \bibinfo {pages}
  {209} (\bibinfo {year} {1992})}\BibitemShut {NoStop}%
\bibitem [{\citenamefont {Murty}(1968)}]{Murty_JQSRT_1968}%
  \BibitemOpen
  \bibfield  {author} {\bibinfo {author} {\bibfnamefont {S.~S.~R.}\
  \bibnamefont {Murty}},\ }\href@noop {} {\bibfield  {journal} {\bibinfo
  {journal} {J. Quant. Spectrosc. Radiat. Transfer}\ }\textbf {\bibinfo
  {volume} {8}},\ \bibinfo {pages} {531} (\bibinfo {year} {1968})}\BibitemShut
  {NoStop}%
\bibitem [{\citenamefont {Marshak}(1958)}]{Marshak_POF_1958}%
  \BibitemOpen
  \bibfield  {author} {\bibinfo {author} {\bibfnamefont {R.~E.}\ \bibnamefont
  {Marshak}},\ }\href@noop {} {\bibfield  {journal} {\bibinfo  {journal} {Phys.
  Fluids}\ }\textbf {\bibinfo {volume} {1}},\ \bibinfo {pages} {24} (\bibinfo
  {year} {1958})}\BibitemShut {NoStop}%
\bibitem [{\citenamefont {Gear}(1971)}]{Gear_book}%
  \BibitemOpen
  \bibfield  {author} {\bibinfo {author} {\bibfnamefont {C.~W.}\ \bibnamefont
  {Gear}},\ }\href@noop {} {\emph {\bibinfo {title} {Numerical
  {I}nitial-{V}alue {P}roblems in {O}rdinary {D}ifferential {E}quations}}}\
  (\bibinfo  {publisher} {Prentice-{H}all},\ \bibinfo {address} {Englewood
  Cliffs, NJ},\ \bibinfo {year} {1971})\BibitemShut {NoStop}%
\bibitem [{\citenamefont {Radhakrishnan}\ and\ \citenamefont
  {Hindmarsh}(1993)}]{lsode_1993}%
  \BibitemOpen
  \bibfield  {author} {\bibinfo {author} {\bibfnamefont {K.}~\bibnamefont
  {Radhakrishnan}}\ and\ \bibinfo {author} {\bibfnamefont {A.~C.}\ \bibnamefont
  {Hindmarsh}},\ }\href@noop {} {\emph {\bibinfo {title} {Description and use
  of {\textsc{lsode}}, the {L}ivermore solver for ordinary differential
  equations}}},\ \bibinfo {type} {NASA Report}\ \bibinfo {number} {1327}\
  (\bibinfo {year} {1993})\BibitemShut {NoStop}%
\bibitem [{\citenamefont {Auer}(1967)}]{Auer_ApJ_1967}%
  \BibitemOpen
  \bibfield  {author} {\bibinfo {author} {\bibfnamefont {L.~H.}\ \bibnamefont
  {Auer}},\ }\href@noop {} {\bibfield  {journal} {\bibinfo  {journal}
  {Astrophys. J.}\ }\textbf {\bibinfo {volume} {150}},\ \bibinfo {pages} {L53}
  (\bibinfo {year} {1967})}\BibitemShut {NoStop}%
\bibitem [{\citenamefont {Belozerov}\ and\ \citenamefont
  {Measures}(1969)}]{Belozerov_JFM_1969}%
  \BibitemOpen
  \bibfield  {author} {\bibinfo {author} {\bibfnamefont {A.~N.}\ \bibnamefont
  {Belozerov}}\ and\ \bibinfo {author} {\bibfnamefont {R.~M.}\ \bibnamefont
  {Measures}},\ }\href@noop {} {\bibfield  {journal} {\bibinfo  {journal} {J.
  Fluid Mech.}\ }\textbf {\bibinfo {volume} {36}},\ \bibinfo {pages} {695}
  (\bibinfo {year} {1969})}\BibitemShut {NoStop}%
\bibitem [{\citenamefont {Pauling}\ and\ \citenamefont
  {Wilson~Jr.}(1935)}]{Pauling_book_1935}%
  \BibitemOpen
  \bibfield  {author} {\bibinfo {author} {\bibfnamefont {L.}~\bibnamefont
  {Pauling}}\ and\ \bibinfo {author} {\bibfnamefont {E.~B.}\ \bibnamefont
  {Wilson~Jr.}},\ }\href@noop {} {\emph {\bibinfo {title} {Introduction to
  {Q}uantum {M}echanics with {A}pplications to {C}hemistry}}}\ (\bibinfo
  {publisher} {Mc{G}raw-{H}ill},\ \bibinfo {address} {New York, NY},\ \bibinfo
  {year} {1935})\BibitemShut {NoStop}%
\end{thebibliography}
\end{document}